\documentclass[11pt]{article}
\usepackage{amsmath,amsfonts,graphics,epsfig}
\usepackage{latexsym}
\usepackage{graphicx}
\usepackage{amssymb}
\usepackage{amsmath}
\usepackage{amsfonts}
\usepackage{color}

\usepackage{url}
\usepackage{epstopdf}
\usepackage{times}
\usepackage{cite}
\usepackage{ifthen}
\usepackage{times}
\usepackage{tabularx}
\usepackage{tikz}
\usetikzlibrary{matrix}
\usepackage{cite}
\usepackage{ifthen}
\usepackage{times}
\usepackage{tabularx}
\usepackage{epstopdf}
\usepackage{bm}
\newcommand{\hide}[1]{\ifthenelse{\boolean{false}}{#1}{}}

\newtheorem{assmp}{Assumption}[section]
\newtheorem{tdef}{Definition}[section]
\newcommand{\barr}{\begin{array}}
\newcommand{\earr}{\end{array}}

\newcommand{\benum}{\begin{enumerate}}
\newcommand{\eenum}{\end{enumerate}}

\newcommand{\bit}{\begin{itemize}}
\newcommand{\eit}{\end{itemize}}

\newcommand{\bdes}{\begin{description}}
\newcommand{\edes}{\end{description}}

\newcommand{\bfig}{\begin{figure}}
\newcommand{\efig}{\end{figure}}

\newcommand{\bemq}{\begin{quote} \begin{em}}
\newcommand{\eemq}{\end{em} \end{quote}}

\newcommand{\brac}[1]{\left({#1}\right)}

\newcommand{\cbrac}[1]{\left\{{#1}\right\}}

\newcommand{\ul}{\underline}

\newcommand{\ol}[1]{\overline{#1}}

\newcommand{\indic}[1]{I_{\cbrac{#1}}}

\newcommand{\given}{\arrowvert}

\newcommand{\ie}{{i.e.}}

\newcommand{\wrt}{w.r.t.}

\newcommand{\iid}{{i.~i.~d.}}

\newcommand{\EX}{\mathbb{E}} % expectation operator
\newcommand{\expect}[1]{\mathbb{E}\left[{#1}\right]}

\newcommand{\bt}{\begin{theorem}}
\newcommand{\bl}{\begin{lemma}}
\newcommand{\bc}{\begin{claim}}
\newcommand{\bp}{\begin{Proposition}}
\newcommand{\bcoro}{\begin{corollary}}
\newcommand{\bres}{\begin{Result}}
\newcommand{\brem}{\begin{Remark}}
\newcommand{\et}{\end{theorem}}
\newcommand{\el}{\end{lemma}}
\newcommand{\ec}{\end{claim}}
\newcommand{\ep}{\end{Proposition}}
\newcommand{\ecoro}{\end{corollary}}
\newcommand{\eres}{\end{Result}}
\newcommand{\erem}{\end{Remark}}

\newcommand{\beq}{\begin{equation}}
\newcommand{\eeq}{\end{equation}}

\newcommand{\UN}[1]{{\mathcal{V}}^{(N)}_{k}}

\newcommand{\norm}[1]{\|{#1}\|}
\newcommand{\abs}[1]{\left \vert {#1} \right \vert}
\newcommand{\mb}[1]{\mathbb{#1}}
\newcommand{\mf}[1]{\mathbf{#1}}
\newcommand{\mc}[1]{\mathcal{#1}}
\newcommand{\bs}[1]{\boldsymbol{#1}}

\newcommand{\be}{\begin{equation}}
\newcommand{\ee}{\end{equation}}

\newtheorem{thm}{Theorem}[section]

\newtheorem{prop}{Proposition}[section]
\newtheorem{lemma}{Lemma}[section]
\newtheorem{lem}{Lemma}[section]

\newtheorem{rem}{Remark}[section]

\newcommand{\tsquare}{\hfill \rule{3mm}{3mm}}

\allowdisplaybreaks
\begin{document}
%\setlength{\abovedisplayskip}{3pt}
%\setlength{\belowdisplayskip}{3pt}
%\abovedisplayshortskip
%\belowdisplayshortskip

\begin{center}
{\Large {\bf Insensitivity of the mean-field Limit 
of Loss Systems Under Power-of-d Routing}}
\end{center}
\vspace{1.5cm}

\begin{center}
{ \bf Thirupathiah VASANTAM\footnotemark[1]  Arpan MUKHOPADHYAY \footnotemark[2] and Ravi R. MAZUMDAR\footnotemark[3]
}

\end{center}
\vspace{1cm}

\noindent \footnotemark[1] Department of Electrical and Computer Engineering, University of Waterloo, Waterloo, ON N2L 3G1, Canada.
E-mail:  tvasantam@uwaterloo.ca

\noindent \footnotemark[2] EPFL, INFCOM LCA2, INF 014, Station 14, CH-1015, Lausanne, Switzerland. E-mail: arpan.mukhopadhyay@epfl.ch

\noindent \footnotemark[3] Department of Electrical and Computer Engineering, University of Waterloo, Waterloo, ON N2L 3G1, Canada.
 E-mail: mazum@uwaterloo.ca
\vspace{1cm}

\begin{center}
\today
\end{center}
\vspace{0.5cm}

\begin{abstract}
In this paper, we study large multi-server loss models under power-of-$d$ routing 
scheme when service time distributions are general with finite mean.  
Previous works have addressed the exponential service time case  when the number of servers goes to infinity giving rise to a mean field model. The fixed point of  limiting mean field equations (MFE) was shown to be 
insensitive to the service time distribution through simulation.
Showing insensitivity to general service time distributions has remained an open problem. Obtaining the MFE in this case poses a challenge due to the resulting Markov description of the system being in positive orthant as opposed to a finite chain in the exponential case. In this paper, we first obtain the MFE and then show that the MFE has a unique fixed point that coincides with the fixed point in the exponential case thus establishing insensitivity. The approach is via a measure-valued Markov process representation and the martingale problem to establish the mean-field limit. The techniques can be applied to other queueing models.

\end{abstract}
\vspace{0.5cm}

\vspace{0.3cm}

%\noindent{\bf Short-title:} Insensitivity of the mean-field for loss systems with power-of-d routing
\vspace{0.5cm}

\noindent{AMS Classification Primary: 60K35}{Secondary 60F10;60J10;62F15}
%\end{frontmatter}

\section{Introduction}
%\subsection{Background and motivation}

We consider a multi-server loss system consisting of $N$ large number of
parallel servers to which jobs arrive according to a Poisson process with rate $N\lambda$ and
the service times are generally distributed with finite mean.
Each server has the capacity to serve up to $C$ number of jobs simultaneously and there is no waiting room.
A central job dispatcher routes an incoming job to one of the servers where the processing
of the job begins immediately if the number of jobs that are already in progress is less than $C$
otherwise, the job gets blocked.
These models appear in practice in cloud computing systems such as Microsoft's Azure \cite{Azure} and Amazon EC2 \cite{AmazonEC2}.

Due to a tremendous growth in Internet applications and the move to externalize storage and computing,
cloud computing systems maintain a large number of parallel servers
to provide service to incoming jobs. In these systems,
jobs are virtual machines(VMs) that request resources such as processor power, I/O bandwidth, disk etc. from a server that is picked from a large set of 
servers. Whenever a job arrives, the central job dispatcher routes an incoming job request
to one of the servers where the job will be processed immediately if the requested amount of resources are
available otherwise it is blocked. The resources allocated to a job will be released once
the service of a job ends. In order to provide good quality of service, the service provider in cloud computing systems uses the routing policy at the job dispatcher that balances loads on servers 
which results in minimum average blocking probability.
In general, load balancing is an efficient method to optimally use
the resources of a system which results in better system performance.
In the large scale cloud computing systems that contain thousands of servers,
the traditional optimal load balancing schemes such as the join-the-shortest-queue (JSQ) results in large computational cost and complexity due to the need to maintain the states of all servers.  One way of overcoming this is by using randomized algorithms that are based on sampling a subset of servers and adopting 
a shortest-queue (SQ) policy amongst them. It has been shown that such algorithms are almost as good as JSQ.

The power-of-$d$ routing policy that routes incoming requests to the shortest of $d$ uniformly sampled servers was first introduced in \cite{Vvedenskaya_inftran_1996} for 
multi-server server systems with FCFS service discipline for the case of $d=2$ and exponential service times. The analysis of a finite $N$ system under the power-of-$d$ routing policy is a difficult problem due to dependence amongst the servers introduced by the random sampling, however using mean-field techniques when $N\to\infty$ provide a tractable way of characterizing the stationary distributions that are accurate when the number $N$ is large.  Indeed the analysis in \cite{Vvedenskaya_inftran_1996} is based on this idea. The results were then extended for the case of $d>2$ in \cite{mit} where it was argued that the case $d=2$ provides most of the gains and whence the term `The power-of-2'  came to be used.

Loss models similar to the one considered in this paper were analyzed in \cite{xie,arpan} under the assumption of exponential
service times for the power-of-$d$ routing policy. They considered the more general heterogeneous case with different server capacities and jobs routed to servers with maximum vacancy among $d$ randomly
chosen servers. It was shown that the power-of-$d$ routing scheme yields almost optimal blocking performance in that the average blocking is very close to the theoretical lower bound on the minimum average blocking achievable by any work conserving policy. 

\begin{figure}
\begin{center}
\begin{tikzpicture}
\centering
  \matrix (m) [matrix of math nodes,row sep=2cm,column sep=3cm,minimum width=3em] {
     \mf{x}^{(N)}(t) & \mf{x}(t) \\
     \mf{x}^{(N)}(\infty) & \bs{\pi} \\};
  \path[-stealth]
    (m-1-1) edge node [left] {$t \to \infty$}node[below,rotate=90] {step 3}(m-2-1)
            edge  node [above] {$N \to \infty$}node [below]{step 1} (m-1-2)
    (m-2-1.east|-m-2-2) edge node [below] {$N \to \infty$}node[above]{step 4}  (m-2-2)
    (m-1-2) edge node [right] {$t \to \infty$} node [above,rotate=90]{step 2}(m-2-2);
\end{tikzpicture}
\end{center}
\caption{Commutativity of limits}
\label{fig:limits}
\end{figure}
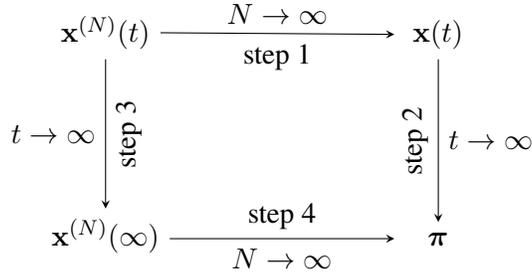

 The complete analysis of queuing systems under the power-of-$d$ routing policy using mean-field techniques can be summarized as in the Figure~\ref{fig:limits}. There are four steps in the complete analysis of the system. The first step is to establish the mean-field limit. In the exponential service time case, the system
 dynamics are first represented as a Markov process $\mf{x}^N(t)=(x^N_l(t),l\geq 0)$
 where $x_l^N(t)$ denotes the fraction of servers with at least $l$ jobs, and then as $N\to\infty$,
 the process  $(\mf{x}^N(t))_{t\geq 0}$ was shown to converge weakly to a system of ordinary differential equations that
 have unique solution called as the mean-field limit (MFE).
 The second step is to show the global stability of the mean-field limt $(\bm{x}(t,\mf{u}))_{t\geq 0}$ where $\mf{u}$ denotes the initial point of the mean-field \ie, $\bm{x}(0,\mf{u})=\bm{u}$. So far, in the literature, step two is shown only for the case when mean-field equations satisfy the quasi-monotonicity\cite{Vvedenskaya_inftran_1996}.  The quasi-monotonicity is described as follows. %Let $\bm{x}(t,\bm{u})$ denotes the mean-field limit with initial point $\bm{u}$ \ie, $\bm{x}(0,\mf{u})=\bm{u}$. 
 The quasi-monotonicity implies that if $\bm{u_1}\geq\bm{u_2}$ by element wise, then $\bm{x}(t,\bm{u_1})\geq\bm{x}(t,\bm{u_2})$ by element wise for every $t\geq 0$. Step two is very difficult to establish when mean-field equations do not satisfy the quasi-monotonicity property. The third step follows from ergodicity when system with finite $N$ servers is stable. The fourth step can be shown by combining step two, step three and Prohorov's theorem \cite{Billing}. The fourth step is crucial to use the fixed-point of the mean-field as an approximation to the steady-state distribution for server occupancies in a system with large $N$. Combining four steps, we have
 \beq
 \label{eq:exchange_limits}
 \lim_{N\to\infty}\lim_{t\to\infty}\mf{x}^N(t)=\lim_{t\to\infty}\lim_{N\to\infty}\mf{x}^N(t).
 \eeq

Further, for multi-server loss sytem with capacity $C$ for each server, if $\mf{P}^{(exp)}=(P_n^{(exp)},0\leq n\leq C)$ such that $P_0^{(exp)}=1$ and $P_n^{(exp)}\geq P_{n+1}^{(exp)}$, then for
exponential service times case, it was shown in \cite{xie,arpan} that the unique fixed-point of the mean-field is 
same as the
unique fixed-point of the mapping $\mf{\theta}(\mf{P}^{exp})$ defined by
\beq
\mf{\theta}(\mf{P}^{(exp)})=\mc{S}(\Lambda(\mf{P}^{(exp)}),\mf{P}^{(exp)}),
\eeq
where for $0\leq n\leq C$ with $P_{C+1}^{(exp)}=0$,
\beq
\Lambda(\mf{P}^{(exp)}):\,\,\,\lambda_n=\lambda\frac{((P_n^{exp})^d-(P_{n+1}^{(exp)})^d)}{(P_n^{(exp)}-P_{n+1}^{(exp)})} 
\eeq
and
\beq
\mc{S}(\Lambda,\mf{P}^{(exp)}): \,\,\,\, \lambda_n (P_n^{(exp)}-P_{n+1}^{(exp)})=(n+1)(P_{n+1}^{(exp)}-P_{n+2}^{(exp)})\mu.
\eeq

The exchange of limits as in equation~\eqref{eq:exchange_limits}, allows us to study the
impact of the power-of-$d$ routing policy by characterizing the fixed-point of the mean-field.
However, the exchange of limits in equation~\eqref{eq:exchange_limits} was established under
the assumption that the service times are exponential.
 In most realistic applications, the service time distributions are not exponential. For example, service times follow Log-normal distributions in call centers \cite{brown}, and  Gamma distributions
 in automatic teller machines (ATMs) \cite{peter} etc.

 For general service times case, the Markovian modeling of the system requires us to
 track the age or residual service time of each job that is in progress in the system.
 Therefore the underlying space on which the Markov process is defined
 is uncountable. This makes establishing the mean-filed limit and then establishing the exchange of limits
 in equation~\eqref{eq:exchange_limits} for general service times a challenging task.
 
 It is well known that the stationary distributions of single loss systems even with prespecified state-dependent arrival rates are insensitive to the service time distribution, i.e., they only depend on the means of the service times   \cite{shelby}. Hence, it is important to investigate insensitivity of large multi-server loss systems with general service time distributions under the power-of-$d$ routing policy where the servers are coupled for finite $N$. Insensitivity was observed in the simulations in \cite{xie,arpan} but there were no proofs provided. The first step is thus to establish the mean-field limit and analyze its equilibrium behavior.

 \subsection{Related Literature}
 \label{sec:related literature}
 %\vspace{0.3cm}
 
Randomized routing schemes were first investigated in \cite{azar} using balls-and-bins models. The power-of-$d$ scheme was considered for FCFS queues with exponential service time distributions in \cite{mit,Vvedenskaya_inftran_1996}. 
 It was shown that in the limiting system, the probability that a queue has atleast $k$ jobs is
 equal to $\lambda^{\frac{d^k-1}{d-1}}$ for $d\geq 2$ while it is equal to $\lambda^k$ for the
 case of $d=1$. This shows that the steady-state tail probabilities decrease double-exponentially with
 queue lengths for $d\geq 2$ whereas it is exponential decay for $d=1$.
 
  The significant improvement in system performance (in terms of buffer occupancy) for FCFS systems under the power-of-$d$ routing policy was also shown for processor sharing (PS) queues in \cite{Arpan_ITC_2014,Arpan_TCNS} when service times are exponentially distributed. In \cite{bramson1}, randomized routing schemes for queueing systems with general service time distributions when service disciplines are FCFS, PS, and LIFO were studied. The steady-state results were characterized by assuming  propagation of chaos (or asymptotic independence of servers) in the system. The propagation of chaos for FCFS systems is established in \cite{bramson2} for the case when service time distributions have decreasing hazard rate functions. In \cite{bramson1} the approach was to study the impact of the power-of-$d$ routing policy by characterizing
  the stationary distribution of the limiting system by considering step~$3$ and step~$4$ of Figure~\ref{fig:limits}. The mean-filed limit and its fixed-point were not studied.

 For general service times case  \cite{rybko} obtained the mean-field for symmetric closed queueing networks with FCFS service discipline
 that consist of $N$ queues and $M$ customers in which a customer that exits a queue joins a queue that is picked
 with probability $\frac{1}{N}$ from $N$ queues. The mean-field was established for the regime when $N\to\infty$, $M\to\infty$ such that $\frac{M}{N}\to\alpha$ using the convergence of infinitesimal generators of Markov processes that represent the system dynamics. However, the equilibrium behavior of the system is not studied. Recently, \cite{reza} considered a system of $N$ FCFS servers and jobs arrive according to a time-inhomogeneous Poisson process with rate $\lambda^{(N)}(\cdot)=a^{(N)}\lambda(\cdot)$ where $\lambda(\cdot)$ is a locally non-negative function with $\frac{a^{(N)}}{N}\to 1$ as $N\to\infty$. The mean-field limit is established for general service time distributions under the power-of-$d$ routing policy for all compact intervals of time. However, the steady-state results were not investigated. 
 
 Multi-server loss models under randomized routing schemes were first studied in \cite{Turner_thesis,Turner_choices_1998} when job lengths are exponentially distributed by using mean-field techniques. The mean-field equations were used to characterize the limiting system and the resulting tail distribution of server occupancies observed to decay rapidly even when there is a small number of routing choices for each arriving job. However, the existence and uniqueness of the fixed-point of the mean-field were not shown. In \cite{xie}, the existence and uniqueness of the fixed-point of the mean-field for homogeneous loss model of \cite{Turner_thesis} was addressed. The heterogeneous case was also treated in \cite{xie} under the asymptotic independence of servers ansatz. The propagation of chaos (or independence on path space) was studied earlier by \cite{Graham_loss,Graham1} in the context of alternate routing in circuit-switched networks.
 
 The complete analysis for heterogeneous loss models under the power-of-$d$ routing scheme when service times are exponential is given in \cite{arpan}. They showed the existence and uniqueness of the stationary point of the mean-field, as well as the global asymptotic stability of the mean-field. Further, the propagation of chaos was shown using intra-type exchangeability of random variables corresponding to server occupancies. The results were then extended to multi-class heterogeneous loss models in \cite{arpan2} where jobs belong to one of the several classes based on the amount of resources they use. All these works were based on the assumption that the job lengths are exponentially distributed. The study of large multi-server loss models with general service time distributions under power-of-$d$ routing and the proof of the insensitivity have not been addressed so far in the literature.

 There is a close connection between the mean-field analysis and the fluid analysis of queues. The fluid limit analysis of complex queuing systems with general service time distributions
 was carried out by representing the system dynamics as a measure-valued process and then the fluid limit
 was established by showing the convergence of measure-valued processes using the theory developed by Dawson in \cite{dawson}. In representing the system dynamics as a measure-valued process, either ages or residual service times of jobs can be used. Fluid limit analysis of different queuing systems using residual times can be found for heavily loaded processor sharing queues in \cite{gromoll_ps}, processor sharing queues with impatient customers in \cite{gromoll_ps_impatient}, M/GI/$\infty$ queue in \cite{moyal}, many-server queues with abandonment under FCFS service discipline in \cite{zhang} etc. Fluid limit analysis using the ages of jobs to construct measure-valued processes can be found for many-server queues with FCFS service discipline in \cite{kaspi2011}, many-server queues with reneging in \cite{kang2010} etc.  In this paper, we use ages of jobs to construct the measure-valued Markov processes that represent the system dynamics and we establish the mean-field limit following the ideas in \cite{moyal,dawson}.

\subsection{Contributions and Organization of the paper}

In this paper, we obtain and show that the mean-field for the {\em power-of-d} routing loss systems is well defined and we characterize the fixed-point or equilibrium of the mean field equation. In particular, we show that the fixed-point is unique and moreover coincides with the fixed point of the MFE in the exponential case. This establishes the insensitivity of the fixed point. In order to interpret the fixed point as the stationary distribution of the limiting model requires us to show that it is a globally asymptotically stable (GAS) equilibrium for the MFE. It appears very difficult to establish this and in the last section, we provide numerical evidence to show that it indeed seems to be true. We thus conjecture that this is true. In which case the results would then establish the insensitivity of the stationary distributions of the limiting loss system (system with $N\to\infty$) to service time distributions.

The rest of the paper is organized as follows: Section~\ref{sec:system model} describes the system model and the power-of-$d$ policy.  In Section~\ref{sec:math}, we introduce the notation and construct various measure spaces required for the analysis. In Section~\ref{sec:state}, we provide a measure-valued representation for the state of the system. The mean-field equations are given in Section~\ref{sec:MFE}. The detailed proofs then follow in Sections~\ref{sec:prelim} to \ref{sec:MFEproof}. In Section~\ref{sec:insensitivity}, we then prove the main result on the uniqueness and characterization of the fixed point of the MFE thus showing that the fixed point is insensitive to the distribution and only depends on the mean service time. In Section~\ref{sec:extensions}, we state the generalization of the results we have obtained to systems with heterogeneous servers. The following Section~\ref{sec:numerics} provides some evidence of the global asymptotic stability of the equilibrium of the MFE that indicates that Step~4 of the commutative diagram is indeed true. Finally, we close with some remarks and observations in Section~\ref{sec:conclusion}.

%%%%%%%%%%%%%%%%%%%%%%%%%%%%%%%%%%%%%%%%%%%%%%%%%%%%%%%%%%%%%%%%%%%%%%%%%%%%%%%%%%%%%%%%%%%%
%%%%%%%%%%%%%%%%%%%%%%%%%%%%%%%%%%%%%%%%%%%%%%%%%%%%%%%%%%%%%%%%%%%%%%%%%%%%%%%%%%%%%%%%%%%%%%%
%%%%%%%%%%%%%%%%%%%%%%%%%%%%%%%%%%%%%%%%%%%%%%%%%%%%%%%%%%%%%%%%%%%%%%%%%%%%%%%%%%%%%%%%%%%%%%%%%%

%##################################################################################
%%##################### System model and Routing policy###################################
\section{System model and the routing policy}

%$$$$$$$$$$$$$$$$$$$$$$$$$$$$$$$System model$$$$$$$$$$$$$$$$$$$$$$$$$$$$$$$$$$$$$$$$$$$$$
\label{sec:system model}
 We consider a system consisting of $N$ large number of parallel
servers that provide service to an incoming sequence of jobs arriving according to a Poisson process with
rate $N\lambda$. The incoming jobs are routed to servers based on the predetermined routing policy implemented at the central job dispatcher. Further, each server is assumed to have capacity to serve up to $C$ number of jobs simultaneously and has no waiting room.
At any time $t$, if a server is currently serving $i \leq C$ jobs, 
then we say that the server has occupancy $i$ and vacancy $C-i$ at time $t$.
If an incoming job is routed to a server with occupancy $C$, then the job is blocked
otherwise the processing of the job begins immediately. 

 Recently it was shown that the power-of-$d$ routing scheme achieves the performance (in terms of the average blocking) close to 
the optimal performance achievable by any work conserving strategy but with much less computational cost \cite{xie,arpan}. We recall the power-of-$d$
routing policy which is the focus of this paper.\\

\begin{tdef}{Power-of-d routing:}
\label{def:power_of_d}
An incoming job is routed to a server with minimum occupancy among $d$ randomly
chosen servers. Ties among servers are broken by choosing a server uniformly at random.
The randomly chosen $d$ servers are called as the potential destination servers and the server
to which a job is routed is called as the destination server.
\end{tdef}

In this paper, we assume the service times are generally distributed with finite mean $\frac{1}{\mu}$ and the central job dispatcher routes an incoming job according to the power-of-$d$ policy. The service requirements of customers
form an $\iid$ sequence with distribution function $G(\cdot)$ on $[0,\infty)$ and
the density function is $g(\cdot)$.
% We also assume that there exists $B_g>0$ such that
%\beq
%\label{eq:pdf_bound}
%sup_{x\in\mc{R}_+}\abs{g(x)}\leq B_g.
%\eeq

The hazard rate function of $G(\cdot)$ is denoted by $\beta(\cdot)$ satisfying $\beta(x)=\frac{g(x)}{\ol{G}(x)}= \frac{g(x)}{1-G(x)}$
for $x\in\mc{R}_+$.
% We assume that there exists $\norm{\beta}>0$ such that 
%\beq
%\label{eq:hazard_bound}
%sup_{x\in\mc{R}_+}\abs{\beta(x)}\leq \norm{\beta}.
%\eeq
Note that the hazard rate function $\beta$ indicates the instantaneous rate at which the
service of a job ends. More precisely, a job with age $y$ (where $y$ denotes the time since its arrival) at time $t$ exits the server in the 
interval $(t,t+dt]$ with probability $\beta(y)dt$. 

\begin{assmp}
The hazard rate function $\beta$ satisfies
\beq
\beta\in\mc{C}_b(\mc{R}_+).
\eeq
where $\mc{C}_b(\mc{R}_+)$ denotes the space of continuous bounded functions on positive real line $\mc{R}_+$
\end{assmp}
%***************************************************************************************************
%****************************************************************************************************
%*******************************************************************************************************
%**********************************************************************************************************
\section{Mathematical framework}
\label{sec:math}
%$$$$$$$$$$$$$$$Notation and terminology$$$$$$$$$$$$$$$$$$$$$$$$$$
\subsection{Notation and terminology}
We begin by introducing the notation which is used throughout the paper. Let $\mathcal{Z}$, $\mathcal{R}$ indicate the set of integers and real numbers, respectively.  Further, let $\mathcal{Z}_+$, $\mathcal{R}_+$ indicate the set of nonnegative integers and nonnegative real numbers, respectively. 
%For $a,b\in\cal{R}$, we define $a\vee b$, $a\wedge b$ to denote the maximum of $a$ and $b$, the minimum of $a$ and $b$, respectively. Also, we use $a^+$ to denote $a\vee 0$. 
%%%%%%%%%%%%%%Function Spaces%%%%%%%%%%%%%%%%%%%%%%%%%%%%%%%%%%%%%
%%%%%%%%%%%%%%%%%%%%%%%%%%%%%%%%%%%%%%%%%%%%%%%%%%%%%%%%%%%%%%%%%%%%%%
\subsubsection{Function and measure spaces.}
We next define the function spaces that are used in the analysis. For any given metric space $\mc{E}$, we define $\mc{K}_{b}(\mc{E}),\mc{C}_{b}(\mc{E}),\mc{C}_{s}(\mc{E})$ to denote the space of bounded measurable real valued functions, the space of bounded continuous real valued functions, and the space of continuous real valued functions with compact support, defined on $\mc{E}$, respectively. Further, let the space of once continuously differentiable real valued functions defined on $\mc{E}$ be denoted by $\mc{C}^1(\mc{E})$ and the subspace of functions in $\mc{C}^1(\mc{E})$ which have compact support is denoted by $\mc{C}^1_{s}(\mc{E})$. The space of bounded functions in $\mc{C}^1(\mc{E})$ whose first derivatives are also bounded is denoted by $\mc{C}^1_{b}(\mc{E})$. %Let $f$ denotes the first derivative of $f\in\mc{C}^1(\mc{E})$.
 We then define,
 for any function $f\in\mc{K}_{b}(\mc{E})$, $h\in\mc{C}^1(\mc{E})$, 
 \begin{align}
 \norm{f}&=\sup_{x\in \mc{E}}\abs{f(x)}\\
  \norm{h}_1&=\norm{h}+\norm{h_{d}}
 \end{align}
 where $h_{d}$ is the first derivative of $h$. In particular, if $h:\mc{R}_+^n\mapsto\mc{R}$, then the $i^{\text{th}}$
 directional derivative denoted by $h_{(d,i)}$ is defined as
 \beq
 h_{(d,i)}(x_1,\ldots,x_n)=\frac{\partial{h(x_1,\ldots,x_n)}}{\partial x_i}
 \eeq and the first derivative $h_d=(h_{(d,i)},1\leq i\leq n)$ has the norm
 \begin{align}
 \norm{h_d}&=\sup_i\sup_{(x_1,\ldots,x_n)\in\mc{R}_+^n}\abs{\frac{\partial{h(x_1,\ldots,x_n)}}{\partial x_i}}\\
 &= \sup_i \norm{h_{(d,i)}}.
 \end{align}
 The space $\mc{C}_{b}(\mc{E})$ is equipped with the uniform topology, $\ie$, we say a sequence of functions $(f_n\in\mc{C}_b(\mc{E}),n\geq 1)$ converges to a function $f\in\mc{C}_b(\mc{E})$ if $\norm{f_n-f}\to 0$ as $n\to\infty$. On the other hand, the space $\mc{C}^1(\mc{E})$ is equipped with the topology induced by the norm $\norm{\cdot}_1$. For a function $f$ defined on $\mc{R}_+^n$,
 we define a function $f'$ such that
 \beq
 f'(x_1,\ldots,x_n)=\sum_{i=1}^n\frac{\partial f(x_1,\ldots,x_n)}{\partial x_i}.
 \eeq

For a given metric space $\mc{E}$, let the Borel $\sigma$-algebra be denoted by $\mc{B}(\mc{E})$. The space of finite non-negative measures on $\mc{E}$ is denoted by $\mc{M}_{F}(\mc{E})$. The measure value with respect to a measure $\nu\in\mc{M}_{F}(\mc{E})$ for a Borel set $B\in\mc{B}(\mc{E})$ is denoted by $\nu(B)$ and at a single element $y\in \mc{E}$ is denoted by $\nu(\{y\})$. The space of probability measures is denoted by $\mc{M}_{1}(\mc{E})$. Also, we define $\mc{M}_{1}^N(\mc{E})$ to denote the space of measures in $\mc{M}_{1}(\mc{E})$ that satisfy
\beq
\mc{M}_{1}^N(\mc{E})=\{\nu\in\mc{M}_{1}(\mc{E}):N\,\nu{(B)}\in\mc{Z}_+,\,\forall B\in\mc{B}(\mc{E})\}.
\eeq
Therefore $\mc{M}_{1}^N(\mc{E})$ is the set of all probability measures $\nu$ that have rational valued measure at every $B\in\mc{B}(\mc{E})$ with the denominator equal to $N$. The set of real valued continuous functions defined
on $\mc{M}_F(\mc{E})$ is denoted by $\mc{C}(\mc{M}_F(\mc{E}))$.
For any $\phi\in\mc{K}_b(\mc{E})$, $\nu\in\mc{M}_{F}(\mc{E})$, we define
\beq
\langle\nu,\phi\rangle=\int_{y\in \mc{E}}\phi(y)\nu(dy).
\eeq 
The space of measures $\mc{M}_{F}(\mc{E})$ is equipped with the weak topology according to which a sequence of measures $\nu_n\in\mc{M}_F(\mc{E})$ converge weakly to a measure $\nu\in\mc{M}_F(\mc{E})$ (denoted by $\nu_n\Rightarrow\nu$) if and only if 
\beq
\langle\nu_n,\phi\rangle\to\langle\nu,\phi\rangle\eeq for every $\phi\in\mc{C}_b(\mc{E})$ as $n\to\infty$. Note that the space of measures $\mc{M}_{F}(\mc{E})$ endowed with the weak topology is a Polish space when $\mc{E}$ is a Polish space. The Dirac measure with unit mass at $x\in \mc{E}$ is denoted by $\delta_x$.

%The vector elements in this paper are of the form $(n,l_1,\ldots,l_C)$ where $0\leq n\leq C$ and $l_i\in\mc{R}_+$.
 To model the dynamics of an Erlang loss system with capacity $C$ for each server as Markov process, we define the state of each server as $(n,a_1,a_2,\ldots,a_n)$ where $n$ denotes the number of
jobs that are in progress at the server and $a_i$ denotes the age of the $i^{\text{th}}$ job in progress. Recall the age of an active job is the time elapsed since its arrival. Therefore we define a metric space $\mc{U}$ such that it contains all the possible server states as elements, namely,
\beq
\mc{U}=\cup_{n=0}^C \mc{U}_n
\eeq 
where $\mc{U}_0=\{0\}$ and an element in $\mc{U}_n$ for $n\geq 1$ is of the form $(n,a_1,\ldots,a_n)$ where $1\leq n\leq C$ and $a_i\in\mc{R}_+$.
%\beq
%l_i\in
%\begin{cases}
%\mc{R}_+ & \text{if }  i\leq n\\
%\{0\}      & \text{Otherwise}.
%\end{cases}
%\eeq
We denote an element of the form $(n,u_1,\ldots,u_{n})\in\mc{U}_n$ by $\ul{u}$. Without loss of generality,
 we also write $\ul{u}\in\mc{U}_0$ to mean that $\ul{u}=0$.
 Further, for all $B\in\mc{B}(\mc{U})$, $\indic{B}$ denotes the indicator function of $B$, $\ie$,
\beq
%\label{eq:metric}
\indic{B}(\ul{u})
=
\begin{cases}
1 & \text{if }  \ul{u}\in B\\
0      & \text{otherwise}.
\end{cases}
\eeq
%\beq
%\indic{B}(\ul{u})=1
%\eeq. 
We define a function $\bm{1}$ that satisfies
\beq
\bm{1}(\ul{u})=1
\eeq
for all $\ul{u}\in\mc{U}.$
The measure $\nu$ restricted to $\mc{U}_0$ is a Dirac measure at $\{0\}$. We say that the measure $\nu$ is continuous at $\ul{x}\in \mc{U}_n$ for $n\geq 1$ if and only if $\nu(\{\ul{x}\})=0$. For any Borel measurable function $f$ that is defined on $\mc{U}$ which is integrable with respect to $\nu\in\mc{M}_F(\mc{U})$, we define
\begin{align}
\langle\nu,f\rangle&=\int_{\ul{y}\in \mc{U}}f(\ul{y})\nu(d\ul{y}) \nonumber  \\
%&=f^{(0)}(0)\nu{(\{0\})}+\sum_{i=1}^C\int_{x_1\in\mc{R}_+}\cdots\int_{x_i\in\mc{R}_+}f^{(i)}(x_1,\ldots,x_n)\,d\nu^{(i)}(x_1,\ldots,x_n),\\
&=f(0)\nu(\{0\})+\sum_{n=1}^C\int_{\ul{z}\in \mc{U}_n}f(\ul{z})\nu(d\ul{z}).
\end{align}

 %The value of $n$ in $\ul{u}\in\mc{U}$ is 
%indicated by mentioning $\ul{u}\in\mc{U}_n$.
%For given $\ul{u}=(n,u_1,\ldots,u_{n})\in\mc{U}$, we define
%\begin{align}
%\mc{N}(\ul{u})&=n,\\
%\mc{A}(\ul{u},i)&=u_{i}.
%\end{align}
%In our analysis, if a server has state $\ul{u}$, then $\mc{N}(\ul{u})$ is defined to indicate the number of
%jobs that are in progress at the server while $\mc{A}(\ul{u},i)$ denotes the age of the $i^{\text{th}}$
%progressing job. 

For $\ul{y}=(n,y_1,\ldots,y_{n}),\ul{z}=(m,z_1,\ldots,z_{m})\in \mc{U}$, we define the metric $d_{\mc{U}}(\ul{y},\ul{z})$ as
\beq
\label{eq:metric}
d_{\mc{U}}(\ul{y},\ul{z})
=
\begin{cases}
\sum_{i=1}^{n}\abs{y_i-z_i} & \text{if }  n=m\\
\infty      & \text{otherwise}.
\end{cases}
\eeq

For any function $f:\mc{U}\to\mc{R}$, we define a function $f^{(i)}$ for $0\leq i\leq C$ referred to as the $i^{\text{th}}$ component of the function $f$ as follows:
\beq
f^{(0)}:\mc{U}_0\mapsto\mc{R}
\eeq
such that
\beq
f^{(0)}(0)=f(0)
\eeq
and
\beq
f^{(i)}:\mc{R}_+^i\mapsto\mc{R}
\eeq
such that
\beq
f^{(i)}(x_1,\ldots,x_i)=f(i,x_1,\ldots,x_i).
\eeq 
Similarly, for any measure $\nu\in\mc{M}_{F}(\mc{U})$,
we define $i^{\text{th}}$ component of measure $\nu$ by $\nu^{(i)}$ such that 
\beq
\nu^{(0)}=\nu{(\{0\})}\delta_{(0)}
\eeq
and for $i\geq 1$, 
\beq
\nu^{(i)}(\{(x_1,\ldots,x_i)\})=\nu{(\{(i,x_1,\ldots,x_i)\})}.
\eeq
Therefore $\nu^{(n)}$ is a Borel measure defined on $(\mc{R}_+^n,\mc{B}(\mc{R}_+^n))$. Therefore, for any Borel measurable function $f$ that is defined on $\mc{U}$ which is integrable with respect to $\nu\in\mc{M}_F(\mc{U})$, we can write
\begin{align}
\langle\nu,f\rangle&=\int_{\ul{y}\in \mc{U}}f(\ul{y})\nu(d\ul{y}) \nonumber  \\
&=f^{(0)}(0)\nu^{(0)}{(\{0\})}+\sum_{i=1}^C\int_{x_1\in\mc{R}_+}\cdots\int_{x_i\in\mc{R}_+}f^{(i)}(x_1,\ldots,x_i)\,d\nu^{(i)}(x_1,\ldots,x_i)
%&=f(0)\nu(\{0\})+\sum_{n=1}^C\int_{\ul{z}\in \mc{U}_n}f(\ul{z})\nu(d\ul{z}).
\end{align}

We say $f$ is differentiable if each component $f^{(i)}$, $i\geq 1$ is
differentiable. For any $f:\mc{U}\mapsto\mc{R}$, we denote the first derivative by $f_d$ whose $i^{\text{th}}$ ($i\geq 1$) component is denoted by $f^{(i)}_d=(f_{(d,j)}^{(i)},1\leq j\leq i)$ where $f_{(d,j)}^{(i)},1\leq j\leq i)$ denotes the $j^{\text{th}}$ directional derivative of $f^{(i)}$ and we consider $f_{(d,0)}^{(0)}=0$ to be the first derivative of $f^{(0)}$. Note that from the definition of first derivative of a function, $f=\indic{\mc{U}_n}$, $n\geq 0$ is differentiable as each $f^{(i)}$, $i\geq 1$ is differentiable.
We define a function $\mc{I}:\mc{U}\mapsto\mc{R}$ as follows
\beq
\mc{I}(n,x_1,\ldots,x_n)=(x_1+\ldots+x_n)
\eeq
for $n\geq 1$ and $\mc{I}(0)=0.$
Hence, we have
\beq
\langle\nu,\mc{I}\rangle=\sum_{n=1}^C\int_{x_1}\cdots\int_{x_n}(x_1+\cdots+x_n)\,d\nu(n,x_1,\ldots,x_n).
\eeq

For any
$\ul{u}\in\mc{U}_n$, $n\geq 1$ and for $y>0$, we define
\beq
%begin{split}
 \tau_y^+(n,u_1,\ldots,u_n)
 =(n,u_1+y,u_2+y,\ldots,u_{n}+y)
 \eeq
 and
 \beq
 \tau_y^+(0)=(0).
 \eeq
% and 
% \beq
% \tau_y^+(\ul{0})=\ul{0}.
%\eeq
For any $y>0, f\in\mc{K}_b(\mc{U})$,  mapping $\tau_y:\mc{K}_b(\mc{U})\to \mc{K}_b(\mc{U})$ denotes,
%for $\ul{z}\in\mc{U}_n,\,n\geq 1$,
\beq
\begin{split}
\tau_yf(\ul{u})&=f(\tau_y^+\ul{u}).
%&=f(n-\sum_{i=1}^n\indic{z_i-y<0},(z_1-y)^+,(z_2-y)^+,\ldots,(z_{n}-y)^+,0,\ldots,0).
\end{split}
\eeq

 For $y>0$, we define a shifted measure $\tau_y\nu\in\mc{M}_F(\mc{U})$ such that for any Borel set $B\in\mc{B}(\mc{U})$, %equation~\eqref{eq:translation_function} such that
\beq
\tau_y\nu{(B)}=\nu(\tau_y^+(B)).
\eeq
For $\nu\in\mc{M}_F(\mc{U})$, the measure $\tau_y\nu\in\mc{M}_F(\mc{U})$  satisfies
\beq
\label{eq:translation_function}
\langle\tau_y\nu,f\rangle=\langle\nu,\tau_yf\rangle
\eeq
for all $f\in\mc{K}_b(\mc{U})$. Existence of the unique measure $\tau_y\nu$ satisfying equation~\eqref{eq:translation_function} follows from Riesz-Markov-Kakutani theorem \cite{Varad,rudin}.

%For any $y\in \mc{R}_+, f\in\mc{K}_b(\mc{U})$, we define
%\beq
%\tau_yf(\ul{z})=
%\begin{cases}
%f(n,z_1+y,z_2+y,\ldots,z_{n}+y)& \text{ for }\ul{z}\in\mc{U}_n,\,n\geq 1,\\
%f(0)   & \text{if } \ul{z}=0.
%\end{cases}
%\eeq 
%For $\nu\in\mc{M}_F(\mc{U})$, we denote a measure $\tau_y\nu\in\mc{M}_F(\mc{U})$ as the measure that satisfy
%\beq
%<f,\tau_y\nu>=<\tau_yf,\nu>
%\eeq
%for all $f\in\mc{K}_b(\mc{U})$.

%%%%%%%%%%%%%%%%%%%%%%%%%%%Measure valued stochastic processes%%%%%%%%
%%%%%%%%%%%%%%%%%%%%%%%%%%%%%%%%%%%%%%%%%%%%%%%%%%%%%%%%%%%%%%%%%%%%%%%%%%%%%

\subsubsection{Measure valued stochastic processes.}

For given Polish space $\mc{H}$, we denote the c\`adl\`ag\footnote{Also referred to as RCLL (right continuous with left limits).} functions that take values in $\mc{H}$ defined on $[0,T]$, $[0,\infty)$ by $\mc{D}_{\mc{H}}([0,T]),\mc{D}_{\mc{H}}([0,\infty))$, respectively. Similarly, we denote the continuous functions that take values in $\mc{H}$ defined on $[0,T]$, $[0,\infty)$ by $\mc{C}_{\mc{H}}([0,T]),\mc{C}_{\mc{H}}([0,\infty))$, respectively. The spaces $\mc{D}_{\mc{H}}([0,T])$,$\mc{D}_{\mc{H}}([0,\infty))$ are equipped with the Skorokhod $J_1$-topology and hence they are Polish spaces. The covariation between of two local martingales $(M^1_t)_{t\geq 0}$ and
$(M^2_t)_{t\geq 0}$ in $\mc{D}_{\mc{R}}([0,T])$ is denoted by $(<M^1,M^2>_t)_{t\geq 0}$ and the (quadratic) variation by 
$(<M^1>_t)_{t\geq 0}=(<M^1,M^1>_t)_{t\geq0}$.

In our analysis, we study $\mc{H}-$valued stochastic process where $\mc{H}=\mc{M}_F(\mc{U})$. The considered stochastic processes are random elements defined on $(\Omega,\mb{F},\mc{P})$ with sample paths in $\mc{D}_{\mc{H}}([0,\infty))$ and are equipped with the Borel $\sigma-$algebra generated by the open sets under the Skorokhod $J_1-$ topology \cite{Billing}. We say a sequence $\{X_n\}$ of $\mc{H}$-valued c\`adl\`ag processes defined on $(\Omega_n,\mb{F}_n,\mc{P}_n)$ converge in distribution to a $\mc{H}$-valued c\`adl\`ag process $X$ defined on $(\Omega,\mb{F},\mc{P})$ if, for every bounded, continuous, real valued functional $F:\mc{D}_{\mc{H}}:[0,\infty)\to\mc{R}$, we have
\beq
\lim_{n\to\infty}\EX_n(F(X_n))=\EX(F(X))
\eeq
where the expectation operators $\EX_n,\EX$ are defined with respect to $\mc{P}_n,\mc{P}$, respectively. We denote the convergence of $\{X_n\}$ in distribution to $X$ by $X_n\Rightarrow X$.

%*****************************************************************************************************
%******************************************************************************************************
%************************************************************************************************************
%**************************************************************************************************************
\section{State descriptor and system dynamics}
\label{sec:state}
For finite $N$, the evolution of the system is obtained by considering the state of each server to be $(n,a_1,\ldots,a_n)\in\mc{U}$ where $n$ denotes the number of jobs that are in progress and $a_i$ denotes the age of the $i^{\text{th}}$ job. Each server with state say $(n,a_1,\ldots,a_n)$ can be viewed as an atom with the given state. Therefore the system evolution can be considered as the evolution of the system with $N$ atoms where the interactions between atoms takes
place while implementing the power-of-$d$ policy when there is an arrival into the system.  
The age of a job that is in service at a server increases linearly with time until its service 
expires.

 We next describe
the possible state for a server at time $t+h$ ($h>0$) given that it has state $(n,a_1,\ldots,a_n)$ at time $t$ by assuming that atmost one event can occur in the interval $(t,t+h]$. In the interval $(t,t+h]$, if there is no arrival into the given server and there is no departure from the given server, then
the server state will be equal to $\tau_h^+(n,a_1,\ldots,a_n)$ at time $t+h$.
Further, if $i^{\text{th}}$ job
expires in the interval $(t,t+h]$, then the server state will be equal to $(n-1,a_1+h,\ldots,a_{i-1}+h,a_{i+1}+h,\ldots,a_n+h)$ at time $t+h$. Considering arrivals, suppose there
is an arrival into the server at time $t+h'$($0<h'\leq h$), then the arriving job
chooses its position uniformly at random out of $n+1$ possible positions and suppose it chooses $j^{\text{th}}$
position, then the server state will be equal to $(n+1,a_1+h,\ldots,a_{j-1}+h,h-h',a_j+h,\ldots,a_n+h)$ at time $t+h$.

Since servers are identical, to model the system evolution by a Markov process, we will show that it is enough to just keep track of the number of servers that lie
in each state $\ul{u}\in\mc{U}$. Precisely, the state descriptor of the system is denoted by
\beq
\label{eq:state_descriptor}
\eta_t^N=\sum_{i=1}^N\delta_{s_t(i)},
\eeq
where $s_t(i)\in\mc{U}$ denotes the state of server $i$ at time $t$. Note that the mass of
$\eta_t^N$ at a state $\ul{u}\in\mc{U}$ is equal to the number of servers with state
$\ul{u}$ at time $t$. Therefore, the mass at state $(n,y_1,\ldots,y_n)$ is 
given by 
\beq
\eta_t^N(\{(n,y_1,\ldots,y_n)\})=\langle\eta_t^N,\indic{(n,y_1,\ldots,y_n)}\rangle.
\eeq
Similarly, the number of servers having $n$ jobs in progress at time $t$
is given by
\beq
\eta_t^N(\mc{U}_n)=\langle\eta_t^N,\indic{\mc{U}_n}\rangle.
\eeq

We next describe the dynamics of $(\eta_t^N)_{t\geq 0}$ over time $t$. Suppose at time $t$, the measure $\eta_t^N$ is given by
\beq
\eta_t^N=\sum_{i=1}^N\delta_{(n_i,a_{i1},\ldots,a_{in_i})}.
\eeq
If there is no arrival into the system or departure from the system in the interval $(t,t+h]$, then
the mass with respect to ($\wrt$) the measure $\eta_t^N$ at any state $(m,y_1,\ldots,y_m)\in\mc{U}$ will be equal to the
mass at $(m,y_1+h,\ldots,y_m+h)$ $\wrt$ the measure $\eta_{t+h}^N$. If there is a departure in the interval $(t,t+h]$ from
a server with state $(m,y_1,\ldots,y_m)$ at time $t$ and the job at position $j$ departs,
then we have
\beq
\eta_{t+h}^N(\{(m,y_1+h,\ldots,y_m+h)\})=\eta_{t}^N(\{(m,y_1,\ldots,y_m)\})-1,
\eeq
%
%
%
%
%
%
%
%
%
%
%
%
%\vspace{-10mm}
\begin{multline}
\eta_{t+h}^N(\{(m-1,y_1+h,\ldots,y_{j-1}+h,y_{j+1}+h,\ldots,y_m+h)\})\\
=\eta_{t}^N(\{(m-1,y_1,\ldots,y_{j-1},y_{j+1},\ldots,y_m)\})+1
\end{multline}
and for all other states of the form $\ul{u}=(r,l_1,\ldots,l_r)\in\mc{U}$ such that $\ul{u}\neq (m,y_1,\ldots,y_m)$ and
$\ul{u}\neq (m-1,y_1,\ldots,y_{j-1},y_{j+1},\ldots,y_m)$, we have
\beq
\eta_{t+h}^N(\{(r,l_1+h,\ldots,l_r+h)\})=\eta_t^N(\{(r,l_1,\ldots,l_r)\}).
\eeq
On the other hand, when there is an arrival into the system at time $t+h'$ ($0< h'\leq h$) and suppose 
the arriving job occupies position $j$ at a server that had state $(m,y_1,\ldots,y_m)$ at time $t$,
then we have
%\begin{align}
\beq
\eta_{t+h}^N(\{(m,y_1+h,\ldots,y_m+h)\})=\eta_{t}^N(\{(m,y_1,\ldots,y_m)\})-1,
\eeq
%\begin{align}
%\vspace{-10mm}
\beq
%\begin{multline}
\eta_{t+h}^N(\{(m+1,y_1+h,\ldots,y_{j-1}+h,h-h',y_{j}+h,y_{j+1}+h,\ldots,y_m+h)\})=1%\\
%=\eta_{t}^N(\{(m+1,y_1,\ldots,y_{j-1},-h',y_{j},\ldots,y_m)\})+1
%\end{multline}
\eeq
%\end{align}
and for all other states of the form $\ul{u}=(r,l_1,\ldots,l_r)\in\mc{U}$ such that $\ul{u}\neq (m,y_1,\ldots,y_m)$, we have,
% and
%$\ul{u}\neq (m+1,y_1,\ldots,y_{j-1},-h',y_{j},\ldots,y_m)$, 
\beq
\eta_{t+h}^N(\{(r,l_1+h,\ldots,l_r+h)\})=\eta_t^N(\{(r,l_1,\ldots,l_r)\}).
\eeq
%Note that $\eta_{t}^N(\{(m+1,y_1,\ldots,y_{j-1},-h',y_{j},\ldots,y_m)\})=0$ for $h'>0$ as the age of a job is nonnegative.
Further, it is easy to see that $\eta_t^N\in\mc{M}_F(\mc{U})$, $(\eta_t^N)_{t\geq 0}\in D_{\mc{M}_F(\mc{U})}([0,\infty)$ and $\eta_t^N(\mc{U})=N$ for all $t\geq 0$.
%%%%%%%%%%%%%%%%%%%%%%%%%%%%%%%%%%%%%%%%%%%%%%%%%%%%%%%%%%%%%%%%%%%%%%%%%%%%%%%%%%%%%%%%%%%%%%%%%%%%%
%%%%%%%%%%%%%%%%%%%%%%%%%%%%%%%%%%%%%%%%%%%%%%%%%%%%%%%%%%%%%%%%%%%%%%%%%%%%%%%%%%%%%%%%%%%%%%%%%%%%%
%%%%%%%%%%%%%%%%%%%%%%%%%%%%%%%%%%%%%%%%%%%%%%%%%%%%%%%%%%%%%%%%%%%%%%%%%%%%%%%%%%%%%%%%%%%%%%%
\section{Mean-field model}
\label{sec:MFE}

In this section we introduce the mean-field model for the system  and state our main results.
In this paper, we study a sequence of systems indexed by $N$ such that a system with
index $N$ has $N$ servers in which jobs arrive according to a Poisson process with rate $N\lambda$
and all other system parameters are identical for all $N$. For given $N$, the process
$(\eta_t^N)_{t\geq 0}$ defined in equation~\eqref{eq:state_descriptor} describes the system dynamics
of a system with index $N$ such that $\eta_t^N(\{\ul{u}\})$ denotes the number of servers lying in
state $\ul{u}$ at time $t$. Our aim is to characterize the limit of the normalized  process $(\ol{\eta}_t^N)_{t\geq 0}$ defined as follows
\beq
%\label{eq:normalized_descriptor}
\ol{\eta}_t^N=\frac{\eta_t^N}{N}.
\eeq

For given system parameters $\lambda,C,d$ and the probability density function $g(\cdot)$ of the  
service time distributions, for analysis purpose, we first define the mean-field model $(\ol{\eta}_t,t\geq 0)$ for the system in Definition~\ref{def:mf_model} and we then show that there exists unique mean-field model solution.  The mean-field model that we define acts as a fluid limit of the 
measure-valued state descriptors $(\eta_t^N,t\geq 0)$ under law of large numbers scaling. Precisely, we show that every  limit point of the sequence of the processes $(\ol{\eta}_t^N)_{t\geq 0}$
has almost surely continuous sample paths that coincide with the unique mean-field model solution.

\textit{Mean-field model:}\\
The dynamics of the mean-field model
$(\ol{\eta}_t,t\geq 0)$ are described by using the set of evolution equations for the real valued process $\langle \ol{\eta}_t,f\rangle_{t\geq 0}$, for all $f\in\mc{C}_b^1(\mc{U})$,
referred to as the mean-field model equations. %By using the mean-field model equations, we then state the result that there
%exists unique solution for mean-field model equations termed as the mean-field model solution.

\begin{tdef}{Mean-field model solution:}
\label{def:mf_model}
 A mean-field model solution for the given system parameters $(\lambda,C,d,g(\cdot))$ is a function $\ol{\eta}:[0,\infty)\mapsto \mc{M}_1(\mc{U})$ that satisfy
%\beq
\begin{enumerate}
\label{def:mf_model}
\item \label{property1}The mapping $t \mapsto \ol{\eta}_t$ is a continuous mapping. This is equivalent to
the continuity of
the mapping $t\mapsto\langle \ol{\eta}_t,\phi\rangle$ for all $\phi\in\mc{C}_b(\mc{U})$ since $\mc{C}_b(\mc{U})$ is a separating class\cite[p. 111]{Ethier_Kurtz_book}.
\item\label{property2} For $\phi\in \mc{C}_b^1(\mc{U})$, the process $(\ol{\eta}_t,t\geq 0)$ satisfies
\begin{multline}
\label{eq:mf_model_eqns}
\langle\ol{\eta}_t,\phi\rangle=\langle\ol{\eta}_0,\phi\rangle+\int_{s=0}^t \langle\ol{\eta}_s,\phi'\rangle\,ds\\
-\int_{s=0}^t\Bigg(\sum_{n=1}^C\sum_{j=1}^n\int_{x_1}\cdots\int_{x_n}\beta(x_j)\\
\times\left(\phi(n-1,x_1,\ldots,x_{j-1},x_{j+1},\ldots,x_n)-\phi(n,x_1,\ldots,x_n)\right)\,d\ol{\eta}_s(n,x_1,\ldots,x_n)\\
+\lambda\bigg[\left(\ol{\eta}_s(\{0\})\frac{(\ol{R}_0(\ol{\eta}_s)^d-\ol{R}_1(\ol{\eta}_s)^d)}{(\ol{R}_0(\ol{\eta}_s)-\ol{R}_1(\ol{\eta}_s))}\left(\phi(1,0)-\phi(0)\right)\right)
+\sum_{n=1}^{C-1}\sum_{j=1}^{n+1}\int_{x_1}\cdots\int_{x_n}\frac{1}{(n+1)}\\
\times\frac{(\ol{R}_n(\ol{\eta}_s)^d-\ol{R}_{n+1}(\ol{\eta}_s)^d)}{(\ol{R}_n(\ol{\eta}_s)-\ol{R}_{n+1}(\ol{\eta}_s))}
(\phi(n+1,x_1,\ldots,x_{j-1},0,x_j,\ldots,x_n)-\phi(n,x_1,\ldots,x_n))\\
\times\,d\ol{\eta}_s(n,x_1,\ldots,x_n)\bigg]\Bigg)ds,
\end{multline}
\end{enumerate}
where $\ol{R}_j(\ol{\eta}_s)=\sum_{n=j}^C\ol{\eta}_s(\mc{U}_n)$.
%\eeq
\end{tdef}
The equation~\eqref{eq:mf_model_eqns} defined for each $\phi\in\mc{C}_b^1(\mc{U})$ is referred to as the mean-field model equation.% By using the properties \ref{property1}-\ref{property2} of the mean-field model, we prove that there exists unique
%mean-field model solution.

The mean-field model equation~\eqref{eq:mf_model_eqns} is defined for class of functions $\phi\in\mc{C}_b^1(\mc{U})$, however, since one would be more interested to understand the fluid limit approximation of the process $(\langle\ol{\eta}_t^N,\indic{B}\rangle, t\geq 0)$ for a Borel set $B\in\mc{B}(\mc{U})$, it would be more useful to obtain the evolution equations for the real valued process of type $(\langle\ol{\eta}_t,\indic{B}\rangle, t\geq 0)$. In this direction, we first
obtain the evolution equations for the real valued process $(\langle\ol{\eta}_t,\psi\rangle,t\geq 0)$ where $\psi\in\mc{C}_b(\mc{U})$.
We later obtain the evolution equations for the process $(\langle\ol{\eta}_t,\psi\rangle,t\geq 0)$ for class of functions $\psi$ that also include functions of type $\indic{B}$
for some
$B\in\mc{B}(\mc{U})$.

\begin{lem}
\label{thm:mf_model_new}
A process $(\nu_t\in\mc{M}_1(\mc{U}),t\geq 0)$ which is a continuous function of $t$ satisfies the
mean-field model equation~\eqref{eq:mf_model_eqns} if and only if it satisfies the equation,
for all $\phi\in\mc{C}_b(\mc{U})$,
\begin{multline}
\label{eq:mean_field_model_new}
\langle\nu_t,\phi\rangle=\langle\nu_0,\tau_t\phi\rangle+\int_{r=0}^t\Bigg(\sum_{n=1}^C\sum_{j=1}^n\int_{x_1}\cdots\int_{x_n}\beta(x_j)\\
\times\left(\tau_{t-r}\phi(n-1,x_1,\ldots,x_{j-1},x_{j+1},\ldots,x_n)-\tau_{t-r}\phi(n,x_1,\ldots,x_n)\right)\,d\nu_r(n,x_1,\ldots,x_n)\\
+\lambda\bigg[\left(\nu_r(\{0\})\frac{(\ol{R}_0(\nu_r)^d-\ol{R}_1(\nu_r)^d)}{(\ol{R}_0(\nu_r)-\ol{R}_1(\nu_r))}\left(\tau_{t-r}\phi(1,0)-\tau_{t-r}\phi(0)\right)\right)
+\sum_{n=1}^{C-1}\sum_{j=1}^{n+1}\int_{x_1}\cdots\int_{x_n}\frac{1}{(n+1)}\\
\times\frac{(\ol{R}_n(\nu_r)^d-\ol{R}_{n+1}(\nu_r)^d)}{(\ol{R}_n(\nu_r)-\ol{R}_{n+1}(\nu_r))}
(\tau_{t-r}\phi(n+1,x_1,\ldots,x_{j-1},0,x_j,\ldots,x_n)-\tau_{t-r}\phi(n,x_1,\ldots,x_n)),\\
\times\,d\nu_r(n,x_1,\ldots,x_n)\bigg]\Bigg)\,dr,
\end{multline}
where $\ol{R}_j(\nu_s)=\sum_{n=j}^C\nu_s(\mc{U}_n)$.
\end{lem}

Using equation~\eqref{eq:mean_field_model_new}, we next state a result that shows that starting with an initial measure $\nu_0$, for
$t\geq 0$, there exists unique measure $\nu_t\in\mc{M}_1(\mc{U})$ satisfying equation~\eqref{eq:mf_model_eqns}.

Since for $\nu\in\mc{M}_F(\mc{U})$, $\langle \nu,\phi\rangle$ is a continuous linear operator on the space
of functions $\phi\in\mc{C}_b(\mc{U})$,
we define
\beq
\norm{\nu}=\sup_{\phi\in\mc{C}_b(\mc{U})}\frac{\abs{\langle \nu,\phi\rangle}}{\norm{\phi}}.
\eeq

\begin{thm}
\label{thm:mf_model_unique}

 There exists unique solution in $\mc{C}_{\mc{M}_1(\mc{U})}([0,\infty))$ satisfying the
mean-field model equations. In particular, if $(\nu_t^1,t\geq 0)$ and $(\nu_t^2,t\geq 0)$
are two mean-field model solutions starting at initial measures $\nu_0^1\in\mc{M}_1(\mc{U})$,\,$\nu_0^2\in\mc{M}_1{\mc{(U)}}$,
respectively, then
 \beq
 \norm{\nu_t^1-\nu_t^2}\leq \norm{\nu_0^1-\nu_0^2}\,e^{(2C\norm{\beta}+8d^2\lambda)t}.
 \eeq
\end{thm}

\textit{Mean-field limit:}\\
We next state the results on convergence of sequence of processes $(\ol{\eta}_t^N,t\geq 0)$. For this, we first make the
following assumption:
\begin{assmp}
 \label{assum:initial_measures}
The sequence of initial measures of the normalized measure-valued processes $(\ol{\eta}_t^N,t\geq 0)$ satisfy
\beq
\ol{\eta}_0^N\Rightarrow \bm{\Theta}
\eeq 
where $\Theta$ is a random measure taking values in $\mc{M}_1(\mc{U})$.
\end{assmp}

\begin{thm}
\label{thm:mean_field_limit}
If the sequence of processes $(\ol{\eta}^N_t,t\geq 0)$ satisfy the assumption~\ref{assum:initial_measures}, then
we have $\ol{\eta}^N\Rightarrow\ol{\eta}$. The process $(\ol{\eta}_t,t\geq 0)$ is referred to as the mean-field
limit that has sample paths almost surely coinciding with the unique mean-field model solution.
\end{thm}

\begin{rem}
For any closed or open subset $B\in \mc{U}$, once we have $\ol{\eta}^N\Rightarrow\ol{\eta}$, if $\ol{\eta}_t$
is absolutely continuous $\wrt$ Lebesgue measure for every $t\geq 0$, then continuous mapping theorem implies that $\langle \ol{\eta}^N,\indic{B}\rangle \Rightarrow \langle\ol{\eta},\indic{B}\rangle$. This shows that for large $N$, the fluid limit approximation of
$\langle \ol{\eta}^N,\indic{B}\rangle$ is given by $\langle\ol{\eta},\indic{B}\rangle$.
\end{rem}

\textbf{Insensitivity:}\\
Before stating the results on the insensitivity of the fixed-point of the mean-field, we first recall the dynamics of probabilities of server occupancies of a single server Erlang loss system
where jobs arrive according to a Poisson process with pre-specified state-dependent arrival rates. We observe an analogy
between the mean-field equations of the considered multi-server Erlang loss system under power-of-$d$ routing policy
and the single server system dynamics. We use this in proving the uniqueness of the fixed-point of the mean-field.

Consider a single server system with capacity $C$ where jobs arrive according to a Poisson process at rate $\alpha_n$ when there are $n$
jobs in service in the system. The service times are generally distributed as considered in the system model. It can be verified that the Kolmogorov equations are given by,
for $\phi\in\mc{C}_b^1(\mc{U})$, let $\nu_t^{(single)}$ denotes the probability measure for server occupancies at time $t$,
then
\begin{multline}
\label{eq:single_server_dynamics}
\langle\nu_t^{(single)},\phi\rangle=\langle\nu_0^{(single)},\phi\rangle+\int_{s=0}^t \langle\nu^{(single)}_s,\phi'\rangle\,ds\\
-\int_{s=0}^t\Bigg(\sum_{n=1}^C\sum_{j=1}^n\int_{x_1}\cdots\int_{x_n}\beta(x_j)\\
\times\left(\phi(n-1,x_1,\ldots,x_{j-1},x_{j+1},\ldots,x_n)-\phi(n,x_1,\ldots,x_n)\right)\,d\nu^{(single)}_s(n,x_1,\ldots,x_n)\\
+\bigg[\left(\alpha_0\nu^{(single)}_s(\{0\})\left(\phi(1,0)-\phi(0)\right)\right)
+\sum_{n=1}^{C-1}\sum_{j=1}^{n+1}\int_{x_1}\cdots\int_{x_n}\frac{1}{(n+1)}\\
\times\alpha_n
(\phi(n+1,x_1,\ldots,x_{j-1},0,x_j,\ldots,x_n)-\phi(n,x_1,\ldots,x_n))\\
\times\,d\nu^{(single)}_s(n,x_1,\ldots,x_n)\bigg]\Bigg)ds.
\end{multline}
%\end{enumerate}
%where $\ol{R}_j(\ol{\eta}_s)=\sum_{n=j}^C\ol{\eta}_s(\mc{U}_n)$.
On comparing mean-field equation~\eqref{eq:mf_model_eqns} with single server Kolomogorov equation~\eqref{eq:single_server_dynamics}, it is clear that both
the dynamics are similar except that $\alpha_i$ in equation~\eqref{eq:single_server_dynamics} is replaced by
$\lambda \frac{(\ol{R}_n(\ol{\eta}_s)^d-\ol{R}_{n+1}(\ol{\eta}_s)^d)}{(\ol{R}_n(\ol{\eta}_s)-\ol{R}_{n+1}(\ol{\eta}_s))}$ when
the probability measure for server occupancies is $\ol{\eta}_s$ at time $s$. This shows that equation~\eqref{eq:single_server_dynamics} represents the evolution of a linear Markov process whereas equation~\eqref{eq:mf_model_eqns} represents the evolution of a non-linear Markov process. 

Furthermore, let the Radon-Nikodym derivative of the measure $\nu_t^{(single)}$ at $\ul{u}\in\mc{U}$ be denoted by $p_t^{(single)}(\ul{u})$.
Then by using differential equations that represent the dynamics of the density function $p_t^{(single)}=(p^{(single)}_t(\ul{u}),\ul{u}\in\mc{U})$ that can be derived by following the analysis in \cite{sevastyanov}, the differential equations
for the process $P^{(single)}_t=(P^{(single)}_t(\ul{u}),\ul{u}\in\mc{U})$ where
\beq
P_t^{(single)}(n,y_1,\ldots,y_n)=\int_{x_1=0}^{y_1}\cdots\int_{x_n=0}^{y_n}p_t^{(single)}(n,x_1,\ldots,x_n)\,dx_1\cdots dx_n,
\eeq
are given by

\beq
\label{eq:single_pde1}
\frac{d P_t^{(single)}(0)}{dt}=\int_{y=0}^{\infty}\beta(y)\left(\frac{\partial P_t^{(single)}(1,y)}{\partial y}\right)\,dy-\alpha_0 P_t^{(single)}(0),
\eeq
for $1\leq n\leq C-1$,
\begin{multline}
\label{eq:single_pde2}
\frac{dP_t^{(single)}(n,y_1,\ldots,y_n)}{dt}=-\sum_{i=1}^n\frac{\partial P_t^{(single)}(n,y_1,\ldots,y_n)}{\partial y_i}\\
+\sum_{j=1}^{n+1}\int_{x_j=0}^{\infty}\beta(x_j)\left(\frac{\partial P_t^{(single)}(n+1,y_1,\ldots,y_{j-1},x_j,y_j,\ldots,y_n)}{\partial x_j}\right)\,dx_j\\
-\sum_{j=1}^n\int_{x_j=0}^{y_j}\beta(x_j)\left(\frac{\partial P_t^{(single)}(n,y_1,\ldots,y_{j-1},x_j,y_{j+1},\ldots,y_n)}{\partial x_j}\right)\,dx_j\\
+\sum_{j=1}^n\left(\frac{\alpha_{n-1}}{n}\right)P_t^{(single)}(n-1,y_1,\ldots,y_{j-1},y_{j+1},\ldots,y_n)\\
-\alpha_nP_t^{(single)}(n,y_1,\ldots,y_{n}),
\end{multline}
and for $n=C$,
\begin{multline}
\label{eq:single_pde3}
\frac{dP_t^{(single)}(n,y_1,\ldots,y_n)}{dt}=-\sum_{i=1}^n\frac{\partial P_t^{(single)}(n,y_1,\ldots,y_n)}{\partial y_i}\\
-\sum_{j=1}^n\int_{x_j=0}^{y_j}\beta(x_j)\left(\frac{\partial P_t^{(single)}(n,y_1,\ldots,y_{j-1},x_j,y_{j+1},\ldots,y_n)}{\partial x_j}\right)\,dx_j\\
+\sum_{j=1}^n\left(\frac{\alpha_{n-1}}{n}\right)P_t^{(single)}(n-1,y_1,\ldots,y_{j-1},y_{j+1},\ldots,y_n).
\end{multline}
%where $R_n(P)=\sum_{j=n}^CP_t(j,\infty,\ldots,\infty)$.

It was shown in \cite{shelby}
that for an Erlang loss system with single server having pre-specified state-dependent arrival rate $\alpha_i$ when
there are $i$ jobs in progress and job lengths are generally distributed with finite mean $\frac{1}{\mu}$, there exists
unique stationary distribution $\bm{\pi}^{(single)}=(\pi^{(single)}(\ul{u}),\ul{u}\in\mc{U})$ given by,
\beq
\label{eq:single_server_fixed_pt}
\pi^{(single)}(n,y_1,\ldots,y_n)=\frac{\left(\prod_{i=1}^n\frac{\alpha_{i-1}}{i\mu}\right)}{1+\sum_{m=1}^C\left(\prod_{i=1}^m\frac{\alpha_{i-1}}{i\mu}\right)}\mu^n\prod_{i=1}^n\int_{x_i=0}^{y_i}\ol{G}(x_i)\,dx_i
\eeq
and
\beq
\pi^{(single)}(0)=\frac{1}{1+\sum_{m=1}^C\left(\prod_{i=1}^m\frac{\alpha_{i-1}}{i\mu}\right)}.
\eeq

We are now ready to state the results on the fixed-point of the mean-field. 
Suppose in equation~\eqref{eq:mean_field_model_new}, if $\nu_0$ is absolutely continuous $\wrt$ Lebesgue measure at all $\ul{u}\in\mc{U}_n$ for $n\geq 1$, then at every $t\geq 0$, we have absolutely continuity of $\nu_t$ at all $\ul{u}\in\mc{U}_n$ for $n\geq 1$, $t\geq 0$ following the fact that $\nu_0$ is absolutely continuous and the mapping $t\mapsto\nu_t$ is continuous.
Suppose $p_t(0)$ denotes $\nu_t(\{0\})$ and $p_t(n,x_1,\ldots,x_n)$ denotes the Radon-Nikodym derivative of 
$\nu_t$ $\wrt$ Lebesgue measure at  $(n,x_1,\ldots,x_n)$. Now we obtain the differential equations
satisfied by the process $P_t=(P_t(\ul{u}),\ul{u}\in\mc{U})$ 
\beq
P_t(n,y_1,\ldots,y_n)=\int_{x_1=0}^{y_1}\ldots\int_{x_n=0}^{y_n}p_t(n,x_1,\ldots,x_n)\,dx_1\cdots dx_n.
\eeq

\begin{lem}
\label{thm:mf_pdes}
The differential equations for the process $P_t=(P_t(\ul{u}),\ul{u}\in\mc{U})$ are given by
\beq
\label{eq:mf_pdes1}
\frac{d P_t(0)}{dt}=\int_{y=0}^{\infty}\beta(y)\left(\frac{\partial P_t(1,y)}{\partial y}\right)\,dy-\lambda\frac{(R_0(P_t)^d-R_1^d(P_t))}{(R_0(P_t)-R_1(P_t))}P_t(0),
\eeq
for $1\leq n\leq C-1$,
\begin{multline}
\label{eq:mf_pdes2}
\frac{dP_t(n,y_1,\ldots,y_n)}{dt}=-\sum_{i=1}^n\frac{\partial P_t(n,y_1,\ldots,y_n)}{\partial y_i}\\
+\sum_{j=1}^{n+1}\int_{x_j=0}^{\infty}\beta(x_j)\left(\frac{\partial P_t(n+1,y_1,\ldots,y_{j-1},x_j,y_j,\ldots,y_n)}{\partial x_j}\right)\,dx_j\\
-\sum_{j=1}^n\int_{x_j=0}^{y_j}\beta(x_j)\left(\frac{\partial P_t(n,y_1,\ldots,y_{j-1},x_j,y_{j+1},\ldots,y_n)}{\partial x_j}\right)\,dx_j\\
+\sum_{j=1}^n\frac{\lambda(R_{n-1}(P_t)^d-R_{n}^d(P_t))}{n(R_{n-1}(P_t)-R_{n}(P_t))}P_t(n-1,y_1,\ldots,y_{j-1},y_{j+1},\ldots,y_n)\\
-\lambda \frac{(R_{n}(P_t)^d-R_{n+1}^d(P_t))}{(R_{n}(P_t)-R_{n+1}(P_t))}P_t(n,y_1,\ldots,y_{n}),
\end{multline}
and for $n=C$,
\begin{multline}
\label{eq:mf_pdes3}
\frac{dP_t(n,y_1,\ldots,y_n)}{dt}=-\sum_{i=1}^n\frac{\partial P_t(n,y_1,\ldots,y_n)}{\partial y_i}\\
-\sum_{j=1}^n\int_{x_j=0}^{y_j}\beta(x_j)\left(\frac{\partial P_t(n,y_1,\ldots,y_{j-1},x_j,y_{j+1},\ldots,y_n)}{\partial x_j}\right)\,dx_j\\
+\sum_{j=1}^n\frac{\lambda(R_{n-1}(P_t)^d-R_{n}^d(P_t))}{n(R_{n-1}(P_t)-R_{n}(P_t))}P_t(n-1,y_1,\ldots,y_{j-1},y_{j+1},\ldots,y_n),
\end{multline}
where $R_n(P_t)=\sum_{j=n}^CP_t(j,\infty,\ldots,\infty)$.
\end{lem}

The proof of Lemma~\ref{thm:mf_pdes} is given in section~\ref{sec:insensitivity}.

We next state the the principal result on the insensitivity of the fixed point of the MFE.
\begin{thm}
\label{thm:insensitivity}
There exists unique fixed-point for the process $P_t=(P_t(\ul{u}),\ul{u}\in\mc{U})$ denoted by $\bm{\pi}$ that satisfies
\beq
\label{eq:mf_fixed_pt}
\pi(n,y_1,\ldots,y_n)=\pi^{(exp)}(n)\mu^n\prod_{i=1}^n\int_{x_i=0}^{y_i}\ol{G}(x_i)\,dx_i.
\eeq
where $\bm{\pi}^{(exp)}=(\pi^{(exp)}(n),0\leq n\leq C)$ denotes the unique fixed-point 
of the mean-field when service times are exponentially distributed with mean $\frac{1}{\mu}$ and $\pi^{(exp)}(n)$ is the
stationary probability that there are $n$ jobs in the limiting system. Further, since
$\int_{x=0}^{\infty}\ol{G}(x)\,dx=\frac{1}{\mu}$, the fixed-point of the mean-field is insensitive as
\beq
\pi(n,\infty,\ldots,\infty)=\pi^{(exp)}(n).
\eeq
\end{thm}
The proof of Theorem~\ref{thm:insensitivity} is given in section~\ref{sec:insensitivity}.

%%%%%%%%%%%%%%%%%%%%%%%%%%%%%%%%%%%%%%%%%%%%%%%%%%%%%%%%%%%%%%%%%%%%%%%%%%%%%%%%%%%%%%%
%%%%%%%%%%%%%%%%%%%%%%%%%%%%%%%%%%%%%%%%%%%%%%%%%%%%%%%%%%%%%%%%%%%%%%%%%%%%%%%%%%%%%%%%%
%%%%%%%%%%%%%%%%%%%%%%%%%%%%%%%%%%%%%%%%%%%%%%%%%%%%%%%%%%%%%%%%%%%%%%%%%%%%%%%%%%%%%%
\section{Preliminary results}
\label{sec:prelim}

In this section, we derive the preliminary results that are needed to establish the convergence of scaled
version of $(\eta_t^N)_{t\geq 0}$ in $(D_{\mc{M}_F(\mc{U})}([0,\infty))$.

\begin{lem}
If $\eta_t^N=\nu$, using the power-of-$d$ routing policy, the probability that a job arriving at time $t$ is routed
to a server with state $(n,z_1,\ldots,z_n)$ is given by
\beq
\label{eq:destination_probability}
p_r(n,z_1,\ldots,z_n;\nu)=\frac{\nu(\{(n,z_1,\ldots,z_n\})}{N}\frac{(R_n(\nu)^d-R_{n+1}(\nu)^d)}{(R_n(\nu)-R_{n+1}(\nu))}
\eeq
where
$R_n(\nu)=\frac{\sum_{j=n}^C\nu(\mc{U}_n)}{N}$ represents the fraction of servers with at least $n$ jobs.
\end{lem}

\noindent{\bf Proof:}
 When a potential destination server is chosen uniformly at random from $N$ servers,
 it will have have state $(n,y_1,\ldots,y_n)$ with probability $\frac{\nu(\{(n,y_1,\ldots,y_n)\})}{N}$. Suppose out of $d$ potential destination servers, say $j$ servers have occupancy $n$ and the remaining $d-j$ servers have
 occupancy at least $n+1$. Further, out of $j$ potential destination servers with occupancy $n$, assume $r$ servers have the state $(n,z_1,\ldots,z_n)$. Then the probability that the destination server is a server with state $(n,z_1,\ldots,z_n)$ is given by 
\begin{multline*}{d\choose j}{j\choose r}\brac{\frac{r}{j}}\left(\frac{\nu(\{(n,z_1,\ldots,z_n)\})}{N}\right)^r
\times\left(\frac{\nu(\{\mc{U}_n\})-\nu(\{(n,z_1,\ldots,z_n)\})}{N}\right)^{j-r} 
\times \left(\sum_{i=n+1}^{C}\nu(\mc{U}_i)\right)^{d-j}.
\end{multline*}
Finally by summing over all the possible values of $j$ and $r$, we get the probability that the destination server lies in state $(n,z_1,\ldots,z_n)$ is as given in equation~\eqref{eq:destination_probability}.
$$\eqno{\tsquare}$$

We next compute the semi-group operator of the Markov process $(\eta_t^N)_{t\geq 0}$. We
consider the filtration
\beq
\mc{F}_t^N=\sigma(\eta_s^N(B):s\leq t,B\in\mc{B}(\mc{U})).
\eeq
We denote the number of 
arrivals in the interval $(0,h]$ by $A_h$ and the event $\{A_h=i\}$ denotes the event
that there are $i$ arrivals in the interval $(0,h]$. Similarly, given the
initial state $\eta_0^N$, we define $D_h$ to
indicate the number of departures that occur in the interval $(0,h]$. Note that a job
with age $x$ at time $t$ departs from the system in the interval $(t,t+h]$ with the probability
$\frac{G(x+h)-G(x)}{\ol{G}(x)}$. Further, from the definition of the hazard rate,
we have
\beq
\label{eq:departure_probability}
\lim_{h\to 0}\frac{1}{h}\frac{G(x+h)-G(x)}{\ol{G}(x)}=\beta(x)
\eeq
and hence
\beq
\label{eq:hazard_rate}
\frac{G(x+h)-G(x)}{\ol{G}(x)}=\beta(x)h+o(h).
\eeq
We next define
\beq
T_h^Nf(\nu)=\expect{f(\eta_h^N)\given \eta_0^N=\nu}
\eeq
where $f$ is a continuous bounded function $f:\mc{M}_{F}(\mc{U})\to \mc{R}$ and the operator
$T_h^N$ is a semigroup operator when $(\eta_t^N)_{t\geq 0}$ is a Markov process. Before computing the expression for $T_h^Nf(\nu)$, we first introduce the following notation. Suppose the measure $\eta_0^N=\nu$ has mass at $m$ points denoted by $\ul{u}^{(l)}=(n_l,u_{1}^{(l)},\ldots,u_{n_l}^{(l)})$
for $1\leq l\leq m$ and the number of servers with state $\ul{u}^{(l)}$ is given by $\nu({\{\ul{u}^{(l)}\}})$.
Let us denote the probability that a job departs from a server with state $\ul{b}=(n,b_1,\ldots,b_n)$ at
time $t$ in the interval $(t,t+h]$ by $p_D(\ul{b};h)$. Then we have
\beq
p_D(\ul{b};h)=\sum_{i=1}^{n}\left(\frac{G(b_i+h)-G(b_i)}{\ol{G}(b_i)}\right)\prod_{j=1;j\neq i}^n\left(\frac{\ol{G}(b_j+h)}{\ol{G}(b_j)}\right).
\eeq
Note that using equation~\eqref{eq:hazard_rate}, we can write
\beq
p_D(\ul{b};h)=\sum_{i=1}^{n}\left(\beta(b_i)h\right)\prod_{j=1;j\neq i}^n\left(1-\beta(b_j)h\right)+o(h).
\eeq
Further, let the probability that there is no departure at a server with state $\ul{b}=(n,b_1,\ldots,b_n)$ at
time $t$ in the interval $(t,t+h]$ be denoted by $p_{ND}(\ul{b};h)$. Then we have
\beq
p_{ND}(\ul{b};h)=\prod_{i=1}^n\frac{\ol{G}(b_i+h)}{\ol{G}(b_i)}.
\eeq
Note that using equation~\eqref{eq:hazard_rate}, we can write
\beq
p_{ND}(\ul{b};h)=\prod_{j=1}^n\left(1-\beta(b_j)h\right)+o(h).
\eeq
%By defining 
%\beq
%p_D^*(\ul{b};h)=\sum_{i=1}^{n}\left(\beta(b_i)h\right)\prod_{j=1;j\neq i}^n\left(1-\beta(b_i)h\right),
%\eeq
%we have
%\beq
%\label{eq:depart_prob_approx}
%p_D(\ul{b};h)=p_D^*(\ul{b};h)+o(h).
%\eeq

\begin{lem}
 Let $f$ be a real valued continuous bounded function defined on $\mc{M}_{F}(\mc{U})$. Then the process $(\eta_t^N)_{t\geq 0}$ is a weak-homogeneous $\mc{M}_F(\mc{U})$-valued Markov process
 with
 semigroup
operator $T_h^N(\cdot)$ given by
\begin{multline}
\label{eq:semigroup_final}
T_h^Nf(\nu)=(1-N\lambda h)
\left(\prod_{j=1,n_j>0}^{m}(p_{ND}(\ul{u}^{(j)};h))^{\nu{(\{\ul{u}^{(j)}\}})}\right)f(\tau_h\nu)
+(1-N\lambda h)\\
 \times\sum_{j=1,n_j>0}^m\sum_{r=1}^{n_j}\nu{(\{\ul{u}^{(j)}\}})
 \left(\frac{G(u_{r}^{(j)}+h)-G(u_{r}^{(j)})}{\ol{G}(u_{r}^{(j)})}\right)
 \Bigg(\prod_{w=1;w\neq r}^{n_j}
 \left(\frac{\ol{G}(u_{w}^{(j)}+h)}{\ol{G}(u_{w}^{(j)})}\right)\Bigg)\\
 \times(p_{ND}(\ul{u}^{(j)};h))^{(\nu{(\{\ul{u}^{(j)}\}})-1)} 
 \left(\prod_{i=1,n_i>0,i\neq j}^{m}(p_{ND}(\ul{u}^{(i)};h))^{\nu{(\{\ul{u}^{(i)}\}})}\right)\\
 \times
f(\tau_h\nu+\delta_{(n_j-1,u_{1}^{(j)}+h,\ldots,u_{r-1}^{(j)}+h,u_{r+1}^{(j)}+h,\ldots,u_{n_j}^{(j)}+h)}-\delta_{(n_j,u_{1}^{(j)}+h,\ldots,u_{n_j}^{(j)}+h)})\\
+\mc{P}(\{D_h=0\})(N\lambda h)\\
\times\mb{E}\Bigg[f\left(\tau_h(\nu+\delta_{(M+1,Z_{1},\ldots,Z_{(L_1-1)},0,Z_{L},\ldots,Z_{M})}-\delta_{(M,Z_{1},\ldots,Z_{M})})\right)\given\eta_0^N=\nu\Bigg]
+\epsilon(\nu,h)
\end{multline}
where by considering $\nu$ as the state of the process $(\eta_t^N)_{t\geq 0}$
at the first arrival instant, $(M,Z_{1},\ldots,Z_{M})$ denotes the random variable representing the destination server
state for the arriving job and $L_1$ is a random variable representing the position
of the arriving job at the destination server and $\epsilon(\nu,h)$ is a $o(h)$ term.
\end{lem}

\noindent{\bf Proof:}
We now consider the expression for $T_h^Nf(\nu)=\expect{f(\eta_h^N)\given \eta_0^N=\nu}$. We can write
\beq
T_h^Nf(\nu)=\sum_{i\geq 0,j\geq o}\mathbb{E}\Bigg[f(\eta_h^N)\indic{A_h=i,D_h=j}\given \eta_0^N=\nu\Bigg].
\eeq

 We next can write
  \begin{multline}
\label{eq:semigroup_expression}
T_h^Nf(\nu)
=\mathbb{E}\Bigg[f(\eta_h^N)\indic{A_h=0,D_h=0}\given \eta_0^N=\nu\Bigg]+\mathbb{E}\Bigg[f(\eta_h^N)\indic{A_h=0,D_h=1}\given \eta_0^N=\nu\Bigg]\\
+\mathbb{E}\Bigg[f(\eta_h^N)\indic{A_h=1,D_h=0}\given \eta_0^N=\nu\Bigg]
+
\sum_{i\geq 1,j\geq 1}\mathbb{E}\Bigg[f(\eta_h^N)\indic{A_h=i,D_h=j}\given \eta_0^N=\nu\Bigg].
\end{multline}
We first simplify the first term on the right side of the equation~\eqref{eq:semigroup_expression}. The probability
that there is no departure in the interval $(0,h]$ is given by
\beq
\mc{P}(\{D_h=0\})=\prod_{j=1,n_j>0}^{m}(p_{ND}(\ul{u}^{(j)};h))^{\nu{(\{\ul{u}^{(j)}\}})}.
\eeq
 We have
 \beq
 %\begin{split}
 \mathbb{E}\Bigg[f(\eta_h^N)\indic{A_h=0,D_h=0}\given \eta_0^N=\nu\Bigg]=\mc{P}(\{A_h=0\})\mc{P}(\{D_h=0\})f(\tau_h\nu).
 \eeq
 We can write
 \begin{multline}
\mathbb{E}\Bigg[f(\eta_h^N)\indic{A_h=0,D_h=0}\given \eta_0^N=\nu\Bigg] =(\mc{P}(\{A_h=0\})+(1-N\lambda h)-(1-N\lambda h))
 \mc{P}(\{D_h=0\})f(\tau_h\nu)
 %\end{split}
 \end{multline}
 Further, we can write
 \begin{multline}
\mathbb{E}\Bigg[f(\eta_h^N)\indic{A_h=0,D_h=0}\given \eta_0^N=\nu\Bigg] =(1-N\lambda h)
\left(\prod_{j=1,n_j>0}^{m}(p_{ND}(\ul{u}^{(j)};h))^{\nu{(\{\ul{u}^{(j)}\}})}\right)\\
\times f(\tau_h\nu)
+\epsilon_1(\nu,h)
 %\end{split}
 \end{multline}
 where 
 \beq
 \epsilon_1(\nu,h)=(\mc{P}(\{A_h=0\})-(1-N\lambda h))
\times \mc{P}(\{D_h=0\})f(\tau_h\nu)
\eeq 
is a $o(h)$ term.
 Similarly we can write the second term of the right side of the equation~\eqref{eq:semigroup_expression}
 as
 \begin{multline}
 \mathbb{E}\Bigg[f(\eta_h^N)\indic{A_h=0,D_h=1}\given \eta_0^N=\nu\Bigg]=(1-N\lambda h)
 \sum_{j=1,n_j>0}^m\sum_{r=1}^{n_j}\nu{(\{\ul{u}^{(j)}\}})\\
 \times\left(\frac{G(u_{r}^{(j)}+h)-G(u_{r}^{(j)})}{\ol{G}(u_{r}^{(j)})}\right)
 \left(\prod_{w=1;w\neq r}^{n_j}\left(\frac{\ol{G}(u_{w}^{(j)}+h)}{\ol{G}(u_{w}^{(j)})}\right)\right)
 (p_{ND}(\ul{u}^{(j)};h))^{(\nu{(\{\ul{u}^{(j)}\}})-1)} \\
 \times\left(\prod_{i=1,n_i>0,i\neq j}^{m}(p_{ND}(\ul{u}^{(i)};h))^{\nu{(\{\ul{u}^{(i)}\}})}\right)\\
 \times
f(\tau_h\nu+\delta_{(n_j-1,u_{1}^{(j)}+h,\ldots,u_{r-1}^{(j)}+h,u_{r+1}^{(j)}+h,\ldots,u_{n_j}^{(j)}+h)}-\delta_{(n_j,u_{1}^{(j)}+h,\ldots,u_{n_j}^{(j)}+h)})+\epsilon_2(\nu,h)
 \end{multline}
 where we use $r$ to denote the index of the departure job at a server with state
 $\ul{u}^{(j)}$ and $\epsilon_2(\nu,h)$ is a $o(h)$ term given by
 \begin{multline}
 \epsilon_2(\nu,h)=(\mc{P}(\{A_h=0\})-(1-N\lambda h))
 \sum_{j=1,n_j>0}^m\sum_{r=1}^{n_j}\nu{(\{\ul{u}^{(j)}\}})\\
\times \left(\frac{G(u_{r}^{(j)}+h)-G(u_{r}^{(j)})}{\ol{G}(u_{r}^{(j)})}\right)\left(\prod_{w=1;w\neq r}^{n_j}\left(\frac{\ol{G}(u_{w}^{(j)}+h)}{\ol{G}(u_{w}^{(j)})}\right)\right)\\
 \times(p_{ND}(\ul{u}^{(j)};h))^{(\nu{(\{\ul{u}^{(j)}\}})-1)} \left(\prod_{i=1,n_i>0,i\neq j}^{m}(p_{ND}(\ul{u}^{(i)};h))^{\nu{(\{\ul{u}^{(i)}\}})}\right)\\
 \times
f(\tau_h\nu+\delta_{(n_j-1,u_{1}^{(j)}+h,\ldots,u_{r-1}^{(j)}+h,u_{r+1}^{(j)}+h,\ldots,u_{n_j}^{(j)}+h)}-\delta_{(n_j,u_{1}^{(j)}+h,\ldots,u_{n_j}^{(j)}+h)})
 \end{multline}

 We next compute the third term on the right side of the equation~\eqref{eq:semigroup_expression}.
 Suppose the job arrives at time $T_1$ which is an exponential random variable 
 with rate $N\lambda$. We can write
 %\vspace{-.5cm}
 \begin{multline}
 \mathbb{E}\Bigg[f(\eta_h^N)\indic{A_h=1,D_h=0}\given \eta_0^N=\nu\Bigg]=\mc{P}(\{D_h=0\})\\
 \times\mathbb{E}\Bigg[f\bigg(\tau_h\nu
+(\delta_{(M+1,Z_{1},\ldots,Z_{(L_1-1)},h-T_1,Z_{L_1},\ldots,Z_{M})}
-\delta_{(M,Z_{1},\ldots,Z_{M})})\bigg)\indic{A_h=1}\given \eta_0^N=\nu\Bigg]
 \end{multline}where
$(M,Z_{1},\ldots,Z_{M})$ denotes the random variable representing the destination server
state for the arriving job and $L_1$ is a random variable representing the position
of the arriving job at the destination server. Note that while choosing the destination server for
the arrival, $\eta_{T_1}^N=\tau_{T_1}\nu$ is used in implementing the power-of-$d$ policy. We
further can write
%\vspace{-.8cm}
\begin{multline}
\label{eq:arrival_term}
\mathbb{E}\Bigg[f(\eta_h^N)\indic{A_h=1,D_h=0}\given \eta_0^N=\nu\Bigg]=\mc{P}(\{D_h=0\})N\lambda h\\
\times\mb{E}\Bigg[f\left(\tau_h(\nu+\delta_{(M+1,Z_{1},\ldots,Z_{(L_1-1)},0,Z_{L},\ldots,Z_{M})}-\delta_{(M,Z_{1},\ldots,Z_{M})})\right)\given\eta_0^N=\nu\Bigg]\\
+\mc{P}(\{D_h=0\})[\mc{P}(\{A_h=1\})-N\lambda h]\\
\times\mb{E}\Bigg[f\left(\tau_h(\nu+\delta_{(M+1,Z_{1},\ldots,Z_{(L_1-1)},0,Z_{L_1},\ldots,Z_{M})}-\delta_{(M,Z_{1},\ldots,Z_{M})})\right)\given\eta_0^N=\nu\Bigg]\\
+\mc{P}(\{D_h=0\})\\
\times\left(\mathbb{E}^{(1)}\Bigg[f\bigg(\tau_h\nu
+(\delta_{(M+1,Z_{1},\ldots,Z_{(L_1-1)},h-T_1,Z_{L_1},\ldots,Z_{M})}
-\delta_{(M,Z_{1},\ldots,Z_{M})})\bigg)\indic{A_h=1}\given \eta_0^N=\nu\Bigg]\right.\\
\left.-\mb{E}^{(2)}\Bigg[f\left(\tau_h(\nu+\delta_{(M+1,Z_{1},\ldots,Z_{(L_1-1)},0,Z_{L_1},\ldots,Z_{M})}-\delta_{(M,Z_{1},\ldots,Z_{M})})\right)\indic{A_h=1}\given \eta_0^N=\nu\Bigg]\right)
\end{multline}
%\vspace{-.12cm}
where, in the first and the second terms on the right side of the equation~\eqref{eq:arrival_term}, the job is considered
as arriving at $T_1=0$ and hence we use $\eta_0^N$ in choosing the destination server for the
first arrival. In the third term, in computing $E^{(1)}$, the arrival occurs at exponentially distributed time $T_1$ while
in computing $E^{(2)}$, the arrival occurs at time $T_1=0$. Since $f$ is a bounded function, it is clear that the second term of
the right side in equation~\eqref{eq:arrival_term}
is a $o(h)$ term. By using the fact that $T_1$ is an exponential random variable
with rate $N\lambda$ and using the l'Hospital's rule, the third term is also a $o(h)$
term. Therefore we can write
\begin{multline}
\label{eq:arrival_term2}
\mathbb{E}\Bigg[f(\eta_h^N)\indic{A_h=1,D_h=0}\given \eta_0^N=\nu\Bigg]=\mc{P}(\{D_h=0\})N\lambda h\\
\times\mb{E}\Bigg[f\left(\tau_h(\nu+\delta_{(M+1,Z_{1},\ldots,Z_{(L_1-1)},0,Z_{L},\ldots,Z_{M})}-\delta_{(M,Z_{1},\ldots,Z_{M})})\right)\given\eta_0^N=\nu\Bigg]\\
+\epsilon_3(\nu,h)
\end{multline}
where $\epsilon_3(\nu,h)$ is a $o(h)$ term equal to the sum of second and third
terms of the right side of equation~\eqref{eq:arrival_term}. 

Finally, by using the fact $f$ is a bounded function, the fourth term on the
right side of equation~\eqref{eq:semigroup_expression} is a $o(h)$ term
denoted by $\epsilon_4(\nu,h)$. By combining expressions for all the four terms
on right side of equation~\eqref{eq:semigroup_expression}, and by defining
\beq
\epsilon(\nu,h)=\epsilon_1(\nu,h)+\epsilon_2(\nu,h)+\epsilon_3(\nu,h)+\epsilon_4(\nu,h),
\eeq
we get expression for $T_h^Nf(\nu)$ as in equation~\eqref{eq:semigroup_final}. Finally, from \cite[p.18]{dawson}, $(\eta_t^N)_{t\geq 0}$ is a weak homogeneous Markov process.
$$\eqno{\square}$$

%Since $(<\eta_t^N,\phi>)_{t\geq 0}\in\mc{D}_{\mc{R}}([0,\infty))$ for all $\phi\in\mc{C}_b(\mc{U})$, the process $%(\eta_t^N)_{t\geq 0}\in \mc{D}_{\mc{M}_F(\mc{U})}([0,\infty))$.

\begin{prop}
The process $(\eta_t^N)_{t\geq 0}$ is a Feller-Dynkin process of $\mc{D}_{\mc{M}_{F}(\mc{U})}([0,\infty))$.
\end{prop}

%\vspace{-0.5cm}
\noindent{\bf Proof:}
%\vspace{-0.5cm}
From Lemma~$3.5.1$ and Corollary~$3.5.2$ of \cite{dawson}, the process $(\eta_t^N)_{t\geq 0}$ has Feller-Dynkin
property if:\\
For $f\in\mc{C}_s^1(U),\nu\in\mc{M}_F(\mc{U})$, let $Q_f:\mc{M}_F(\mc{U})\longmapsto\mc{R}$ be defined by $Q_f(\nu)=e^{-\langle\nu,f\rangle}$, then we must have
\begin{enumerate}
\label{eq:feller}
%\vspace{-.4cm}
\item The mapping $\nu\longmapsto\expect{Q_f(\eta_h^N)\given\eta_0^N=\nu}$ is continuous for all $f\in\mc{C}_s^1(U)$ and $h>0$.
%\vspace{-.45cm}
\item For all $h>0$, we have $\expect{Q_{\bf{1}}(\eta_h^N)\given\eta_0^N=\nu}\to 0$ as $\nu(\mc{U})\to\infty$.
%\vspace{-0.45cm}
\item For all $\nu\in\mc{M}_F(\mc{U})$ and $f\in\mc{C}_s^1(\mc{U})$, we have $\expect{Q_f(\eta_h^N)\given\eta_0^N=\nu}\to Q_f(\nu)$ as $h\to 0$. 
\end{enumerate}
By using equation~\eqref{eq:semigroup_final}, we have
\begin{multline}
\label{eq:simplifier}
\expect{Q_{f}(\eta_h^N)\given\eta_0^N=\nu}=e^{-\langle\tau_h\nu,f\rangle}\Bigg\{(1-N\lambda h)
\left(\prod_{j=1,n_j>0}^{m}(p_{ND}(\ul{u}^{(j)};h))^{\nu{(\{\ul{u}^{(j)}\}})}\right)\\
+(1-N\lambda h)
 \sum_{j=1,n_j>0}^m\sum_{r=1}^{n_j}\nu{(\{\ul{u}^{(j)}\}})
 \left(\frac{G(u_{r}^{(j)}+h)-G(u_{r}^{(j)})}{\ol{G}(u_{r}^{(j)})}\right)\\
 \times\left(\prod_{w=1;w\neq r}^{n_j}\left(\frac{\ol{G}(u_{w}^{(j)}+h)}{\ol{G}(u_{w}^{(j)})}\right)\right)
 (p_{ND}(\ul{u}^{(j)};h))^{(\nu{(\{\ul{u}^{(j)}\}})-1)}\\
 \times \left(\prod_{i=1,n_i>0,i\neq j}^{m}(p_{ND}(\ul{u}^{(i)};h))^{\nu{(\{\ul{u}^{(i)}\}})}\right)\\
 \times
Q_f\left(\tau_h(\delta_{(n_j-1,u_{1}^{(j)},\ldots,u_{r-1}^{(j)},u_{r+1}^{(j)},\ldots,u_{n_j}^{(j)})}-\delta_{(n_j,u_{1}^{(j)},\ldots,u_{n_j}^{(j)})})\right)
+\mc{P}(\{D_h=0\})(N\lambda h)\\
\times\mb{E}\Bigg[Q_f\left(\tau_h(\delta_{(M+1,Z_{1},\ldots,Z_{(L_1-1)},0,Z_{L},\ldots,Z_{M})}-\delta_{(M,Z_{1},\ldots,Z_{M})})\right)\given\eta_0^N=\nu\Bigg]
+\epsilon_f(\nu,h)
\Bigg\}
\end{multline}
where $\epsilon_f(\nu,h)$ is given by
\beq
\label{eq:error2}
\epsilon_f(\nu,h)=\epsilon_{1f}(\nu,h)+\epsilon_{2f}(\nu,h)+\epsilon_{3f}(\nu,h)+\epsilon_{4f}(\nu,h)\eeq
such that\\
%\vspace{-1cm}
\beq
%\vspace{-1cm}
\epsilon_{1f}(\nu,h)=(\mc{P}(\{A_h=0\})-(1-N\lambda h))\mc{P}(\{D_h=0\}),
\eeq
\begin{multline}
 \epsilon_{2f}(\nu,h)=(\mc{P}(\{A_h=0\})-(1-N\lambda h))
 \sum_{j=1,n_j>0}^m\sum_{r=1}^{n_j}\nu{(\{\ul{u}^{(j)}\}})\\
\times \left(\frac{G(u_{r}^{(j)}+h)-G(u_{r}^{(j)})}{\ol{G}(u_{r}^{(j)})}\right)
 \left(\prod_{w=1;w\neq r}^{n_j}\left(\frac{\ol{G}(u_{w}^{(j)}+h)}{\ol{G}(u_{w}^{(j)})}\right)\right)\\
 \times(p_{ND}(\ul{u}^{(j)};h))^{(\nu{(\{\ul{u}^{(j)}\}})-1)} \left(\prod_{i=1,n_i>0,i\neq j}^{m}(p_{ND}(\ul{u}^{(i)};h))^{\nu{(\{\ul{u}^{(i)}\}})}\right)\\
 \times
Q_f(\delta_{(n_j-1,u_{1}^{(j)}+h,\ldots,u_{r-1}^{(j)}+h,u_{r+1}^{(j)}+h,\ldots,u_{n_j}^{(j)}+h)}-\delta_{(n_j,u_{1}^{(j)}+h,\ldots,u_{n_j}^{(j)}+h)}),
 \end{multline}
 \begin{multline}
\epsilon_{3f}(\nu,h)=\mc{P}(\{D_h=0\})[\mc{P}(\{A_h=1\})-N\lambda h]\\
\times\mb{E}\Bigg[Q_f\left(\tau_h(\delta_{(M+1,Z_{1},\ldots,Z_{(L_1-1)},0,Z_{L_1},\ldots,Z_{M})}-\delta_{(M,Z_{1},\ldots,Z_{M})})\right)\given\eta_0^N=\nu\Bigg]
+\mc{P}(\{D_h=0\})\\
\times\left(\mathbb{E}^{(1)}\Bigg[Q_f\bigg((\delta_{(M+1,Z_{1},\ldots,Z_{(L_1-1)},h-T_1,Z_{L_1},\ldots,Z_{M})}
-\delta_{(M,Z_{1},\ldots,Z_{M})})\bigg)\indic{A_h=1}\given \eta_0^N=\nu\Bigg]\right.\\
\left.-\mb{E}^{(2)}\Bigg[Q_f\left(\tau_h(\delta_{(M+1,Z_{1},\ldots,Z_{(L_1-1)},0,Z_{L_1},\ldots,Z_{M})}-\delta_{(M,Z_{1},\ldots,Z_{M})})\right)\indic{A_h=1}\given \eta_0^N=\nu\Bigg]\right)
\end{multline}
and
\begin{multline}
\label{eq:4th_term}
\epsilon_{4f}(\nu,h)=\\
\sum_{i\geq 1,j\geq 1}\mathbb{E}\Bigg[Q_f\left(\sum_{r=1}^i(\delta_{(M_r+1,Z_{1}^{(r)},\ldots,Z_{L_r-1}^{(r)},h-T_r,Z_{L_r+1}^{(r)},\ldots,Z_{M_r}^{(r)})}-\delta_{(M_r,Z_{1}^{(r)},\ldots,\ldots,Z_{M_r}^{(r)})})\right.\\
\left.+\sum_{l=1}^j
(\delta_{(n_l-1,X_{1}^{(l)},\ldots,X_{J_l-1}^{(l)},X_{J_l+1}^{(l)},\ldots,X_{n_l}^{(l)}}-\delta_{(n_l,X_{1}^{(l)},\ldots,X_{n_l}^{(l)}}))\right)\\
\times\indic{A_h=i,D_h=j}\given \eta_0^N=\nu\Bigg].
\end{multline}

In equation~\eqref{eq:4th_term}, $T_r$ denotes the arrival time of $r^{\text{th}}$ job which is routed to a server with state $\ul{v}^{(r)}$
such that $\tau_{(h-T_r)}^+\ul{v}^{(r)}=(M_r,Z_{1}^{(r)},\ldots,Z_{M_r}^{(r)})$ and $L_r$ is the position of $r^{\text{th}}$
arriving job at the destination server. Corresponding to departures, suppose $l^{\text{th}}$ departure
occurs at a server with state $\ul{q}^{(l)}$ say at time $t_l$ and the position of the job is $J_l$, then $\tau_{(h-t_l)}^+\ul{q}^{(l)}=(n_l,X_{1}^{(l)},\ldots,X_{n_l}^{(l)})$. By using the same arguments as for $\epsilon(\nu,h)$ in equation~\eqref{eq:semigroup_final}, $\epsilon_f(\nu,h)$ is 
also a $o(h)$ term.

To prove the first condition required for Feller property, we write equation~\eqref{eq:simplifier} as
\beq
\expect{Q_{f}(\eta_h^N)\given\eta_0^N=\nu}=(e^{-\langle\tau_h\nu,f\rangle})U(\nu,h).
\eeq
Clearly, $(e^{-\langle\tau_h\nu,f\rangle})$ is a continuous mapping of $\nu$. We next need to prove $U(\nu,h)$ is
a continuous mapping of $\nu$. Since $\nu$ is a point measure at finite $N$, the routing probabilities under power-of-$d$ policy as shown in equation~\eqref{eq:destination_probability}
and the departure probabilities are continuous mappings of $\nu$ and hence $U(\nu,h)$
is a continuous mapping of $\nu$. The second condition follows directly since $\tau_h\nu(\mc{U})=\nu(\mc{U})=N$.
The third condition follows from the fact that $\langle\tau_h\nu,f\rangle=\langle\nu,\tau_hf\rangle$ and then by applying the dominated convergence theorem we have $\langle\tau_h\nu,f\rangle\to\langle\nu,f\rangle$ as $h\to 0$. Hence the process $(\eta_t^N)_{t\geq 0}$ is a Feller process.
$$\eqno{\tsquare}$$

%%%%%%%%%%%%%%%%%%%%%%%%%%%%%%%%%%%%%%%%%%%%%%%%%%%%%%%%%%%%%%%%%%%%%%%%%%%%%%%%%%%%%%%%%%%%%%%%%%%
%%%%%%%%%%%%%%%%%%%%%%%%%%%%%%%%%%%%%%%%%%%%%%%%%%%%%%%%%%%%%%%%%%%%%%%%%%%%%%%%%%%%%%%%%%%%%%%%%%%%%%%
%%%%%%%%%%%%%%%%%%%%%%%%%%%%%%%%%%%%%%%%%%%%%%%%%%%%%%%%%%%%%%%%%%%%%%%%%%%%%%%%%%%%%%%%%%%%%%%%%%%%%%%%%%%%
\section{Existence and uniqueness of mean-field model solution}
\label{sec:MFE_anal}

In this section, we prove that there exists unique solution to the mean-field
model equations. Uniqueness
of the mean-field model solution is used in proving the convergence of the sequence of processes
$(\frac{\eta^N_t}{N},{t\geq 0})$ as $N\to\infty$. In this proof, we repeatedly use the Fundamental theorem of
calculus.
%Note that the mean-field equations are same Kolomogorov equations for a single server loss
%%system with prespecified state-dependent arrival rate $\lambda_n$ when there are $n$ jobs in progress
%whereas in the mean-field equations, the term $\lambda_n$ is replaced by
%$\lambda \frac{(\ol{R}_n(\ol{\eta}_t)^d-\ol{R}_{n+1}(\ol{\eta}_t)^d)}{(\ol{R}_n(\ol{\eta}_t)-\ol{R}_{n+1}(\ol{\eta}_t))}$.

\vspace{0.2cm}

\noindent\textbf{Proof of Lemma}~\ref{thm:mf_model_new}:

We first show that any process $(\nu_t,t\geq 0)$ satisfying the equation~\eqref{eq:mf_model_eqns} also satisfies
the equation~\eqref{eq:mean_field_model_new}. By using the fundamental theorem of calculus, for $\phi\in\mc{C}_b^1(\mc{U})$, a real valued process $(\langle \nu_t,\phi\rangle,t\geq 0)$ satisfying the equation~\eqref{eq:mf_model_eqns} is a solution
to the following differential equation~\eqref{eq:mean_field_pde} if the integrand in equation~\eqref{eq:mf_model_eqns} is a continuous function
of $s$,
\begin{multline}
\label{eq:mean_field_pde}
%\begin{multline}
%\label{eq:martingale}
\frac{d\langle\nu_t,\phi\rangle}{dt}= \langle\nu_t,\phi'\rangle
+\Bigg(\sum_{n=1}^C\sum_{j=1}^n\int_{x_1}\cdots\int_{x_n}\beta(x_j)\\
\times\left(\phi(n-1,x_1,\ldots,x_{j-1},x_{j+1},\ldots,x_n)-\phi(n,x_1,\ldots,x_n)\right)\,d\nu_t(n,x_1,\ldots,x_n)\\
+\lambda\bigg[\left(\nu_t(\{0\})\frac{(\ol{R}_0(\nu_t)^d-\ol{R}_1(\nu_t)^d)}{(\ol{R}_0(\nu_t)-\ol{R}_1(\nu_t))}\left(\phi(1,0)-\phi(0)\right)\right)
+\sum_{n=1}^{C-1}\sum_{j=1}^{n+1}\int_{x_1}\cdots\int_{x_n}\frac{1}{(n+1)}\\
\times\frac{(\ol{R}_n(\nu_t)^d-\ol{R}_{n+1}(\nu_t)^d)}{(\ol{R}_n(\nu_t)-\ol{R}_{n+1}(\nu_t))}
(\phi(n+1,x_1,\ldots,x_{j-1},0,x_j,\ldots,x_n)-\phi(n,x_1,\ldots,x_n))\\
\times\,d\nu_t(n,x_1,\ldots,x_n)\bigg]\Bigg).
\end{multline}
It is equivalent to proving the two terms on the right side of equation~\eqref{eq:mean_field_pde} are continuous functions of $t$. Since $\phi\in\mc{C}_b^1(\mc{U})$ and the mapping $t\mapsto\nu_t$ is continuous, the first term $\langle \nu_t,\phi'\rangle$ is a continuous function of $t$. In the second term, the expression related to the case of departures can be written as
\begin{multline}
\sum_{n=1}^C\sum_{j=1}^n\int_{x_1}\cdots\int_{x_n}\beta(x_j)
\left(\phi(n-1,x_1,\ldots,x_{j-1},x_{j+1},\ldots,x_n)-\phi(n,x_1,\ldots,x_n)\right)\\
\times d\nu_t(n,x_1,\ldots,x_n)
=\langle\nu_t,\psi_1\rangle,
\end{multline}
where the function $\psi_1$ is defined such that
\beq
\psi_1(0)=0
\eeq
and for $n\geq 1$
\begin{multline}
\psi_1(n,x_1,\ldots,x_n)\\
=\sum_{j=1}^n\beta(x_j)((\phi(n-1,x_1,\ldots,x_{j-1},x_{j+1},\ldots,x_n)-\phi(n,x_1,\ldots,x_n)).
\end{multline}
Since $\phi\in\mc{C}_b^1(\mc{U})$ and $\beta\in\mc{C}_b^1(\mc{R}_+)$, we have that $\psi_1\in\mc{C}_b(\mc{U})$. Therefore
$\langle \nu_t,\psi_1\rangle$ is a continuous function of $t$. Now consider the expression that corresponds to the case of arrivals, we can write
\begin{multline}
\lambda\bigg[\left(\nu_t(\{0\})\frac{(\ol{R}_0(\nu_t)^d-\ol{R}_1(\nu_t)^d)}{(\ol{R}_0(\nu_t)-\ol{R}_1(\nu_t))}\left(\phi(1,0)-\phi(0)\right)\right)
+\sum_{n=1}^{C-1}\sum_{j=1}^{n+1}\int_{x_1}\cdots\int_{x_n}\frac{1}{(n+1)}\\
\times\frac{(\ol{R}_n(\nu_t)^d-\ol{R}_{n+1}(\nu_t)^d)}{(\ol{R}_n(\nu_t)-\ol{R}_{n+1}(\nu_t))}
(\phi(n+1,x_1,\ldots,x_{j-1},0,x_j,\ldots,x_n)-\phi(n,x_1,\ldots,x_n))\\
\times\,d\nu_t(n,x_1,\ldots,x_n)\\
=\langle \nu_t,\psi_{(\nu_t)}\rangle,
\end{multline}
where $\psi_{(\nu_t)}$ is defined as,
for $n=C$,
\beq
\psi_{(\nu_t)}(n,x_1,\ldots,x_n)=0
\eeq
and for $n<C$,
\begin{multline}
\psi_{(\nu_t)}(n,x_1,\ldots,x_n)=\frac{\lambda}{(n+1)}
\frac{(\ol{R}_n(\nu_t)^d-\ol{R}_{n+1}(\nu_t)^d)}{(\ol{R}_n(\nu_t)-\ol{R}_{n+1}(\nu_t))}\\
\times (\phi(n+1,x_1,\ldots,x_{j-1},0,x_j,\ldots,x_n)-\phi(n,x_1,\ldots,x_n)).
\end{multline}
Therefore, for given $\nu_t$, since $\phi\in\mc{C}_b(\mc{U})$, the above defined function $\psi_{\nu_t}\in\mc{C}_b(\mc{U})$.
Hence for some $a\geq 0$ such that $t\neq a$, the function $\langle \nu_t,\psi_{(\nu_a)}\rangle$ is a continuous function of $t$. We next prove that the mapping $t\mapsto\langle \nu_t,\psi_{(\nu_t)}\rangle$ is continuous, $\ie$, we need to prove that
$\langle \nu_{t+b},\psi_{(\nu_{t+b})}\rangle\to\langle \nu_t,\psi_{(\nu_t)}\rangle$ if $b\to 0$. We have
\beq
\label{eq:norm_bound}
\abs{\langle \nu_{t+b},\psi_{(\nu_{t+b})}\rangle-\langle \nu_t,\psi_{(\nu_t)}\rangle}\leq \abs{\langle \nu_{t+b},\psi_{(\nu_{t+b})}\rangle-\langle \nu_{t+b},\psi_{(\nu_t)}\rangle}+\abs{\langle \nu_{t+b},\psi_{(\nu_t)}\rangle-\langle \nu_t,\psi_{(\nu_t)}\rangle}.
\eeq
Since $\psi_{(\nu_t)}\in\mc{C}_b(\mc{U})$, we have
\beq
\lim_{b\to 0}\abs{\langle \nu_{t+b},\psi_{(\nu_t)}\rangle-\langle \nu_t,\psi_{(\nu_t)}\rangle}=0.
\eeq
We next prove that 
\beq
\lim_{b\to 0}\abs{\langle \nu_{t+b},\psi_{(\nu_{t+b})}-\psi_{(\nu_t)}\rangle}=0.
\eeq
For $L>0$, let
\beq
U^{(L)}=\{(n,x_1,\ldots,x_n)\in U_n: n\geq 1, x_i\geq L \text{ for all } 1\leq i\leq n\}.
\eeq
For given $\epsilon>0$, we can find some $L>0$ such that 
\beq
\langle \nu_t,\indic{U^{(L)}}\rangle<\epsilon.
\eeq
Furthermore, from continuity of $t\mapsto\nu_t$, we can find some $h_1>0$ such that
for all $b\in [-\min{(t,h_1)},h_1]$,
\beq
\label{eq:measure_bound}
\langle \nu_{t+b},\indic{U^{(L)}}\rangle<\epsilon. 
\eeq
Since $\ol{R}_n(\nu_t)=\langle \nu_t,\indic{\cup_{j=n}^CU_j}\rangle$ is a continuous function of $t$, $\psi_{(\nu_t)}$ is a continuous function of $t$.
Therefore, $\psi_{(\nu_{t+b})}$ is uniformly continuous on the interval $b\in [-\min{(t,h_1)},h_1]$ and $\ul{u}\in \ol{U}^{(L)}$ (the complement of $U^{(L)}$). Therefore there exists some $h_2\in(0,h_1)$ such that for $b\in[-\min(t,h_2),h_2]$,
$\ul{u}\in\ol{U}^{(L)}$, we have 
\beq
\label{eq:fn_bound}
\abs{\psi_{(\nu_{t+b})}(\ul{u})-\psi_{(\nu_{t})}(\ul{u})}<\epsilon.
\eeq
Using equations~\eqref{eq:measure_bound}-\eqref{eq:fn_bound}, for $b\in[-\min(t,h_2),h_2]$, we have,
\begin{multline}
\abs{\langle \nu_{t+b},\psi_{(\nu_{t+b})}-\psi_{(\nu_t)}\rangle}\leq \epsilon\langle \nu_{t+b},\indic{\ol{U}^{(L)}}\rangle+2d\lambda\norm{\phi}\epsilon\\
\leq \epsilon+2d\lambda\norm{\phi}\epsilon.
\end{multline}
By letting $b\to 0$ and then $\epsilon\to 0$ in equation~\eqref{eq:norm_bound}, we have continuity of the mapping
$t\mapsto\langle\nu_t,\psi_{(\nu_t)}\rangle$.

We next obtain an equivalent form of the equations that are satisfied by the solution to the equation~\eqref{eq:mean_field_pde} using the change of variables. Let us define a function $\tilde{\phi}$ from $\phi\in\mc{C}_b^1(\mc{U})$ as follows:
For $r\leq t$, let
\begin{align}
\tilde{\phi}(n,x_1,\ldots,x_n)&=\phi(n,x_1+t-r,\ldots,x_n+t-r)\\
&=\phi(\tau_{t-r}^+(n,x_1,\ldots,x_n))\\
&=\tau_{t-r}\phi(n,x_1,\ldots,x_n)
\end{align}
and $\tilde{\phi}(0)=\phi(0)$.
Now let us look at the change of $\langle \nu_r,\tilde{\phi}\rangle$ $\wrt$ the variable `$r$'. We can write
\beq
\frac{d\langle \nu_r,\tilde{\phi}\rangle}{dr}=\frac{d\langle \nu_r,\tilde{\phi}\rangle}{dr}\given (\text{fixed }\tilde{\phi})+\frac{d\langle \nu_r,\tilde{\phi}\rangle}{dr}\given (\text{fixed } \nu_r)
\eeq
where the first term on the right side considers the change in $\langle \nu_r,\tilde{\phi}\rangle$
due to change in $\nu_r$ as a function of $r$ at fixed $\tilde{\phi}$ while the second term considers 
the change in $\langle \nu_r,\tilde{\phi}\rangle$ due to change in $\tilde{\phi}$ as a function
of $r$ at fixed $\nu_r$. Therefore the first term is computed using equation~\eqref{eq:mean_field_pde} and
the second term is equal to $-\langle\nu_t,\phi'\rangle$. Hence, on combining two terms
we have
\begin{multline}
\frac{d\langle \nu_r,\tilde{\phi}\rangle}{dr}=\Bigg(\sum_{n=1}^C\sum_{j=1}^n\int_{x_1}\cdots\int_{x_n}\beta(x_j)\\\times\left(\tilde{\phi}(n-1,x_1,\ldots,x_{j-1},x_{j+1},\ldots,x_n)-\tilde{\phi}(n,x_1,\ldots,x_n)\right)\,d\nu_r(n,x_1,\ldots,x_n)\\
+\lambda\bigg[\left(\nu_r(\{0\})\frac{(\ol{R}_0(\nu_r)^d-\ol{R}_1(\nu_r)^d)}{(\ol{R}_0(\nu_r)-\ol{R}_1(\nu_r))}\left(\tilde{\phi}(1,0)-\tilde{\phi}(0)\right)\right)
+\sum_{n=1}^{C-1}\sum_{j=1}^{n+1}\int_{x_1}\cdots\int_{x_n}\frac{1}{(n+1)}\\
\times\frac{(\ol{R}_n(\nu_r)^d-\ol{R}_{n+1}(\nu_r)^d)}{(\ol{R}_n(\nu_r)-\ol{R}_{n+1}(\nu_r))}
(\tilde{\phi}(n+1,x_1,\ldots,x_{j-1},0,x_j,\ldots,x_n)-\tilde{\phi}(n,x_1,\ldots,x_n))\\
\times\,d\nu_r(n,x_1,\ldots,x_n)\bigg]\Bigg).
\end{multline}
Now integrating $\frac{d\langle \nu_r,\tilde{\phi}\rangle}{dr}$ with respect to $r$ from $0$ to $t$,
we get equation~\eqref{eq:mean_field_model_new}.
%\begin{multline}
%\label{eq:mean_field_pde_new}
%\langle\nu_t,\phi\rangle=\langle\nu_0,\tau_t\phi\rangle+\int_{r=0}^t\Bigg(\sum_{n=1}^C\sum_{j=1}^n\int_{x_1}\cdots\int_{x_n}\beta(x_j)\\
%\times\left(\tilde{\phi}(n-1,x_1,\ldots,x_{j-1},x_{j+1},\ldots,x_n)-\tilde{\phi}(n,x_1,\ldots,x_n)\right)\,d\nu_r(n,x_1,\ldots,x_n)\\
%+\lambda\bigg[\left(\nu_r(\{0\})\frac{(\ol{R}_0(\nu_r)^d-\ol{R}_1(\nu_r)^d)}{(\ol{R}_0(\nu_r)-\ol{R}_1(\nu_r))}\left(\tilde{\phi}(1,0)-\tilde{\phi}(0)\right)\right)
%+\sum_{n=1}^{C-1}\sum_{j=1}^{n+1}\int_{x_1}\cdots\int_{x_n}\frac{1}{(n+1)}\\
%\times\frac{(\ol{R}_n(\nu_r)^d-\ol{R}_{n+1}(\nu_r)^d)}{(\ol{R}_n(\nu_r)-\ol{R}_{n+1}(\nu_r))}
%(\tilde{\phi}(n+1,x_1,\ldots,x_{j-1},0,x_j,\ldots,x_n)-\tilde{\phi}(n,x_1,\ldots,x_n))\\
%\times\,d\nu_r(n,x_1,\ldots,x_n)\bigg]\Bigg)\,dr.
%\end{multline}

We next prove that for $\phi\in\mc{C}_b^1(\mc{U})$, the solution $(\langle \ol{\eta}_t,\phi\rangle,t\geq 0)$ of the 
equation~\eqref{eq:mean_field_model_new} is a solution to the equation~\eqref{eq:mf_model_eqns}. This is equivalent to
proving that the differentiation of $\langle \ol{\eta}_t,\phi\rangle$ with respect to $t$ exists. Since $\phi\in\mc{C}_b^1(\mc{U})$, the existence of $\frac{d\langle \ol{\eta}_0,\tau_t\phi\rangle}{dt}$ follows from bounded convergence theorem. By using Leibniz integral rule, we verify the existence of the differentiation of the second term on the right side
of equation~\eqref{eq:mean_field_model_new} with respect to $t$. According to this rule, the first condition is that the integrand needs to be continuous with respect to both the variables $r$ and $t$. This follows from the same arguments that we used to prove the continuity of the integrand in equation~\eqref{eq:mf_model_eqns}. The second condition is that the differentiation of the integrand with respect to $t$ must exist and the differential should be continuous with respect to both $r$ and $t$. The differentiation of the integrand exists from the bounded convergence theorem as $\phi\in\mc{C}_b^1(\mc{U})$ and it is continuous with respect to $r$ and $t$ from the same arguments that we used to prove the continuity of the integrand in equation~\eqref{eq:mf_model_eqns}.
Therefore any process $\nu_t\in\mc{C}_{\mc{M}_1(\mc{U})}([0,\infty))$ is
a solution to the equation~\eqref{eq:mf_model_eqns} if and only if it
is solution to the equation~\eqref{eq:mean_field_model_new}. Further, note that $\phi$ need not be differentiable
in equation~\eqref{eq:mean_field_model_new}.
$$\eqno{\tsquare}$$
\vspace{0.2cm}
 
\noindent\textbf{Proof of Theorem}~\ref{thm:mf_model_new}:

From equation~\eqref{eq:mean_field_model_new}, we first make it clear that for all $\phi\in\mc{C}_b(\mc{U})$, 
the operator $\mc{\phi}\mapsto\langle\nu_t,\phi\rangle$ is a linear operator with $\nu_t(\mc{U})=1$. Hence from Riesz-Markov-Kakutani theorem \cite{Varad,rudin} by assuming $\nu_t\in\mc{M}_1(\mc{U})$ (since we are interested in studying the limit of a sequence of probability measures $\{\ol{\eta}_t^N\}$), existence of unique operator $\phi\mapsto\langle \nu_t,\phi\rangle$ implies the existence of the unique probability measure $\nu_t$.

Given an initial measure $\nu_0$, we next prove that there exists atmost one mean-field model solution by showing that there exists atmost one real valued process $\langle \nu_t,\phi\rangle$ corresponding to the mean-field model. Suppose $(\nu_t^1)_{t\geq 0},(\nu_t^2)_{t\geq 0}$ are two solutions satisfying the mean-field model equations
with initial points $\nu_0^1,\nu_0^2$, respectively.
Then we have, for $\phi\in\mc{C}_b(\mc{U})$,
\begin{multline}
\label{eq:difference}
\langle\nu_t^1-\nu_t^2,\phi\rangle=\langle\nu_0^1-\nu_0^2,\tau_t\phi\rangle
+\int_{s=0}^t\Bigg(\sum_{n=1}^C\sum_{j=1}^n\int_{x_1}\cdots\int_{x_n}\beta(x_j)\\
\times\left(\tau_{t-s}\phi(n-1,x_1,\ldots,x_{j-1},x_{j+1},\ldots,x_n)-\tau_{t-s}\phi(n,x_1,\ldots,x_n)\right)\,\\
\times d(\nu_s^1-\nu_s^2)(n,x_1,\ldots,x_n)\Bigg)\,ds\\
+\int_{s=0}^t\Bigg(\lambda\bigg[\left(\nu_s^1(\{0\})\frac{(\ol{R}_0(\nu_s^1)^d-\ol{R}_1(\nu_s^1)^d)}{(\ol{R}_0(\nu_s^1)-\ol{R}_1(\nu_s^1))}\left(\tau_{t-s}\phi(1,0)-\tau_{t-s}\phi(0)\right)\right)\\
+\sum_{n=1}^{C-1}\sum_{j=1}^{n+1}\int_{x_1}\cdots\int_{x_n}\frac{1}{(n+1)}\frac{(\ol{R}_n(\nu_s^1)^d-\ol{R}_{n+1}(\nu_s^1)^d)}{(\ol{R}_n(\nu_s^1)-\ol{R}_{n+1}(\nu_s^1))}\\
\times
(\tau_{t-s}\phi(n+1,x_1,\ldots,x_{j-1},0,x_j,\ldots,x_n)-\tau_{t-s}\phi(n,x_1,\ldots,x_n))\\
\times\,d\nu_s^1(n,x_1,\ldots,x_n)\bigg]\\
-\lambda\bigg[\left(\nu_s^2(\{0\})\frac{(\ol{R}_0(\nu_s^2)^d-\ol{R}_1(\nu_s^2)^d)}{(\ol{R}_0(\nu_s^2)-\ol{R}_1(\nu_s^2))}\left(\tau_{t-s}\phi(1,0)-\tau_{t-s}\phi(0)\right)\right)\\
-\sum_{n=1}^{C-1}\sum_{j=1}^{n+1}\int_{x_1}\cdots\int_{x_n}\frac{1}{(n+1)}
\frac{(\ol{R}_n(\nu_s^2)^d-\ol{R}_{n+1}(\nu_s^2)^d)}{(\ol{R}_n(\nu_s^2)-\ol{R}_{n+1}(\nu_s^2))}\\
\times(\tau_{t-s}\phi(n+1,x_1,\ldots,x_{j-1},0,x_j,\ldots,x_n)-\tau_{t-s}\phi(n,x_1,\ldots,x_n))\\
\times\,d\nu_s^2(n,x_1,\ldots,x_n)\bigg]\Bigg)ds.
\end{multline}
We next need to show the following result
\beq
\label{eq:gronwall}
\norm{\nu_t^1-\nu_t^2}\leq b+c\int_{s=0}^t\norm{\nu_s^1-\nu_s^2} ds
\eeq
for some $b,c>0$, $t\in[0,T]$ and then from the Gronewall's inequality it follows that
\beq
\norm{\nu_t^1-\nu_t^2}\leq b\, e^{ct}
\eeq
for $t\in[0,T]$.
In this direction, the first term on the right side of equation~\eqref{eq:difference} can be written as
\beq
\abs{\langle\nu_0^1-\nu_0^2,\tau_t\phi\rangle}\leq\norm{\nu_0^1-\nu_0^2}\norm{\phi}.
\eeq
To simplify the second term, we define a function $h_{t,s}$ as follows:
\begin{multline}
h_{t,s}(n,x_1,\ldots,x_n)=\\
\sum_{k=1}^n\beta(x_k)\tau_{t-s}(\phi(n-1,x_1,\ldots,x_{j-1},x_{j+1},\ldots,x_n)
-\phi(n,x_1,\ldots,x_n))
\end{multline}
and $h_{t,s}(0)=0$. Then since $\phi\in\mc{C}_b(\mc{U})$ and $\beta\in\mc{C}_b(\mc{R}_+)$,
we have $h_{t,s}\in\mc{C}_b(\mc{U})$. Further, we have
\beq
\norm{h_{t,s}}\leq 2C\norm{\beta}\norm{\phi}.
\eeq
Using the definition of $h_{t,s}$, we have
\begin{multline}
\int_{s=0}^t\Bigg(\sum_{n=1}^C\sum_{j=1}^n\int_{x_1}\cdots\int_{x_n}\beta(x_j)\bigg(\tau_{t-s}\phi(n-1,x_1,\ldots,x_{j-1},x_{j+1},\ldots,x_n)\\
-\tau_{t-s}\phi(n,x_1,\ldots,x_n)\bigg)\,d(\nu_s^1-\nu_s^2)(n,x_1,\ldots,x_n)\,ds=\int_{s=0}^t\langle\nu_s^1-\nu_s^2,h_{t,s}\rangle ds.
\end{multline}
To simplify the third term, we define a function $f_{t,s,\nu}$ as follows,
for $0\leq n\leq C-1$,
\begin{multline}
f_{t,s,\nu}(n,x_1,\ldots,x_n)=\sum_{j=1}^{n+1}\frac{1}{(n+1)}
\frac{(\ol{R}_n(\nu)^d-\ol{R}_{n+1}(\nu)^d)}{(\ol{R}_n(\nu)-\ol{R}_{n+1}(\nu))}\\
\times(\tau_{t-s}\phi(n+1,x_1,\ldots,x_{j-1},0,x_j,\ldots,x_n)-\tau_{t-s}\phi(n,x_1,\ldots,x_n))
\end{multline}
and $f_{t,s,\nu}(C,x_1,\ldots,x_C)=0$ for $x_i\geq 0$ for all $i$. Then the third term is
equal to\\
 $\int_{s=0}^t\lambda\left(\langle\nu_s^1,f_{t,s,\nu_s^1}\rangle-\langle\nu_s^2,f_{t,s,\nu_s^2}\rangle\right)\,ds$.
 Further, we can write
 \begin{align}
 \abs{\langle\nu_s^1,f_{t,s,\nu_s^1}\rangle-\langle\nu_s^2,f_{t,s,\nu_s^2}\rangle}&\leq \abs{\langle\nu_s^1-\nu_s^2,f_{t,s,\nu_s^1}\rangle}+\abs{\langle\nu_s^2,f_{t,s,\nu_s^1}-f_{t,s,\nu_s^2}\rangle}\\
 &\leq \norm{\nu_s^1-\nu_s^2}\norm{f_{t,s,\nu_s^1}}+\norm{\nu_s^2}\norm{f_{t,s,\nu_s^1}-f_{t,s,\nu_s^1}}.
 \end{align}
 Since $\nu_s^2$ is a probability measure $\norm{\nu_s^2}=1$ and also we have $\norm{f_{t,s,\nu_s^1}}\leq 2d\norm{\phi}$. We also have 
  \begin{multline}
  \abs{f_{t,s,\nu_s^1}(n,x_1,\ldots,x_n)-f_{t,s,\nu_s^2}(n,x_1,\ldots,x_n)}\\
  \leq 2d^2\norm{\phi}\left(\abs{\ol{R}_n(\nu_s^1)-\ol{R}_n(\nu_s^2)}+\abs{\ol{R}_{n+1}(\nu_s^1)-\ol{R}_{n+1}(\nu_s^2)}\right).
  \end{multline}
 We next write
 \beq
 \ol{R}_n(\nu_s^1)=\langle\nu_s^1,f^*\rangle
 \eeq
 where $f^*$ is a function defined as
 \beq
 f^*(m,x_1,\ldots,x_m)=1
 \eeq
 for $m\geq n$ and  $f^*(m,x_1,\ldots,x_m)=0$ for $m<n$. Then we have
 \beq
 \abs{\ol{R}_n(\nu_s^1)-\ol{R}_n(\nu_s^2)}\leq \norm{\nu_s^1-\nu_s^2}\norm{f^*}=\norm{\nu_s^1-\nu_s^2}.
 \eeq
 Therefore by using bounds for all the terms, we get
 \begin{multline}
 \abs{\langle\nu_t^1-\nu_t^2,\phi\rangle}\leq \left(\norm{\nu_0^1-\nu_0^2}+\int_{s=0}^t2\norm{\beta}C\norm{\nu_s^1-\nu_s^2}\,ds\right.\\
 \left.+\int_{s=0}^t8d^2\lambda\norm{\nu_s^1-\nu_s^2}\,ds\right)\norm{\phi}.
 \end{multline}
 Hence we have
 \beq
 \norm{\nu_t^1-\nu_t^2}\leq \norm{\nu_0^1-\nu_0^2}+(2C\norm{\beta}+8d^2\lambda)\int_{s=0}^t\norm{\nu_s^1-\nu_s^2}\,ds.
 \eeq
 Therefore, from equation~\eqref{eq:gronwall}, we have
 \beq
 \norm{\nu_t^1-\nu_t^2}\leq \norm{\nu_0^1-\nu_0^2}\,e^{(2C\norm{\beta}+8d^2\lambda)t}.
 \eeq
 Hence starting from an initial measure $\nu_0$, there exists atmost one solution for the mean-field model equations.
 
 We next prove that there exists a process $(\nu_t,t\geq 0)\in\mc{C}_{\mc{M}_1(\mc{U})}([0,\infty))$ satisfying the mean-field
 model equations. This follows from the relative compactness of the sequence $\{\ol{\eta}_t^N,t\geq 0\}$ in $\mc{D}_{\mc{M}_1(\mc{U})}([0,\infty))$ from the proof of Theorem~\ref{thm:mean_field_limit}. In particular, we have that every limit
 point of the sequence $\{\ol{\eta}_t^N,t\geq 0\}$ satisfies the  equation~\eqref{eq:mean_field_model_new}. Further, each limiting point is almost surely continuous. This concludes that there exists a solution to the mean-field model equations.

$$\eqno{\tsquare}$$

%%%%%%%%%%%%%%%%%%%%%%%%%%%%%%%%%%%%%%%%%%%%%%%%%%%%%%%%%%%%%%%%%%%%%%%%%%%%%%%%%%%%%%%%%%%%%%%%
%%%%%%%%%%%%%%%%%%%%%%%%%%%%%%%%%%%%%%%%%%%%%%%%%%%%%%%%%%%%%%%%%%%%%%%%%%%%%%%%%%%%%%%%%%%%%%%%%%%
%%%%%%%%%%%%%%%%%%%%%%%%%%%%%%%%%%%%%%%%%%%%%%%%%%%%%%%%%%%%%%%%%%%%%%%%%%%%%%%%%%%%%%%%%%%%%%%%%%%%%%

\section{Martingale construction}
\label{sec:martingale}

In this section, by using the infinitesimal generator of the process $(\eta_t^N)_{t\geq 0}$, we construct a martingale $(M_t^N(\phi))_{t\geq 0}\in D_{\mc{R}}([0,\infty))$ where $\phi\in\mc{C}_b^1(\mc{U})$. We then show that the scaled version of the process $(M_t^N(\phi))_{t\geq 0}$ converges in distribution to the null process based on which we later establish convergence
of the scaled version of the process $(\eta_t^N)_{t\geq 0}$.

Since the set of linear combinations of $Q_f$ for $f\in\mc{C}_s^1(\mc{U})$ is dense in
the set $\mc{C}(\mc{M}_F(\mc{U}))$\cite[proposition~$7.10$]{roberts}, for any continuous function $F\in\mc{C}(\mc{M}_F(\mc{U}))$, the infinitesimal generator $$A^NF(\nu)=\lim_{h\to 0}\frac{\mb{E}[F(\eta_h^N)\given\eta_0^N=\nu]-F(\nu)}{h}$$
is given by,
\begin{multline}
\label{eq:generator}
A^NF(\nu)=\lim_{h\to 0}\frac{F(\tau_h\nu)-F(\nu)}{h}-N\lambda F(\nu)\\
-F(\nu)\sum_{n=1}^C\sum_{j=1}^n\int_{x_1}\cdots\int_{x_n}\beta(x_j)\,d\nu(n,x_1,\ldots,x_n)
+\sum_{n=1}^C\sum_{j=1}^n\int_{x_1}\cdots\int_{x_n}\beta(x_j)\\
\times\left(F(\nu+\delta_{(n-1,x_1,\ldots,x_{j-1},x_{j+1},\ldots,x_n)}-\delta_{(n,x_1,\ldots,x_n)}\right)\,d\nu(n,x_1,\ldots,x_n)\\
+N\lambda\Bigg[\left(\frac{\nu(\{0\})}{N}\frac{(R_0(\nu)^d-R_1(\nu)^d)}{(R_0(\nu)-R_1(\nu))}\left(F(\nu+\delta_{(1,0)}-\delta_{(0)})\right)\right)\\
+\sum_{n=1}^{C-1}\sum_{j=1}^{n+1}\int_{x_1}\cdots\int_{x_n}\frac{1}{N(n+1)}
\frac{(R_n(\nu)^d-R_{n+1}(\nu)^d)}{(R_n(\nu)-R_{n+1}(\nu))}\\
\times F(\nu+\delta_{(n+1,x_1,\ldots,x_{j-1},0,x_j,\ldots,x_n)}-\delta_{(n,x_1,\ldots,x_n)})\,d\nu(n,x_1,\ldots,x_n)\\
+\int_{x_1}\cdots\int_{x_C}\frac{1}{N}\frac{(R_C(\nu)^d-R_{C+1}(\nu)^d)}{(R_C(\nu)-R_{C+1}(\nu))}F(\nu)\,d\nu(C,x_1,\ldots,x_C)\Bigg],
\end{multline}
where $F$ is such that the limit exists and $R_{C+1}(\nu)=0$.

\begin{prop}
\label{thm:martingale}
For all $\phi\in\mc{C}_b^1(\mc{U})$, the process $(M_t^N(\phi))_{t\geq 0}$ given by
%\vspace{-0mm}
\beq
%\vspace{-8mm}
\label{eq:martingale}
M_t^N(\phi)=\langle\eta_t^N,\phi\rangle-\langle\eta_0^N,\phi\rangle-\int_{s=0}^tA^N\langle\eta^N_s,\phi\rangle ds
\eeq
is a RCLL (process that is right continuous with left limits) square integrable \\$\mc{F}_t^N-$martingale.
For $\phi,\psi\in\mc{C}_b^1(\mc{U})$, the mutual variation of $(M_t^N(\phi))_{t\geq 0}$ with $(M_t^N(\psi))_{t\geq 0}$ is
given by
%\vspace{-5mm}
\begin{multline}
\label{eq:mutual_variation}
<M^N_.(\phi),M^N_.(\psi)>_t=\int_{s=0}^t\Bigg(\sum_{n=1}^C\sum_{j=1}^n\int_{x_1}\cdots\int_{x_n}\beta(x_j)\\
\times\left(\phi(n-1,x_1,\ldots,x_{j-1},x_{j+1},\ldots,x_n)-\phi(n,x_1,\ldots,x_n)\right)\\
\times\left(\psi(n-1,x_1,\ldots,x_{j-1},x_{j+1},\ldots,x_n)-\psi(n,x_1,\ldots,x_n)\right)\,d\eta_s^N(n,x_1,\ldots,x_n)\\
+N\lambda\bigg[\left(\frac{\eta_s^N(\{0\})}{N}\frac{(R_0(\eta_s^N)^d-R_1(\eta_s^N)^d)}{(R_0(\eta_s^N)-R_1(\eta_s^N))}\left(\phi(1,0)-\phi(0)\right)\left(\psi(1,0)-\psi(0)\right)\right)\\
+\sum_{n=1}^{C-1}\sum_{j=1}^{n+1}\int_{x_1}\cdots\int_{x_n}\frac{1}{N(n+1)}
\frac{(R_n(\eta_s^N)^d-R_{n+1}(\eta_s^N)^d)}{(R_n(\eta_s^N)-R_{n+1}(\eta_s^N))}\\
\times(\phi(n+1,x_1,\ldots,x_{j-1},0,x_j,\ldots,x_n)-\phi(n,x_1,\ldots,x_n))\\
\times (\psi(n+1,x_1,\ldots,x_{j-1},0,x_j,\ldots,x_n)-\psi(n,x_1,\ldots,x_n))\,d\eta_s^N(n,x_1,\ldots,x_n)\bigg]\Bigg)ds
\end{multline}
\end{prop}
%\vspace{-.9cm}

\noindent{\bf Proof:}
For $\phi\in\mc{C}_b^1(\mc{U})$, it is clear that the function $(\langle\eta_t^N,\phi\rangle)_{t\geq 0}$ belongs to the domain of $A^N$.
Therefore, by using the Dynkin's formula \cite{Ethier_Kurtz_book}, the process
$(M_t^N(\phi))_{t\geq 0}$ defined by
\beq
\label{eq:martingale2}
M_t^N(\phi)=\langle\eta_t^N,\phi\rangle-\langle\eta_0^N,\phi\rangle-\int_{s=0}^tA^N\langle\eta^N_s,\phi\rangle ds
\eeq
is a RCLL $\mc{F}_t^N-$local martingale. Therefore, by simplification, we get
\begin{multline}
%\label{eq:martingale}
M_t^N(\phi)=\langle\eta_t^N,\phi\rangle-\langle\eta_0^N,\phi\rangle-\int_{s=0}^t \langle\eta_s^N,\phi'\rangle\,ds
-\int_{s=0}^t\Bigg(\sum_{n=1}^C\sum_{j=1}^n\int_{x_1}\cdots\int_{x_n}\beta(x_j)\\
\times\left(\phi(n-1,x_1,\ldots,x_{j-1},x_{j+1},\ldots,x_n)-\phi(n,x_1,\ldots,x_n)\right)
d\eta_s^N(n,x_1,\ldots,x_n)\\
+N\lambda\bigg[\left(\frac{\eta_s^N(\{0\})}{N}\frac{(R_0(\eta_s^N)^d-R_1(\eta_s^N)^d)}{(R_0(\eta_s^N)-R_1(\eta_s^N))}\left(\phi(1,0)-\phi(0)\right)\right)\\
+\sum_{n=1}^{C-1}\sum_{j=1}^{n+1}\int_{x_1}\cdots\int_{x_n}\frac{1}{N(n+1)}
\frac{(R_n(\eta_s^N)^d-R_{n+1}(\eta_s^N)^d)}{(R_n(\eta_s^N)-R_{n+1}(\eta_s^N))}\\
\times
(\phi(n+1,x_1,\ldots,x_{j-1},0,x_j,\ldots,x_n)-\phi(n,x_1,\ldots,x_n))\,d\eta_s^N(n,x_1,\ldots,x_n)\bigg]\Bigg)ds,
\end{multline}
where 
\begin{equation}
\langle\eta_s^N,\phi'\rangle=\sum_{n=1}^C\sum_{i=1}^n\int_{x_1}\cdots\int_{x_n}\frac{\partial \phi(n,x_1,\ldots,x_n)}{\partial x_i}\,d\eta_s^N(n,x_1,\ldots,x_n). 
\end{equation}
Let $\psi\in\mc{C}_b^1(\mc{U})$, then the mapping $(\langle\eta_t^N,\phi\rangle\langle\eta_t^N,\psi\rangle)_{t\geq 0}$ also belongs to the
domain of $A^N$. Let the martingale $\tilde{M}_t^N(\phi,\psi)$ be defined by
\beq
\label{eq:joint_martingale}
\tilde{M}_t^N(\phi,\psi)=\langle\eta_t^N,\phi\rangle\langle\eta_t^N,\psi\rangle-\langle\eta_0^N,\phi\rangle\langle\eta_0^N,\psi\rangle-\int_{s=0}^tA^N\langle\eta^N_s,\phi\rangle\langle\eta^N_s,\psi\rangle ds
\eeq
is a RCLL $\mc{F}_t^N-$local martingale. It is verified that, we have
\begin{multline}
\label{eq:generator_relaton}
A^N\langle\nu,\phi\rangle\langle\nu,\psi\rangle=\langle\nu,\phi\rangle A^N\langle\nu,\psi\rangle+\langle\nu,\psi\rangle A^N\langle\nu,\phi\rangle\\
+\sum_{n=1}^C\sum_{j=1}^n\int_{x_1}\cdots\int_{x_n}\beta(x_j)\left(\phi(n-1,x_1,\ldots,x_{j-1},x_{j+1},\ldots,x_n)-\phi(n,x_1,\ldots,x_n)\right)\\
\times\left(\psi(n-1,x_1,\ldots,x_{j-1},x_{j+1},\ldots,x_n)-\psi(n,x_1,\ldots,x_n)\right)\,d\nu(n,x_1,\ldots,x_n)\\
+N\lambda\bigg[\left(\frac{\nu(\{0\})}{N}\frac{(R_0(\nu)^d-R_1(\nu)^d)}{(R_0(\nu)-R_1(\nu))}\left(\phi(1,0)-\phi(0)\right)\right)
+\sum_{n=1}^{C-1}\sum_{j=1}^{n+1}\int_{x_1}\cdots\int_{x_n}\frac{1}{N(n+1)}\\
\times\frac{(R_n(\nu)^d-R_{n+1}(\nu)^d)}{(R_n(\nu)-R_{n+1}(\nu))}
(\phi(n+1,x_1,\ldots,x_{j-1},0,x_j,\ldots,x_n)-\phi(n,x_1,\ldots,x_n))\\
\times (\psi(n+1,x_1,\ldots,x_{j-1},0,x_j,\ldots,x_n)-\psi(n,x_1,\ldots,x_n))\,d\nu(n,x_1,\ldots,x_n)\bigg].
\end{multline}
By using It$\hat{o}$'s formula, we have
\begin{multline}
\langle\eta_t^N,\phi\rangle\langle\eta_t^N,\psi\rangle=\langle\eta_0^N,\phi\rangle\langle\eta_0^N,\psi\rangle+\int_{s=0}^t\langle\eta_s^N,\phi\rangle\,dM_s^N(\psi)\\
+\int_{s=0}^t\langle\eta_s^N,\psi\rangle\,dM_s^N(\phi)+\int_{s=0}^t\langle\eta_s^N,\phi\rangle A^N\langle\eta_s^N,\psi\rangle\,ds\\
+\int_{s=0}^t\langle\eta_s^N,\psi\rangle\,A^N\langle\eta_s^N,\phi\rangle\,ds
+<\langle\eta_{\cdot}^N,\phi\rangle,\langle\eta_{\cdot}^N,\psi\rangle>_t.
\end{multline}
Further, by using equations~\eqref{eq:joint_martingale}-\eqref{eq:generator_relaton}, we have
\begin{multline}
\label{eq:variation_process_relation}
\int_{s=0}^t\langle\eta_s^N,\phi\rangle\,dM_s^N(\psi)
+\int_{s=0}^t\langle\eta_s^N,\psi\rangle\,dM_s^N(\phi)\\
+\int_{s=0}^t\langle\eta_s^N,\psi\rangle\,A^N\langle\eta_s^N,\phi\rangle\,ds
+<\langle\eta_{\cdot}^N,\phi\rangle,\langle\eta_{\cdot}^N,\psi\rangle>_t
=\\
\tilde{M}_t^N(\phi,\psi)+\int_{s=0}^t\Bigg(\sum_{n=1}^C\sum_{j=1}^n\int_{x_1}\cdots\int_{x_n}\beta(x_j)\\
\times\left(\phi(n-1,x_1,\ldots,x_{j-1},x_{j+1},\ldots,x_n)-\phi(n,x_1,\ldots,x_n)\right)\\
\times\left(\psi(n-1,x_1,\ldots,x_{j-1},x_{j+1},\ldots,x_n)-\psi(n,x_1,\ldots,x_n)\right)\,d\eta_s^N(n,x_1,\ldots,x_n)\\
+N\lambda\bigg[\left(\frac{\eta_s^N(\{0\})}{N}\frac{(R_0(\eta_s^N)^d-R_1(\eta_s^N)^d)}{(R_0(\eta_s^N)-R_1(\eta_s^N))}\left(\phi(1,0)-\phi(0)\right)\left(\psi(1,0)-\psi(0)\right)\right)\\
+\sum_{n=1}^{C-1}\sum_{j=1}^{n+1}\int_{x_1}\cdots\int_{x_n}\frac{1}{N(n+1)}\\
\times\frac{(R_n(\eta_s^N)^d-R_{n+1}(\eta_s^N)^d)}{(R_n(\eta_s^N)-R_{n+1}(\eta_s^N))}
(\phi(n+1,x_1,\ldots,x_{j-1},0,x_j,\ldots,x_n)-\phi(n,x_1,\ldots,x_n))\\
\times (\psi(n+1,x_1,\ldots,x_{j-1},0,x_j,\ldots,x_n)-\psi(n,x_1,\ldots,x_n))\,d\eta_s^N(n,x_1,\ldots,x_n)\bigg]\Bigg)ds.
\end{multline}
By identifying the finite variation process, $\mc{P}-$a.s. we have 
\begin{multline}
<\langle\eta_{\cdot}^N,\phi\rangle,\langle\eta_{\cdot}^N,\psi\rangle>_t=\\
\int_{s=0}^t\Bigg(\sum_{n=1}^C\sum_{j=1}^n\int_{x_1}\cdots\int_{x_n}\beta(x_j)\left(\phi(n-1,x_1,\ldots,x_{j-1},x_{j+1},\ldots,x_n)-\phi(n,x_1,\ldots,x_n)\right)\\
\times\left(\psi(n-1,x_1,\ldots,x_{j-1},x_{j+1},\ldots,x_n)-\psi(n,x_1,\ldots,x_n)\right)\,d\eta_s^N(n,x_1,\ldots,x_n)\\
+N\lambda\bigg[\left(\frac{\eta_s^N(\{0\})}{N}\frac{(R_0(\eta_s^N)^d-R_1(\eta_s^N)^d)}{(R_0(\eta_s^N)-R_1(\eta_s^N))}\left(\phi(1,0)-\phi(0)\right)\left(\psi(1,0)-\psi(0)\right)\right)\\
+\sum_{n=1}^{C-1}\sum_{j=1}^{n+1}\int_{x_1}\cdots\int_{x_n}\frac{1}{N(n+1)}
\frac{(R_n(\eta_s^N)^d-R_{n+1}(\eta_s^N)^d)}{(R_n(\eta_s^N)-R_{n+1}(\eta_s^N))}\\
\times(\phi(n+1,x_1,\ldots,x_{j-1},0,x_j,\ldots,x_n)-\phi(n,x_1,\ldots,x_n))\\
\times (\psi(n+1,x_1,\ldots,x_{j-1},0,x_j,\ldots,x_n)-\psi(n,x_1,\ldots,x_n))\,d\eta_s^N(n,x_1,\ldots,x_n)\bigg]\Bigg)ds.
\end{multline}
From equation~\eqref{eq:martingale2}, we have $<\langle\eta_{\cdot}^N,\phi\rangle,\langle\eta_{\cdot}^N,\psi\rangle>_t=<M_{\cdot}^N(\phi),M_{\cdot}^N(\psi)>_t$. Therefore since $\phi,\psi\in\mc{C}_b^1(\mc{U})$ and $\beta\in\mc{C}_b(\mc{R}_+)$, we have 
\beq
\expect{<M_{\cdot}^N(\phi),M_{\cdot}^N(\psi)>_t}<\infty\eeq and hence
$(M_t^N(\phi))_{t\geq 0}$
 is a square integrable martingale.
$$\eqno{\tsquare}$$

%%%%%%%%%%%%%%%%%%%%%%%%%%%%%%%%%%%%%%%%%%%%%%%%%%%%%%%%%%%%%%%%%%%%%%%%%%%%%%%%%%%%%%%%%%%%%%%%%%%%%
%%%%%%%%%%%%%%%%%%%%%%%%%%%%%%%%%%%%%%%%%%%%%%%%%%%%%%%%%%%%%%%%%%%%%%%%%%%%%%%%%%%%%%%%%%%%%%%%%%%%%%
%%%%%%%%%%%%%%%%%%%%%%%%%%%%%%%%%%%%%%%%%%%%%%%%%%%%%%%%%%%%%%%%%%%%%%%%%%%%%%%%%%%%%%%%%%%%%%%%%%%%%%
\section{Mean-field limit}
\label{sec:MFEproof}

In this section we consider a sequence of systems indexed by $N$ such that a system with
index $N$ has $N$ servers in which jobs arrive according to a Poisson process with rate $N\lambda$
and all other system parameters are identical for all $N$. For given $N$, the process
$(\eta_t^N)_{t\geq 0}$ defined in equation~\eqref{eq:state_descriptor} describes the system dynamics
of a system with index $N$ such that $\eta_t^N(\{\ul{u}\})$ denotes the number of servers lying in
state $\ul{u}$ at time $t$. We now construct another process $(\ol{\eta}_t^N)_{t\geq 0}$ as follows
\beq
\label{eq:normalized_descriptor}
\ol{\eta}_t^N=\frac{\eta_t^N}{N}.
\eeq
Therefore $\ol{\eta}_t^N(\{\ul{u}\})$ denotes the fraction of servers lying in state $\ul{u}$ at time $t$.
Let $(\ol{\mc{F}}_t^N)_{t\geq 0}$ denotes the filtration associated with the process
 $(\ol{\eta}_t^N)_{t\geq 0}$.
Note that we have $(\ol{\eta}_t^N)_{t\geq 0}\in\mc{D}_{\mc{M}_1^N(\mc{U})}([0,\infty))$. We first show that
the sequence of processes $(\ol{\eta}_t^N,t\geq 0)$ is relatively compact and then we prove
that every limit point $(\bm{\chi}_t,t\geq 0)$ has sample path that is almost surely continuous with respect to $t$ and 
coincides with the mean-field model solution of the
system. Since for every limit point $(\bm{\chi}_t,t\geq 0)$, $\bm{\chi}_0$ is almsot surely same as the random measure $\mf{\Theta}$ from assumption~\ref{assum:initial_measures} and the mean-field model solution is unique for given initial measure, then we have that almost surely all
limit points are identical referred to as the the mean-field 
limit denoted by $(\ol{\eta}_t,t\geq 0)$. 

For $\phi\in\mc{C}_b^1(\mc{U})$, from Proposition~\ref{thm:martingale}, the process
$(\ol{M}_t^N(\phi))_{t\geq 0}$ defined as follows is an RCLL square integrable $\ol{\mc{F}}_t^N-$martingale
 \begin{multline}
\label{eq:normalized_martingale}
%\begin{multline}
%\label{eq:martingale}
\ol{M}_t^N(\phi)=\langle\ol{\eta}_t^N,\phi\rangle-\langle\ol{\eta}_0^N,\phi\rangle-\int_{s=0}^t \langle\ol{\eta}_s^N,\phi'\rangle\,ds\\
-\int_{s=0}^t\Bigg(\sum_{n=1}^C\sum_{j=1}^n\int_{x_1}\cdots\int_{x_n}\beta(x_j)\\
\times\left(\phi(n-1,x_1,\ldots,x_{j-1},x_{j+1},\ldots,x_n)-\phi(n,x_1,\ldots,x_n)\right)\,d\ol{\eta}_s^N(n,x_1,\ldots,x_n)\\
+\lambda\bigg[\left(\ol{\eta}_s^N(\{0\})\frac{(\ol{R}_0(\ol{\eta}_s^N)^d-\ol{R}_1(\ol{\eta}_s^N)^d)}{(\ol{R}_0(\ol{\eta}_s^N)-\ol{R}_1(\ol{\eta}_s^N))}\left(\phi(1,0)-\phi(0)\right)\right)
+\sum_{n=1}^{C-1}\sum_{j=1}^{n+1}\int_{x_1}\cdots\int_{x_n}\frac{1}{(n+1)}\\
\times\frac{(\ol{R}_n(\ol{\eta}_s^N)^d-\ol{R}_{n+1}(\ol{\eta}_s^N)^d)}{(\ol{R}_n(\ol{\eta}_s^N)-\ol{R}_{n+1}(\ol{\eta}_s^N))}
(\phi(n+1,x_1,\ldots,x_{j-1},0,x_j,\ldots,x_n)-\phi(n,x_1,\ldots,x_n))\\
\times\,d\ol{\eta}_s^N(n,x_1,\ldots,x_n)\bigg]\Bigg)ds,
\end{multline}
where
\beq
\ol{R}_n(\ol{\eta}_s^N)=\sum_{i=n}^C\ol{\eta}_s^N(\mc{U}_i). 
\eeq
We further have
\begin{multline}
\label{eq:normalized_variation_process}
<\ol{M}_\cdot^N(\phi),\ol{M}_\cdot^N(\psi)>_t=
\frac{1}{N}\Bigg[\int_{s=0}^t\Bigg(\sum_{n=1}^C\sum_{j=1}^n\int_{x_1}\cdots\int_{x_n}\beta(x_j)\\
\times\left(\phi(n-1,x_1,\ldots,x_{j-1},x_{j+1},\ldots,x_n)-\phi(n,x_1,\ldots,x_n)\right)\\
\times \left(\psi(n-1,x_1,\ldots,x_{j-1},x_{j+1},\ldots,x_n)-\psi(n,x_1,\ldots,x_n)\right)\,d\ol{\eta}_s^N(n,x_1,\ldots,x_n)\\
+\lambda\bigg[\left(\ol{\eta}_s^N(\{0\})\frac{(\ol{R}_0(\ol{\eta}_s^N)^d-\ol{R}_1(\ol{\eta}_s^N)^d)}{(\ol{R}_0(\ol{\eta}_s^N)-\ol{R}_1(\ol{\eta}_s^N))}\left(\phi(1,0)-\phi(0)\right)\left(\psi(1,0)-\psi(0)\right)\right)\\
+\sum_{n=1}^{C-1}\sum_{j=1}^{n+1}\int_{x_1}\cdots\int_{x_n}\frac{1}{(n+1)}\\
\times\frac{(\ol{R}_n(\ol{\eta}_s^N)^d-\ol{R}_{n+1}(\ol{\eta}_s^N)^d)}{(\ol{R}_n(\ol{\eta}_s^N)-\ol{R}_{n+1}(\ol{\eta}_s^N))}
(\phi(n+1,x_1,\ldots,x_{j-1},0,x_j,\ldots,x_n)-\phi(n,x_1,\ldots,x_n))\\
\times (\psi(n+1,x_1,\ldots,x_{j-1},0,x_j,\ldots,x_n)-\psi(n,x_1,\ldots,x_n))\\
\times\,d\ol{\eta}_s^N(n,x_1,\ldots,x_n)\bigg]\Bigg)ds\Bigg].
\end{multline}

We are now ready to establish the convergence of $(\ol{\eta}_t^N)_{t\geq 0}$. Before proving the convergence
of $(\ol{\eta}_t^N)_{t\geq 0}$, we first state the crux of the proof. We first establish
that the sequence of the processes $\{(\ol{\eta}_t^N)_{t\geq 0}\}$ is relatively compact
in $\mc{D}_{\mc{M}_1(\mc{U})}([0,\infty))$. Since the space $\mc{M}_1(\mc{U})$ endowed with
the weak topology is complete and separable, by Prohorov's theorem \cite{Billing}, establishing the
relative compactness of the sequence of the processes $\{(\ol{\eta}_t^N)_{t\geq 0}\}$ is equivalent
to proving the tightness of the processes $\{(\ol{\eta}_t^N)_{t\geq 0}\}$. From Theorem~$4.6$ of \cite{jakubowski},  Jakubowski's criteria that we recall below can be used to establish the
relative compactness of the sequence of the processes $\{(\ol{\eta}_t^N)_{t\geq 0}\}$.\\
\vspace{0.2cm}

\textbf{Jakubowski's criteria:}
\vspace{0.2cm}

A sequence of $\{X^N\}$ of $\mc{D}_{\mc{M}_1(\mc{U})}([0,\infty))-$ valued random elements
defined on $(\Omega,\mb{F},\mb{P})$ is tight if and only if the following
two conditions are satisfied:
\begin{enumerate}
\item[J1:]
For each $T>0$ and $\gamma>0$, there exists a compact set $\mb{K}_{T,\gamma}\subset\mc{M}_1(\mc{U})$ such that
\beq
\label{eq:J1}
\liminf_{N\to\infty}\mb{P}(X^N_t\in\mb{K}_{T,\gamma} \forall t\in[0,T])>1-\gamma.
\eeq
This condition is called as the compact-containment condition.
\item[J2:]
There exists a family $\mc{Q}$ of real valued continuous functions $F$ defined on $\mc{M}_1(\mc{U})$ that
separates points in $\mc{M}_1(\mc{U})$ and is closed under addition such that
for every $F\in \mc{Q}$, the sequence $\{(F(X_t^N))_{t\geq 0}\}$ is tight in
$\mc{D}_{\mc{R}}([0,\infty))$.
\end{enumerate}

To show condition J2, we define a class of functions $\mc{Q}$ as follows.
\beq
\label{def:classQ}
\mc{Q}\triangleq \{F:\exists f\in\mc{C}_b^1(\mc{U}) \text{ such that } F(\nu)=\langle\nu,f\rangle,\, \forall \nu\in\mc{M}_1(\mc{U}) \}
\eeq
Clearly every function $F\in\mc{Q}$ is continuous \wrt\,the weak topology on $\mc{M}_1(\mc{U})$
and further the class of functions $\mc{Q}$ separates points in $\mc{M}_1(\mc{U})$ and also closed under addition.

We next state the following sufficient condition (From Theorem~$C.9$,\cite{roberts}) to prove condition J2.\\
\vspace{0.2cm}

\textbf{Tightness in $\mc{D}_{\mc{R}}([0,T])$:} If $S=\mc{D}_{\mc{R}}([0,T])$ and $(\mb{P}_n)$
is a sequence of probability distributions on $S$, then $(\mb{P}_n)$ is tight if for any $\epsilon>0$,
\begin{enumerate}
\item[C1:]
There exists $b$ such that 
\beq
\mb{P}_n(\abs{X(0)}>b)\leq\epsilon
\eeq for all $n\in\mc{Z}_+$
\item[C2:]
For any $\gamma>0$, there exists $\rho>0$ such that 
\beq
\mb{P}_n(w_X(\rho)>\gamma)\leq\epsilon
\eeq
for $n$ sufficiently large, where
\beq
w_X(\rho)=\sup\{\abs{X(t)-X(s)}:s,t\leq T, \abs{s-t}\leq\rho\}
\eeq
and any limiting point $\mb{P}$ satisfies $\mb{P}(\mc{C}_{\mc{R}}([0,T]))=1$.
\end{enumerate}

By using conditions J1, J2, C1, and C2, we next give proof of Theorem~\ref{thm:mean_field_limit}.  
\vspace{0.2cm}

\textbf{Proof of Theorem~\ref{thm:mean_field_limit}:}

%\begin{thm}
%\label{thm:meanfield}
%If $\ol{\eta}_0^N\Rightarrow\ol{\eta}_0$ as $N\to\infty$, then we have 
%\beq
%\ol{\eta}^N\Rightarrow \ol{\eta}
%\eeq
%in $\mc{D}_{\mc{M}_1(\mc{U})}([0,\infty))$ as $N\to\infty$.
%\end{thm}
We first establish the relative compactness of the sequence $(\ol{\eta}^N_t)_{t\geq 0}$. For this, we next
prove the conditions C1 and C2 that are sufficient to prove the relative compactness of
$(\langle\ol{\eta}^N_t,\phi\rangle)_{t\geq 0}$ for $\phi\in\mc{C}_b^1(\mc{U})$ in $D_{\mc{R}}([0,\infty))$.
For any $T>0$, $t\in[0,T]$, we have
\beq
\langle\ol{\eta}_t^N,\phi\rangle\leq \norm{\phi}_1 \langle\ol{\eta}_t^N,\mf{1}\rangle
\eeq
and since $\langle\ol{\eta}_t^N,\mf{1}\rangle=1$, the condition C1 is trivially satisfied with $b=\norm{\phi}_1$.

We next prove condition C2. For $\epsilon>0$, by using equation~\eqref{eq:normalized_variation_process} and
Doob's inequality, we have
\begin{align}
%\end{eqnarray}
\mc{P}\left(\sup_{t\leq T}\abs{\ol{M}_t^N(\phi)}\geq \epsilon\right)&\leq \frac{4}{\epsilon^2}\expect{<\ol{M}^N_{\cdot}(\phi)>_T}\\
&\leq 4T\norm{\phi}^2\frac{1}{N}(\norm{\beta}+d\lambda)\to 0
\end{align}
as $N\to\infty$. Therefore the sequence of processes $(\ol{M}_t^N(\phi))_{t\geq 0}$ converges in distribution
to the null process from standard convergence criterion in $\mc{D}_{\mc{R}}([0,T])$. Further,
the sequence of processes $(\ol{M}_t^N(\phi))_{t\geq 0}$ is tight in $\mc{D}_{\mc{R}}([0,T])$ and
hence, there exists $\rho'>0$ and $N'>0$ such that for all $N\geq N'$, we have
\beq
\label{eq:martingale_tight}
\mc{P}\left(\sup_{u,v\leq T,\abs{u-v}\leq\rho'}\abs{\ol{M}^N_v(\phi)-\ol{M}^N_u(\phi)}\geq \frac{\gamma}{2}\right)\leq \frac{\epsilon}{2}
\eeq

For any $u<v\leq T$, from equation~\eqref{eq:normalized_martingale}, we have
\begin{multline}
\abs{\langle\ol{\eta}_v^N,\phi\rangle-\langle\ol{\eta}_u^N,\phi\rangle}\leq \int_{s=u}^v\abs{\langle\ol{\eta}_s^N,\phi'\rangle}ds+2\norm{\beta}\norm{\phi}C\abs{u-v}+2\norm{\phi}\lambda\abs{u-v}\\
+\abs{\ol{M}_v^N(\phi)-\ol{M}_u^N(\phi)}.
\end{multline}
We further can write
\beq
\label{eq:change}
\abs{\langle\ol{\eta}_v^N,\phi\rangle-\langle\ol{\eta}_u^N,\phi\rangle}\leq \abs{v-u}C\norm{\phi}_1(1+2\norm{\beta}+2d\lambda)+\abs{\ol{M}_v^N(\phi)-\ol{M}_u^N(\phi)}.
\eeq
Therefore by using equations $\eqref{eq:martingale_tight}$ and $\eqref{eq:change}$, there exists $\rho>0$
and $N_1>0$ such that for $N\geq N_1$
\beq
\label{eq:process_tight}
\mc{P}\left(\sup_{u,v\leq T,\abs{u-v}\leq\rho}\abs{\langle\ol{\eta}^N_v,\phi\rangle-\langle\ol{\eta}^N_u,\phi\rangle}\geq \gamma\right)\leq \epsilon.
\eeq
This proves condition C2. Combining C1 and C2, we have condition J2.

We next prove compact containment condition J1. For this, let us consider
\beq
\langle\ol{\eta}_t^N,\mc{I}\rangle=\sum_{n=1}^C\int_{x_1}\cdots\int_{x_n}(x_1+\ldots+x_n)\,d\ol{\eta}^N_t(n,x_1,\ldots,x_n).
\eeq
Let $(n_{i}(t),x_{i1}(t)\ldots,x_{in_i(t)}(t))$ denotes the state of the $i^{\text{th}}$ server at time $t$ where $x_{ij}(t)$ denotes the age of the $j^{\text{th}}$ job at $i^{\text{th}}$
server. Clearly, we have 
\beq
\langle\ol{\eta}_t^N,\mc{I}\rangle=\frac{1}{N}\sum_{i=1,n_i(t)>0}^N(x_{i1}(t)+\cdots+x_{in_i(t)}(t)).
\eeq
Suppose $Y_t$ is the random variable representing the age of a job that is in progress at time $t$, and
$X$ is a random variable with job length distribution $G$, then for any $b\geq 0$, we have
\beq
\label{eq:coupling}
\mc{P}(Y_t\geq b)\leq \mc{P}(X\geq b).
\eeq
Therefore using equation~\eqref{eq:coupling}, since each server has capacity $C$, for any time $t\geq 0$, we can write
\beq
\mc{P}(\langle\ol{\eta}_t^N,\mc{I}\rangle\geq b)\leq \mc{P}(\frac{1}{N}\sum_{i=1}^N(Y_{i1}+\ldots+Y_{iC})\geq b),
\eeq
where $(Y_{ij},1\leq i\leq N,1\leq j\leq C)$ are i.i.d random variables with distribution $G$.
Further, by weak law of large numbers, we have
\beq
\frac{1}{N}\sum_{i=1}^N(Y_{i1}+\ldots+Y_{iC})\Rightarrow \frac{C}{\mu}
\eeq
as $N\to\infty$.
Therefore, we have
\beq
\label{eq:compact_set}
\mc{P}\left(\sup_{t\in[0,T]}\langle\ol{\eta}_t^N,\mc{I}\rangle \,> \frac{2C}{\mu}\right)\to 0
\eeq
as $N\to\infty$. For all $0<\gamma<1$, let us define
\beq
\mc{L}_{T,\gamma}\triangleq \left\{\zeta\in\mc{M}_1(\mc{U}):\,\langle\zeta,\mc{I}\rangle\leq \frac{2C}{\mu}\right\}.
\eeq
Since $\langle\zeta,\mc{I}\rangle\leq \frac{2C}{\mu}$ for $\zeta\in\mc{L}_{T,\gamma}$, for any Borel
set $B=([0,y_1),\ldots,[0,y_n))\in\mc{B}(\mc{U}_n)$ with $n\geq 1$, we have
\beq
\zeta([y_1,\infty),\ldots,[y_n,\infty))\leq \zeta([0,\infty),\ldots,[0,\infty),[y_{i},\infty),[0,\infty),\ldots,[0,\infty)).
\eeq
We further have for any $i$,
\beq
\zeta([y_1,\infty),\ldots,[y_n,\infty))\leq \frac{2C}{\mu y_i}
\eeq
and hence 
\beq
\lim_{y_1\to\infty}\cdots\lim_{y_n\to\infty}\sup_{\zeta\in\mc{L}_{T,\gamma}}\zeta([y_1,\infty),\ldots,[y_n,\infty))=0.
\eeq

Therefore from Lemma~A$7.5$ of \cite{kallenberg}, $\mc{L}_{T,\gamma}$ is relatively compact in $\mc{M}_1(\mc{U})$.
Further, from equation~\eqref{eq:compact_set}, we have
\beq
\liminf_{N\to\infty}\mc{P}(\ol{\eta}_t^N\in \mc{L}_{T,\gamma} \forall t\in[0,T])\geq 1-\gamma.
\eeq
Suppose $\mb{K}_{T,\gamma}$ is the closure of $\mc{L}_{T,\gamma}$, then we
have a compact set $\mb{K}_{T,\gamma}$ such that 
\beq
\liminf_{N\to\infty}\mc{P}(\ol{\eta}_t^N\in \mb{K}_{T,\gamma} \forall t\in[0,T])\geq 1-\gamma.
\eeq
This establishes the condition J1 and hence the proof of tightness of the sequence
of processes $(\ol{\eta}^N_t)_{t\geq 0}$ is completed. 

Let $(\bm{\chi}_t)_{t\geq 0}$ be
a limit of a converging subsequence of $(\ol{\eta}_t^N)_{\geq 0}$ such that $\bm{\chi}_0$ is almost surely same
as the random measure $\Theta$. Then by the continuous mapping theorem, the sample path evolution
satisfies
\begin{multline}
\label{eq:mf_limit_point}
\langle\chi_t,\phi\rangle=\langle\chi_0,\phi\rangle+\int_{s=0}^t \langle\chi_s,\phi'\rangle\,ds\\
-\int_{s=0}^t\Bigg(\sum_{n=1}^C\sum_{j=1}^n\int_{x_1}\cdots\int_{x_n}\beta(x_j)\\
\times\left(\phi(n-1,x_1,\ldots,x_{j-1},x_{j+1},\ldots,x_n)-\phi(n,x_1,\ldots,x_n)\right)\,d\chi_s(n,x_1,\ldots,x_n)\\
+\lambda\bigg[\left(\chi_s(\{0\})\frac{(\ol{R}_0(\chi_s)^d-\ol{R}_1(\chi_s)^d)}{(\ol{R}_0(\chi_s)-\ol{R}_1(\chi_s))}\left(\phi(1,0)-\phi(0)\right)\right)
+\sum_{n=1}^{C-1}\sum_{j=1}^{n+1}\int_{x_1}\cdots\int_{x_n}\frac{1}{(n+1)}\\
\times\frac{(\ol{R}_n(\chi_s)^d-\ol{R}_{n+1}(\chi_s)^d)}{(\ol{R}_n(\chi_s)-\ol{R}_{n+1}(\chi_s))}
(\phi(n+1,x_1,\ldots,x_{j-1},0,x_j,\ldots,x_n)-\phi(n,x_1,\ldots,x_n))\\
\times\,d\chi_s(n,x_1,\ldots,x_n)\bigg]\Bigg)ds.
\end{multline}

We next prove that if every limit point $(\bm{\chi}_t,t\geq 0)$ with the random measure $\bm{\chi}_0$ almost surely same as
the random measure $\bm{\Theta}$, then the sample paths coincide almost surely with the unique mean-field model
solution. The first property that the mapping $t\mapsto\chi_t$ is continuous follows from equation~\eqref{eq:process_tight} from which
we have
\beq
\langle\chi_t,\phi\rangle=\langle\chi_{t-},\phi\rangle
\eeq
almost surely for all $\phi\in\mc{C}_b^1(\mc{U})$. Since $\mc{C}_b^1(\mc{U})$ is a separating class and $(\chi_t,t\geq 0)\in\mc{D}_{\mc{M}_1(\mc{U})}([0,\infty))$, we have almost surely
\beq
\chi_t=\chi_{t-}
\eeq
and
\beq
\chi_t=\chi_{t+}.
\eeq
Hence the mapping $t\mapsto\chi_t$ is continuous almost surely.
%However, from Theorem~\ref{thm:unique_meanfield}, there exists unique solution satisfying the mean-field
%equations. Hence, if $(\ol{\eta}_t)_{t\geq 0}$ is the unique solution of the mean-field
%equations, then the limit point of every converging subsequence is the unique process $(\ol{\eta}_t)_{t\geq 0}$
%satisfying the mean-field equations.
% Since $\mc{C}_b^1(\mc{U})$ is a separating class of $\mc{M}_1(\mc{U})$,
%we have convergence of $(\ol{\eta}_t^N)_{t\geq 0}$ to $(\ol{\eta}_t)_{t\geq 0}$ in $\mc{D}_{\mc{M}_1(\mc{U})}([0,\infty))$.
From equation~\eqref{eq:mf_limit_point}, the process $(\chi_t,t\geq 0)$ satisfies the mean-field model equation.
This shows that the sample path of every limit point is almost surely same as the mean-field model solution.
Since every limit point $(\bm{\chi}_t,t\geq 0)$ has the property that the random measure $\bm{\chi}_0$ is
same as the random measure $\bm{\Theta}$ almost surely, all the limit points have almost surely identical sample paths
such that the sample paths coincide with the unique mean-field model solution. This completes the proof.
$$\eqno{\tsquare}$$
%%%%%%%%%%%%%%%%%%%%%%%%%%%%%%%%%%%%%%%%%%%%%%%%%%%%%%%%%%%%%%%%%%%%%%%%%%%%%%%%%%%%%%%%%%%%%%%%%%%%%%%%%%
%%%%%%%%%%%%%%%%%%%%%%%%%%%%%%%%%%%%%%%%%%%%%%%%%%%%%%%%%%%%%%%%%%%%%%%%%%%%%%%%%%%%%%%%%%%%%%%%%%%%%%%%%%%%%55
%%%%%%%%%%%%%%%%%%%%%%%%%%%%%%%%%%%%%%%%%%%%%%%%%%%%%%%%%%%%%%%%%%%%%%%%%%%%%%%%%%%%%%%%%%%%%%%%%%%%%%%%%%%%%%%%%5
\section{Insensitivity}
\label{sec:insensitivity}
\textbf{Proof of Lemma~\ref{thm:mf_pdes}:}

Suppose $\ol{\eta}_0$ is absolutely continuous $\wrt$ Lebesgue measure at all $\ul{u}\in\mc{U}_n$ for $n\geq 1$.
Then at every $t\geq 0$, we have absolutely continuity of $\ol{\eta}_t$ at all $\ul{u}\in\mc{U}_n$ for $n\geq 1$.
Suppose $p_t(0)$ denotes $\ol{\eta}_t(\{0\})$ and $p_t(n,x_1,\ldots,x_n)$ denotes the Radon-Nikodym derivative of 
$\ol{\eta}_t$ $\wrt$ Lebesgue measure at  $(n,x_1,\ldots,x_n)$. Now we obtain differential equations
satisfied by the process $P_t=(P_t(\ul{u}),\ul{u}\in\mc{U})$, 
\beq
P_t(n,y_1,\ldots,y_n)=\int_{x_1=0}^{y_1}\ldots\int_{x_n=0}^{y_n}p_t(n,x_1,\ldots,x_n)\,dx_1\cdots dx_n.
\eeq
Let us consider the function $\hat{\phi}=\indic{\ul{l}\in \mc{U}_n:\,0\leq l_i\leq y_i,\, \forall i}$.
For a absolutely continuous measure $\nu_s$ which has no atoms, we have
\beq
\langle \nu_s,\hat{\phi}\rangle=\langle\nu_s,\psi\rangle,
\eeq
where
$\psi=\indic{\ul{u}\in \mc{U}_n:\,0< l_i< y_i,\, \forall i}$. Since there exists a sequence of functions $\{f_n\}\in\mc{C}_b(\mc{U})$ that increase point wise to $\indic{O}$ where $O$ is a open set in $\mc{U}_n$, $n\geq 1$, by using
monotone convergence theorem and equation~\eqref{eq:mean_field_model_new}, we have that the equation~\eqref{eq:mean_field_model_new} is true even for the function $\psi$ (Indicators on open sets). Furthermore, since the measure $\nu_s$
is absolutely continuous for all $s\geq 0$, we have that equation~\eqref{eq:mean_field_model_new} is true even for the function $\hat{\phi}$ (Indicators on closed sets). Therefore we can obtain the evolution equations for the process $(P_t)_{t\geq 0}$ that is defined as
$P_t(n,y_1,\ldots,y_n)=\langle\nu_t,\hat{\phi}\rangle$ using equation~\eqref{eq:mean_field_model_new}.
We further can simplify expression for the process $(P_t(\ul{u}),\ul{u}\in\mc{U})_{t\geq 0}$ with the evolution
given by equation~\eqref{eq:mean_field_model_new} using the fact that
\begin{align}
\langle \nu_s, \tau_b \indic{\ul{l}\in \mc{U}_n:\,0\leq l_i\leq y_i,\, \forall i}\rangle&=\langle \nu_s, \indic{\ul{l}\in \mc{U}_n:\,0\leq l_i+b\leq y_i,\, \forall i}\rangle\\
&=\langle \nu_s, \indic{\ul{l}\in \mc{U}_n:\,0\leq l_i\leq y_i-b,\, \forall i}\rangle.
\end{align}
By differentiating $P_t(n,y_1,\ldots,y_n)$ with respect to $t$, by simple calculations, it is verified
that the process $P_t=(P_t(\ul{u}),\ul{u}\in\mc{U})$ satisfies the following system of differential equations
\beq
\frac{d P_t(0)}{dt}=\int_{y=0}^{\infty}\beta(y)\left(\frac{\partial P_t(1,y)}{\partial y}\right)\,dy-\lambda\frac{(R_0(P_t)^d-R_1^d(P_t))}{(R_0(P_t)-R_1(P_t))}P_t(0),
\eeq
for $1\leq n\leq C-1$,
\begin{multline}
\frac{dP_t(n,y_1,\ldots,y_n)}{dt}=-\sum_{i=1}^n\frac{\partial P_t(n,y_1,\ldots,y_n)}{\partial y_i}\\
+\sum_{j=1}^{n+1}\int_{x_j=0}^{\infty}\beta(x_j)\left(\frac{\partial P_t(n+1,y_1,\ldots,y_{j-1},x_j,y_j,\ldots,y_n)}{\partial x_j}\right)\,dx_j\\
-\sum_{j=1}^n\int_{x_j=0}^{y_j}\beta(x_j)\left(\frac{\partial P_t(n,y_1,\ldots,y_{j-1},x_j,y_{j+1},\ldots,y_n)}{\partial x_j}\right)\,dx_j\\
+\sum_{j=1}^n\frac{\lambda(R_{n-1}(P_t)^d-R_{n}^d(P_t))}{n(R_{n-1}(P_t)-R_{n}(P_t))}P_t(n-1,y_1,\ldots,y_{j-1},y_{j+1},\ldots,y_n)\\
-\lambda \frac{(R_{n}(P_t)^d-R_{n+1}^d(P_t))}{(R_{n}(P_t)-R_{n+1}(P_t))}P_t(n,y_1,\ldots,y_{n}),
\end{multline}
and for $n=C$,
\begin{multline}
\frac{dP_t(n,y_1,\ldots,y_n)}{dt}=-\sum_{i=1}^n\frac{\partial P_t(n,y_1,\ldots,y_n)}{\partial y_i}\\
-\sum_{j=1}^n\int_{x_j=0}^{y_j}\beta(x_j)\left(\frac{\partial P_t(n,y_1,\ldots,y_{j-1},x_j,y_{j+1},\ldots,y_n)}{\partial x_j}\right)\,dx_j\\
+\sum_{j=1}^n\frac{\lambda(R_{n-1}(P_t)^d-R_{n}^d(P_t))}{n(R_{n-1}(P_t)-R_{n}(P_t))}P_t(n-1,y_1,\ldots,y_{j-1},y_{j+1},\ldots,y_n),
\end{multline}
where $R_n(P_t)=\sum_{j=n}^CP_t(j,\infty,\ldots,\infty)$. 
$$\eqno{\tsquare}$$
%Note that for the given state
%$(n,y_1,\ldots,y_n)$, the $j^{\text{th}}$ arrivals case contributes to $P_t(n,y_1,\ldots,y_n)$ only if $y_j\leq t$
%as the age of a new arrival into the system is always less than the elapsed time $t$.
%\beq
%\eeq

\begin{rem}
If we specialize the equations above to the exponential case with mean $\frac{1}{\mu}$, we note that $\beta(x)=\mu$, and denoting $Q_t(n) = P_t(n,\infty,\ldots,\infty)$, $Q_t=(Q_t(n),0\leq n\leq C)$ and noting that:
\begin{multline}
\sum_{j=1}^n\int_{x_j=0}^{\infty}\beta(x_j)\left(\frac{\partial P_t(n,\infty,\ldots,\infty,x_j,y_{j+1},\ldots,\infty)}{\partial x_j}\right)\,dx_j 
=  n \mu P_t(n,\infty,\ldots,\infty)\\ = n \mu Q_t(n)
\end{multline}

Therefore, for $n=0$,
\beq
\frac{d Q_t(0)}{dt}= \mu Q_t(1)-\lambda\frac{(R_0(Q_t)^d-R_1^d(Q_t))}{(R_0(Q_t)-R_1(Q_t))}Q_t(0),
\eeq
for $1\leq n\leq C-1$
\begin{multline}
\frac{dQ_t(n)}{dt}=(n+1)\mu Q_t(n+1)-n \mu Q_t(n)\\
+\frac{\lambda(R_{n-1}(Q_t)^d-R_{n}^d(Q_t))}{(R_{n-1}(Q_t)-R_{n}(Q_t))}Q_t(n-1)
-\lambda \frac{(R_{n}(Q_t)^d-R_{n+1}^d(Q_t))}{(R_{n}(Q_t)-R_{n+1}(Q_t))}Q_t(n),
\end{multline}
and for $n=C$,
\begin{equation}
\frac{dQ_t (C)}{dt}=-C\mu Q_t(C)
+\frac{\lambda(R_{n-1}(Q_t)^d-R_{n}^d(Q_t))}{(R_{n-1}(Q_t)-R_{n}(Q_t))}Q_t(C-1),
\end{equation}
where $R_n(Q_t)=\sum_{j=n}^CQ_t(j)$. 

It can be readily seen that these equations correspond to the corresponding equations given in \cite{arpan} for the case of exponential distributions with rate $\mu=1$  where the corresponding equations are expressed in terms of the tail distributions $R_n(Q_t) = \sum_{k=n}^C Q_t(k)$ in the notation of that paper.
\end{rem}

\textbf{Proof of Theorem~\ref{thm:insensitivity}:}\\

%\begin{thm}
%\label{thm:insensitivity}
%There exists unique fixed-point for the process $P=(P_t(\ul{u}),t\geq 0,\ul{u}\in\mc{U})$ denoted by $\bm{\pi}$ that satisfies
%\beq
%\label{eq:fixed_pt}
%\pi(n,y_1,\ldots,y_n)=\pi^{(exp)}(n)\mu^n\prod_{i=1}^n\int_{x_i=0}^{y_i}\ol{G}(x_i)\,dx_i.
%\eeq
%where $\bm{\pi}^{(exp)}=(\pi^{(exp)}(n),0\leq n\leq C)$ denotes the unique fixed-point 
%of the mean-field when service times are exponentially distributed with mean $\frac{1}{\mu}$ and $\pi^{(exp)}(n)$ is the
%stationary probability that there are $n$ jobs in the limiting system. Further, since
%$\int_{x=0}^{\infty}\ol{G}(x)\,dx=\frac{1}{\mu}$, the fixed-point of the mean-field is insensitive as
%\beq
%\pi(n,\infty,\ldots,\infty)=\pi^{(exp)}(n).
%\eeq
%\end{thm}
\noindent{\bf Proof:}
 We next prove that there
exists unique fixed-point $\bm{\pi}=(\pi(\ul{u}),\ul{u}\in\mc{U})$. Suppose $\bm{\theta}=(\theta(\ul{u}),\ul{u}\in\mc{U})$
denotes a fixed-point for the process $(P_t)_{t\geq 0}$. We first prove that for any fixed-point
$\bm{\theta}$ of the process $(P_t)_{t\geq 0}$, we have
\beq
\label{eq:fixed_pt_condition1}
\theta(n,y_1,\ldots,y_n)=
\frac{\left(\prod_{i=1}^n\frac{\lambda^{(GEN)}_{i-1}(\bm{\theta})}{i\mu}\right)}{1+\sum_{m=1}^C\left(\prod_{i=1}^m\frac{\lambda^{(GEN)}_{i-1}(\bm{\theta})}{i\mu}\right)}\mu^n\prod_{i=1}^n\int_{x_i=0}^{y_i}\ol{G}(x_i)\,dx_i
\eeq
and
\beq
\label{eq:fixed_pt_condition2}
\theta(0)=\frac{1}{1+\sum_{m=1}^C\left(\prod_{i=1}^m\frac{\lambda^{(GEN)}_{i-1}(\bm{\theta})}{i\mu}\right)}
\eeq
where $\lambda^{(GEN)}_n(\bm{\theta})=\lambda\frac{R_n(\bm{\theta})^d-R_{n+1}(\bm{\theta})^d}{R_n(\bm{\theta})-R_{n+1}(\bm{\theta})}$.
We prove this by contradiction. Suppose there exists a fixed-point $\bm{\gamma}$ that does not satisfy equations~\eqref{eq:fixed_pt_condition1}-\eqref{eq:fixed_pt_condition2}. Using this fixed-point $\bm{\gamma}$, we first compute the set of arrival rates
$(\lambda^{(GEN)}_i(\bm{\gamma}),\ \ 0\leq i\leq C)$. Now consider a single server loss system where prespecified
state-dependent arrival rate is equal to $\lambda^{(GEN)}_i(\bm{\gamma})$ when there are $i$ jobs in progress and service time distributions are same as considered in the system model. Then the unique stationary distribution is given by equation~\eqref{eq:single_server_fixed_pt} where $\alpha_i$ is replaced by $\lambda^{(GEN)}_i(\bm{\gamma})$. However, by comparing the stationary evolution equations corresponding to single server dynamics given in equations~\eqref{eq:single_pde1}-\eqref{eq:single_pde3} and mean-field dynamics given in equations~\eqref{eq:mf_pdes1}-\eqref{eq:mf_pdes3}, it
is easy to see that $\bm{\gamma}$ is also another stationary distribution for single server system with pre-specified arrival rates $(\lambda^{(GEN)}_i(\bm{\gamma}),0\leq i\leq C)$ which contradicts the fact there exists unique stationary distribution for single server system with pre-specified state-dependent arrival rates established in \cite{shelby}. Therefore equations~\eqref{eq:fixed_pt_condition1}-\eqref{eq:fixed_pt_condition2} must be true.

%\eeq

For any fixed-point $\bm{\theta}$ of the mean-field, let $\Gamma=(\Gamma_n,0\leq n\leq C)$ such that $\Gamma_n=\theta(n,\infty,\ldots,\infty))$ and $\Gamma_0=\theta(0)$. Then from equations~\eqref{eq:fixed_pt_condition1}-\eqref{eq:fixed_pt_condition2}, we have
\beq
\label{eq:exp_fixed_pt_condition1}
\Gamma_n=
\frac{\left(\prod_{i=1}^n\frac{\lambda^{(exp)}_{i-1}(\bm{\Gamma})}{i\mu}\right)}{1+\sum_{m=1}^C\left(\prod_{i=1}^m\frac{\lambda^{(exp)}_{i-1}(\bm{\Gamma})}{i\mu}\right)}
\eeq
and
\beq
\label{eq:exp_fixed_pt_condition2}
\Gamma_0=\frac{1}{1+\sum_{m=1}^C\left(\prod_{i=1}^m\frac{\lambda^{(exp)}_{i-1}(\bm{\Gamma})}{i\mu}\right)}
\eeq
where
\beq
\label{eq:exp_arrival_rate}
\lambda^{(exp)}_n(\Gamma)=\lambda\frac{(\sum_{j=n}^C\Gamma_j)^d-(\sum_{j=n+1}^C\Gamma_j)^d}{(\sum_{j=n}^C\Gamma_j)-(\sum_{j=n+1}^C\Gamma_j)}.
\eeq
From equation~\eqref{eq:exp_fixed_pt_condition1}, it is clear that
\beq
\label{eq:exp_iter}
\lambda^{(exp)}_n(\Gamma)\Gamma_n=(n+1)\mu\Gamma_{n+1}.
\eeq
It has been shown in \cite{arpan} that the only probability measure satisfying equations~\eqref{eq:exp_arrival_rate}-\eqref{eq:exp_iter} is the unique fixed-point $\bm{\pi}^{(exp)}$ of the mean-field when job lengths are exponentially 
distributed with mean $\frac{1}{\mu}$. Therefore from equations \eqref{eq:fixed_pt_condition1} and \eqref{eq:exp_fixed_pt_condition1}, we have for every fixed point $\bm{\theta}$, 
\beq
\theta(n,\infty,\ldots,\infty)=\pi^{(exp)}(n)
\eeq
concluding the insensitivity of the fixed-point of the mean-field. Further, from equation~\eqref{eq:fixed_pt_condition1},
for every fixed point $\bm{\theta}$, we have
\beq
\theta(n,y_1,\ldots,y_n)=
\pi^{(exp)}(n)\mu^n\prod_{i=1}^n\int_{x_i=0}^{y_i}\ol{G}(x_i)\,dx_i
\eeq
concluding uniqueness of the fixed-point of the mean-field as $\bm{\pi}^{(exp)}$ is unique.
$$\eqno{\tsquare}$$
%***********************************************************************************************
%*************************************************************************************************
%**************************************************************************************************
%*******************************************************************************************************
\section{Extensions}
\label{sec:extensions}

Although we have considered a homogeneous system where all servers have same capacity $C$, at the cost of more complex notation, the analysis
can be easily extended to a heterogeneous system where servers are classified into different types based on their capacities. We only state the results without proofs.

Suppose servers are classified into $K$ types such that the fraction of type~$k$ servers is $\gamma_k$ and each type~$k$ server has capacity $C_k$. In other words there are $N\gamma_k$ servers with capacity $C_k$. Assume without loss of generality $C_1\leq C_2\leq\cdots\leq C_K$. This model with exponential service times was studied in \cite{arpan}.\\

\noindent\textbf{Power-of-$d$ policy for heterogeneous systems:}

In order to account for the heterogeneity in servers' capacities, the power-of-$d$ policy defined in Definition~\ref{def:power_of_d} needs slight modification. Now, an arriving job is routed
to a server with maximum vacancy (available number or servers) among $d$ randomly chosen servers. Ties among servers of the same type are broken
by choosing a server uniformly at random and among different types are broken by choosing a server
with maximum capacity.

For $n\geq 1$, let 
\beq
\mc{U}_n^{(k)}=\{(k,n,x_1,\ldots,x_n):x_i\in \mc{R}_+,\forall i\}
\eeq
denotes the set of all possible
type~k server states when there are $n$ jobs in progress
and 
\beq
\mc{U}_0^{(k)}=\{(k,0)\}.
\eeq
%denotes the type~$k$ server state when there are no jobs in progress.
%\beq
%\mc{U}^{(k)}=\{(k,n,x_1,\ldots,x_n):x_i\geq 0 \text{ for } 0\leq i\leq n, 0\leq n\leq C_k\}
%\eeq
% contains all possible
%type~k server states and $(k,0)$ is used to denote the state when there are no jobs. 
Then
\beq
\mc{U}^{(k)}=\cup_{n=0}^{C_k}\mc{U}_n^{(k)}
\eeq
denotes the set of all possible states of a type~$k$ server.
%and
%\beq
%\mc{U}=\prod_{k=1}^K\mc{U}^{(k)}.
%\eeq
%Further, let
%
%denotes the set of type~$k$ server states when there are $n$ jobs in progress. Then 
%\beq
%\mc{U}=\prod_{k=1}^K(\cup_n\mc{U}^{(k)}_n).
%\eeq

Similar to the homogeneous case, for the Markovian modelling of the system, it is enough for us to track for each type~$k$, 
the number of servers lying in each state $\ul{u}\in\mc{U}^{(k)}$. For fine $N$ systems, the system evolution is described
through the dynamics of the process, for every $k\in\{1,2,\ldots,K\}$,
\beq
\ol{\eta}^N_{k,t}=\frac{1}{N\gamma_k}\sum_{j=1}^{N\gamma_k}\delta_{(\ul{u}^{(j)})},
\eeq
where $\ul{u}^{(j)}$ denotes the state of $j^{\text{th}}$ server of type~$k$ at time $t$. Therefore, at any time $t$,
the system state is defined through the set of measures $(\ol{\eta}^N_{k,t},k\in{\{1,\ldots,K\}})$ where $\ol{\eta}^N_{k,t}$
is a probability measure defined on $\mc{U}^{(k)}$. Note that for $\ul{u}\in\mc{U}^{(k)}$, $\ol{\eta}_{k,t}^N(\{\ul{u}\})=\langle \ol{\eta}_{k,t}^N,\indic{\ul{u}}\rangle$ denotes the fraction of type~$k$ servers lying in state $\ul{u}$. We refer
$\ol{\eta}_{k,t}^N$ as type~$k$ probability measure at time $t$. Further, let
\beq
\eta^N_{k,t}=\sum_{j=1}^{N\gamma_k}\delta_{(\ul{u}^{(j)})}
\eeq
and hence
\beq
\ol{\eta}^N_{k,t}=\frac{1}{N\gamma_k}\eta^N_{k,t}.
\eeq
 Note that servers of different types or equivalently the probability measures of different types interact at the
 arrival instants while applying the power-of-$d$ routing policy.

To simplify the analysis, we next model the system evolution by defining a single
measure $\ol{\eta}^N_t$ such that it is equivalent to modeling the system evolution by
the set of measures $(\ol{\eta}^N_{k,t},k\in{\{1,\ldots,K\}})$. For this, we first define
\beq
\mc{U}=\cup_{k=1}^K\mc{U}^{(k)}.
\eeq
The space $\mc{U}$ is equipped with the metric $d_{\mc{U}}$, for $\ul{u}=(k,n,x_1,\ldots,x_n)$ $($\ie$, \ul{u}\in\mc{U}^{(k)})$
and $\ul{v}=(j,m,y_1,\ldots,y_m)$ (\ie, $\ul{v}\in\mc{U}^{(j)}$),
\beq
d_{\mc{U}}(\ul{u},\ul{v})=
\begin{cases}
\sum_{r=1}^n\abs{x_r-y_r} & \text{ if } j=k \text{ and }n=m,\\
\infty             & \text{otherwise}.
\end{cases}
\eeq
We next consider the measures $\ol{\eta}_t^N$, $\eta_t^N$ defined on $\mc{U}$
satisfying, for $\ul{u}\in\mc{U}^{(k)}$
\beq
\ol{\eta}_t^N(\{\ul{u}\})=\ol{\eta}_{k,t}^N(\{\ul{u}\})
\eeq
and 
\beq
\eta_t^N(\{\ul{u}\})=\eta_{k,t}^N(\{\ul{u}\}).
\eeq
Therefore the measures $\ol{\eta}_t^N$, $\eta_t^N$ restricted to the space $\mc{U}^{(k)}$
are same as the measures $\ol{\eta}_{k,t}^N$, $\eta_{k,t}^N$, respectively. Further, note that
\beq
\eta_t^N(\mc{U})=N
\eeq
and 
\beq
\ol{\eta}_t^N(\mc{U})=K.
\eeq

%\beq
%%\label{eq:metric}
%\indic{B}(\ul{u})
%=
%\begin{cases}
%1 & \text{if }  \ul{u}\in B\\
%0      & \text{otherwise}.
%\end{cases}
%\eeq

%Further, we denote
%\beq
%\ol{\eta}^N_t=(\ol{\eta}^N_{k,t},k\in{\{1,\ldots,K\}}).
%\eeq

Now for any function $\psi:\mc{U}\mapsto \mc{R}$, the type~$k$ function $\psi^{(k)}:\mc{U}^{(k)}\mapsto\mc{R}$
is defined by
\beq
\psi^{(k)}(\ul{u})=\psi(\ul{u})
\eeq 
for $\ul{u}\in\mc{U}^{(k)}$. The $n^{\text{th}}$ component of $\psi^{(k)}$ is denoted by $\psi^{(k,n)}$. Then we define
\beq
\langle \eta_t^N,\psi\rangle=\sum_{k=1}^K\langle \eta_{k,t}^N,\psi^{(k)}\rangle.
\eeq

The process $(\eta_t^N,t\geq 0)$ that models the heterogeneous system defined on $\mc{U}$ is a Markov process.
The analysis is then follows from the same arguments as that of the homogeneous case. We first need to obtain the
result stated in Proposition~\ref{thm:martingale} for the heterogeneous system by using the fact that
the process $(\eta_t^N,t\geq 0)$ is a Markov process and then we need to repeat the steps stated in Section~\ref{sec:MFEproof}. Note that if we know $\langle \ol{\eta}_t,\psi\rangle$ for all $\psi\in\mc{C}_b^1(\mc{U})$,
then we can obtain $\langle \ol{\eta}_{k,t},\phi\rangle$ for all $\phi\in\mc{C}_b^1(\mc{U}^{(k)})$ from $\langle \ol{\eta}_t,\psi\rangle$ by simply choosing $\psi^{(k)}=\phi$ and $\psi^{(j)}=0$ for $j\neq k$.
Then for $\phi\in\mc{C}_b^1(\mc{U}^{(k)})$, the Kolmogorov equations are
%\item\label{property2} For $\phi\in \mc{C}_b^1(\mc{U})$, the process $(\ol{\eta}_t,t\geq 0)$ satisfies
\begin{multline}
\label{eq:mf_model_eqns_heterogeneous}
\langle\ol{\eta}_{k,t},\phi\rangle=\langle\ol{\eta}_{k,0},\phi\rangle+\int_{s=0}^t \langle\ol{\eta}_{k,s},\phi'\rangle\,ds\\
-\int_{s=0}^t\Bigg(\sum_{n=1}^C\sum_{j=1}^n\int_{x_1}\cdots\int_{x_n}\beta(x_j)\\
\times\left(\phi(k,n-1,x_1,\ldots,x_{j-1},x_{j+1},\ldots,x_n)-\phi(k,n,x_1,\ldots,x_n)\right)\,d\ol{\eta}_{k,s}(k,n,x_1,\ldots,x_n)\\
+\bigg[\left(\ol{\eta}_{k,s}(\{k,0\})\lambda_{k,0}(\ol{\eta}_s)\left(\phi(k,1,0)-\phi(k,0)\right)\right)
+\sum_{n=1}^{C-1}\sum_{j=1}^{n+1}\int_{x_1}\cdots\int_{x_n}\frac{1}{(n+1)}\\
\times\lambda_{k,n}(\ol{\eta}_s)
(\phi(k,n+1,x_1,\ldots,x_{j-1},0,x_j,\ldots,x_n)-\phi(k,n,x_1,\ldots,x_n))\\
\times\,d\ol{\eta}_{k,s}(k,n,x_1,\ldots,x_n)\bigg]\Bigg)ds,
\end{multline}
%\end{enumerate}
where 
\begin{multline}\lambda_{k,n}(\ol{\eta}_s)=\frac{\lambda}{\gamma_k\ol{\eta}_{k,s}(\mc{U}_n^{(k)})}\\
\times\left( \left(\sum_{i=1}^k\gamma_i\left(\sum_{j=n}^{C_i}\ol{\eta}_{i,s}(\mc{U}_{j+C_i-C_k}^{(i)})\right)+\sum_{i=k+1}^K\gamma_i\left(\sum_{j=n+1}^{C_i}\ol{\eta}_{i,s}(\mc{U}_{j+C_i-C_k}^{(i)})\right)\right)^d\right.\\
\left.-\left(\sum_{i=1}^{k-1}\gamma_i\left(\sum_{j=n}^{C_i}\ol{\eta}_{i,s}(\mc{U}_{j+C_i-C_k}^{(i)})\right)+\sum_{i=k}^K\gamma_i\left(\sum_{j=n+1}^{C_i}\ol{\eta}_{i,s}(\mc{U}_{j+C_i-C_k}^{(i)})\right)\right)^d
\right).
\end{multline}

Suppose the Radon-Nikodym derivative of the measure $\ol{\eta}_{k,t}$ with respect to Lebesgue measure at $(k,n,x_1,\ldots,x_n)$
is $p_t(k,n,x_1,\ldots,x_n)$. Let  \[\bm{P}_t=(P_t(k,n,y_1,\ldots,y_n),0\leq n\leq C_k, y_i\in \mc{R}_+ \text{ for all } i, 1\leq k\leq K)\]
where 
\beq P_t(k,n,y_1,\ldots,y_n)=\int_{x_1=0}^{y_1}\cdots\int_{x_n=0}^{y_n}p_t(k,n,x_1,\ldots,x_n)\,dx_1\cdots dx_n.
\eeq
Then the process $\bm{P}_t$ has unique fixed-point $\bm{\pi}=(\pi(k,n,y_1,\ldots,y_n), 0\leq n\leq C_k, y_i \text{ for all } i, 1\leq k\leq K)$
that satisfies
\beq
\pi(k,n,y_1,\ldots,y_n)=\pi^{(exp)}_{k,n}\mu^n\prod_{i=1}^n\int_{x_i=0}^{y_i}\ol{G}(x_i)\,dx_i
\eeq
where $\pi^{(exp)}_{k,n}$ denotes the stationary probability that a type~$k$ server has $n$ jobs
in the limiting system under the assumption of exponential service
time distributions with mean $\frac{1}{\mu}$.

\section{Numerical results}
\label{sec:numerics}

One of the results we have not established in the paper is a proof of Step 4 of the commutative diagram. If one can establish that the equilibrium or fixed-point of the MFE is globally asymptotically stable, then Step 4 would follow from Prohorov's theorem, see  \cite{Billing}. The difficulty in this case is that the MFE equation does not possess any monotonicity properties. Nevertheless in numerical studies the global asymptotic stability property always seems to hold thus indicating that the fixed point is indeed the stationary distribution of the asymptotic limit of the system. Thus in this section we provide some numerical evidence for the global asymptotic stability of the fixed point.

%Since mixed-Erlang distributions are dense in the class of general distributions on $(0,\infty)$\cite{asmussen},
For computational tractibility we assume that the service time distributions are mixed-Erlang. These distributions are sufficiently rich as they are dense in the class of general distributions \cite{asmussen}. The advantage is that the MFEs reduce to ODEs that canbe solved by using numerical methods to obtain approximate solutions.

We consider the system parameters as follows: The capacities of servers are assumed to be $C=5$. The average job length is assumed to be equal to one, \ie\, $\mu=1$. The service times follow a Mixed-Erlang distribution given by sums of independent exponentially distributed random variables (known as an Erlang distribution) where the number of exponential phases (or independent random exponentials) is equal to $i\in \{1,2,\ldots,M\}$ with probability $p_i$ such that $\sum_{i=1}^Mp_i=1$. Each exponential phase is assumed to have rate $\mu_p$. Therefore, we have,
 \beq\frac{1}{\mu}=\frac{\sum_{i=1}^Mip_i}{\mu_p}.\label{eq:mean}\eeq 
 We choose $M=3$, $p_1=.3,p_2=0.3,p_3=0.4$.
 
 Under mixed-Erlang service time distribution assumptions, let $S$ be the set of all
 possible server states given by
 \beq
 S=\cup_{n=0}^CS_n
 \eeq 
 where
 $S_0=\{(0,0,\ldots,0)\}$ and $S_n=\{(l_1,\ldots,l_n,0,\ldots,0):1\leq l_i\leq M, 1\leq i\leq n\}$. We
 define $\ul{l}=(l_1,\ldots,l_C)$ and for $\ul{l}\in S_n$, we have $l_i=0$ for $i>n$.
 The system dynamics can be modeled as a Markov process $\mf{x}^N(t)=(x^N_{\ul{l}}(t),\ul{l}\in S)$ where $x^N_{(l_1,\ldots,l_n,0,\ldots,0)}(t)$ denotes the fraction of servers
 with $n$ jobs such that $i^{\text{th}}$ job has $l_i$ remaining phases at time $t$. Since the Markov process $\mf{x}^N(t)$
 is defined on finite dimensional space, we can establish the mean-field limit $\mf{x}(t)=(x_{\ul{l}}(t),\ul{l}\in S)$
 by using the same procedure as that of the exponential service times case in \cite{arpan}. Hence we recall the following result without proof from \cite{Thiru}.
\begin{thm}
\label{thm:mean-field}
If $\mf{x}^N(0)$ converges in distribution to a state $\mf{u}$, then
the process $\mf{x}^N(\cdot)$ converges in distribution to a deterministic process $\mf{x}(\cdot,\mf{u})$ as $N\to\infty$ called the mean-field. The process $\mf{x}(\cdot,\mf{u})$ is the unique solution of the following system of differential equations.
\begin{align}
\mathbf{x}(0,\mathbf{u})&=\mathbf{u},\label{eq:fmean-field}\\
{{\dot{x}_{\ul{l}}}}(t,\mathbf{u})&={h_{\ul{l}}}(\mathbf{x}(t,\mathbf{u})),
\end{align} 
and $\mf{h}=(h_{\ul{l}},\ul{l}\in S)$ with the mapping $h_{\ul{l}}$ given by
\begin{multline}
\label{eq:smean-field}
h_{\underline{l}}\mathbf{(x)}=\sum_{b=1}^{Z(\ul{l})}\left(\frac{ p_{l_b}}{Z(\ul{l})}\right)x_{(l_1,l_2,\ldots,\,l_{b-1},l_{b+1},\ldots,\,l_{C},0,)}\\
\times \lambda^{(ME)}_{Z(\ul{l})-1}(\mf{x})
- x_{\ul{l}}\lambda^{(ME)}_{Z(\ul{l})}(\mf{x})
\indic{Z(\ul{l})<C}\\
+\sum_{b=1}^{Z(\ul{l})+1}\mu_p\indic{Z(\ul{l})<C}x_{(l_1,\ldots,l_{b-1},1,l_b,\ldots,\,l_{C-1})}\\
+\sum_{b=1}^{Z(\ul{l})}\mu_px_{(l_1,\ldots,l_{b-1},l_b+1,l_{b+1},\ldots,\,l_{C})}-Z(\ul{l})\mu_px_{\ul{l}},
\end{multline}
\end{thm}
where $Z(\ul{l})$ denotes the number of nonzero elements in $\ul{l}\in S$ and
\begin{multline}
\label{eq:arrival_rate}
\lambda^{(ME)}_{n}(\mf{u})=\frac{\lambda}{(\sum_{\ul{l}:Z(\ul{l})=n}u_{\ul{l}})}\\\times\left[ \left(\sum_{i=n}^C\sum_{\ul{l}:Z(\ul{l})=i}u_{\ul{l}}\right)^d-\left(\sum_{i=n+1}^C\sum_{\ul{l}:Z(\ul{l})=i}u_{\ul{l}}\right)^d
\right].
\end{multline}

\begin{figure}[h]
 \centering
 \includegraphics[height=0.4\columnwidth]{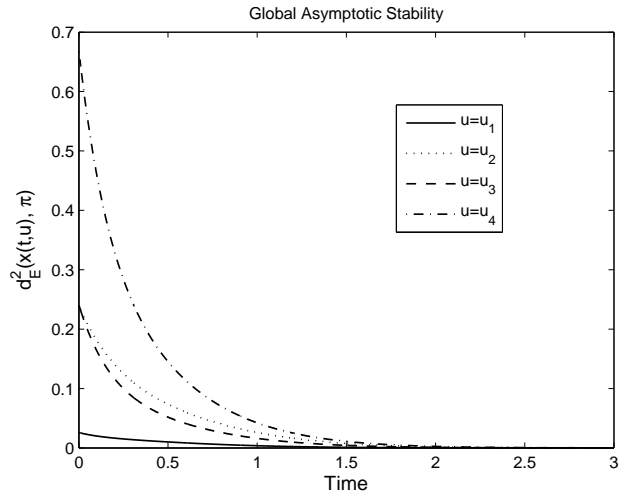}
 \caption{Convergence of mean-field to the fixed-point}
 \label{fig:global_stability}
 \end{figure}

In Figure~\ref{fig:global_stability}, we plot $d_{E}^2(\mf{x}(t,\mf{u}),\mf{\pi})$ as a function of $t$
where $d_{E}$ is the euclidean distance defined by
\beq
d_{E}(\mf{u},\mf{v})=\left(\sum_{\ul{l}\in S}\abs{u_{\ul{l}}-v_{\ul{l}}}^2\right)^{\frac{1}{2}}.
\eeq

It is observed that for $d=2$, $\lambda=1$, and for four different initial points $\mf{u}_1,\mf{u}_2,\mf{u}_3$, and $\mf{u}_4$, the mean-field $\mf{x}(t,\mf{u})$ for mixed-Erlang service time distribution converges to its unique fixed-point $\bm{\pi}$. Note that the computed $\bm{\pi}$ depends on the chosen value of $d$. This provides evidence that $\bm{\pi}$ is globally stable.

\section{Concluding Remarks}
\label{sec:conclusion}.

In this paper we have provided a measure-valued process approach to establish the mean-field behavior of loss systems with Power-of-$d$ routing and general service time requirements. The extension of these results to multi-class systems is also of interest and these follow in a similar manner {\em mutatis mutandis} from the approach used here. The extensions to other disciplines such as processor sharing are also of interest and the measure-valued approach used here is most appropriate. One open problem in these classes of problems is establishing the global asymptotic stability of the fixed point when unique. The standard theory does not apply as the MFEs describe a class of non-linear Markov processes on $\mc{R}_+^n$ without any obvious monotonicity properties and perhaps one way is to study the dissipative properties of the non-linear semi-groups of the Markov processes.
\bibliographystyle{apt}

\begin{thebibliography}{10}

\bibitem{reza}
{\sc {Aghajani}, R. and {Ramanan}, K.} (2017).
\newblock {The hydrodynamic limit of a randomized load balancing network}.
\newblock {\em ArXiv e-prints\/}.

\bibitem{AmazonEC2}
{\sc Amazon}.
\newblock Amazon {EC2}.
\newblock \url{http://aws.amazon.com/ec2/}.

\bibitem{azar}
{\sc Azar, Y., Broder, A.~Z., Karlin, A.~R. and Upfal, E.} (1999).
\newblock Balanced allocations.
\newblock {\em SIAM J. Comput.\/} {\bf 29,} 180--200.

\bibitem{Billing}
{\sc Billingsley, P.} (1999).
\newblock {\em Convergence of probability measures} second~ed.
\newblock Wiley Series in Probability and Statistics: Probability and
  Statistics. John Wiley \& Sons, Inc., New York.
\newblock A Wiley-Interscience Publication.

\bibitem{bramson1}
{\sc Bramson, M., Lu, Y. and Prabhakar, B.} (2010).
\newblock {Randomized load balancing with general service time distributions}.
\newblock In {\em Proceedings of ACM SIGMETRICS}.
\newblock pp.~275--286.

\bibitem{bramson2}
{\sc Bramson, M., Lu, Y. and Prabhakar, B.} (2012).
\newblock Asymptotic independence of queues under randomized load balancing.
\newblock {\em Queueing Systems\/} {\bf 71,} 247--292.

\bibitem{brown}
{\sc Brown, L., Gans, N., Mandelbaum, A., Sakov, A., Shen, H., Zeltyn, S. and
  Zhao, L.} (2005).
\newblock Statistical analysis of a telephone call center.
\newblock {\em Journal of the American Statistical Association\/} {\bf 100,}
  36--50.

\bibitem{shelby}
{\sc Brumelle, S.~L.} (1978).
\newblock A generalization of erlang's loss system to state dependent arrival
  and service rates.
\newblock {\em Mathematics of Operations Research\/} {\bf 3,} 10--16.

\bibitem{dawson}
{\sc Dawson, D.~A.} (1993).
\newblock {\em Measure-valued Markov processes} vol.~1541 of {\em {\'Ecole
  d'\'Et\'e de Probabilit\'es de Saint-Flour XXI---1991}}.
\newblock Springer, Berlin.

\bibitem{moyal}
{\sc Decreusefond, L. and Moyal, P.} (2008).
\newblock A functional central limit theorem for the m/gi/$\infty$queue.
\newblock {\em Ann. Appl. Probab.\/} {\bf 18,} 2156--2178.

\bibitem{Ethier_Kurtz_book}
{\sc Ethier, S.~N. and Kurtz, T.~G.} (1985).
\newblock {\em Markov Processes: Characterization and Convergence}.
\newblock John Wiley and Sons Ltd.

\bibitem{Graham_loss}
{\sc Graham, C. and M\'el\'eard, S.} (1993).
\newblock Propagation of chaos for a fully connected loss network with
  alternate routing.
\newblock {\em Stochastic Processes and their Applications\/} {\bf 44,}
  159--180.

\bibitem{Graham1}
{\sc Graham, C. and M\'el\'eard, S.} (1997).
\newblock Stochastic particle approximations for generalized boltzmann models
  and convergence estimates.
\newblock {\em The Annals of Probability\/} {\bf 28,} 115--132.

\bibitem{gromoll_ps}
{\sc Gromoll, H.~C., Puha, A.~L. and Williams, R.~J.} (2002).
\newblock The fluid limit of a heavily loaded processor sharing queue.
\newblock {\em Ann. Appl. Probab.\/} {\bf 12,} 797--859.

\bibitem{gromoll_ps_impatient}
{\sc Gromoll, H.~C., Robert, P. and Zwart, B.} (2008).
\newblock Fluid limits for processor-sharing queues with impatience.
\newblock {\em Math. Oper. Res.\/} {\bf 33,} 375--402.

\bibitem{jakubowski}
{\sc Jakubowski, A.} (1986).
\newblock On the skorokhod topology.
\newblock {\em Annales de l'I.H.P. Probabilit\'es et Statistiques\/} {\bf 22,}
  263--285.

\bibitem{kallenberg}
{\sc Kallenberg, O.} (1983).
\newblock {\em Random measures}.
\newblock Akademie-Verlag.

\bibitem{kang2010}
{\sc Kang, W. and Ramanan, K.} (2010).
\newblock Fluid limits of many-server queues with reneging.
\newblock {\em Ann. Appl. Probab.\/} {\bf 20,} 2204--2260.

\bibitem{rybko}
{\sc Karpelevich, F.~I. and Rybko, A.~N.} (2000).
\newblock Thermodynamic limit for the mean field model of simple symmetrical
  closed queueing network.
\newblock {\em Markov Processes and Related Fields\/} {\bf 6,} 89--105.

\bibitem{kaspi2011}
{\sc Kaspi, H. and Ramanan, K.} (2011).
\newblock Law of large numbers limits for many-server queues.
\newblock {\em Ann. Appl. Probab.\/} {\bf 21,} 33--114.

\bibitem{peter}
{\sc Kolesar, P.} (1984).
\newblock Stalking the endangered cat: A queueing analysis of congestion at
  automatic teller machines.
\newblock {\em Interfaces\/} {\bf 14,} 16--26.

\bibitem{Azure}
{\sc Microsoft}.
\newblock Microsoft {A}zure.
\newblock \url{http://www.microsoft.com/windowsazure/}.

\bibitem{mit}
{\sc Mitzenmacher, M.} (1996).
\newblock The power of two choices in randomized load balancing.
\newblock {\em PhD Thesis, Berkeley\/}.

\bibitem{Arpan_ITC_2014}
{\sc Mukhopadhyay, A. and Mazumdar, R.~R.} (2014).
\newblock Rate-based randomized routing in large heterogeneous processor
  sharing systems.
\newblock In {\em Proceedings of 26th International Teletraffic Congress (ITC
  26)}.

\bibitem{Arpan_TCNS}
{\sc Mukhopadhyay, A. and Mazumdar, R.~R.} (2016).
\newblock Analysis of randomized join-the-shortest-queue (jsq) schemes in large
  heterogeneous processor sharing systems.
\newblock {\em IEEE Transactions on Control of Network Systems\/} {\bf 3(2),}
  116--126.

\bibitem{arpan}
{\sc Mukhopadhyay, A., Mazumdar, R.~R. and Guillemin, F.} (2015).
\newblock The power of randomized routing in heterogeneous loss systems.
\newblock In {\em Teletraffic Congress (ITC 27), 2015 27th International}.
\newblock pp.~125--133.

\bibitem{arpan2}
{\sc Mukhopadhyayay, A., Karthik, A., Mazumdar, R.~R. and Guillemin, F.~M.}
  (September 2015).
\newblock Mean field and propagation of chaos in multi-class heterogeneous loss
  models.
\newblock {\em Performance Evaluation\/} {\bf 91,} 117--131.

\bibitem{roberts}
{\sc Robert, P.} (2003).
\newblock Stochastic Modelling and Applied Probability Series. Springer-Verlag.

\bibitem{rudin}
{\sc Rudin, W.} (1987).
\newblock {\em Real and complex analysis} third~ed.
\newblock McGraw-Hill Book Co., New York.

\bibitem{sevastyanov}
{\sc Sevasta��yanov, B.~A.} (1957).
\newblock An ergodic theorem for markov processes and its application to
  telephone systems with refusals.
\newblock {\em Theory of Probability \& Its Applications\/} {\bf 2,} 104--112.

\bibitem{Turner_thesis}
{\sc Turner, S.~R.~E.} (1996).
\newblock Resource pooling in stochastic networks.
\newblock {\em Ph.D. dissertation, University of Cambridge\/}.

\bibitem{Turner_choices_1998}
{\sc Turner, S.~R.~E.} (1998).
\newblock The effect of increasing routing choice on resource pooling.
\newblock {\em Probability in the Engineering and Informational Sciences\/}
  {\bf 12,} 109--124.

\bibitem{Varad}
{\sc Varadarajan, V.} (1959)).
\newblock On a theorem of f. riesz concerning the form of linear functionals.
\newblock {\em Fund. Math.\/} {\bf 46,} 209--220.

\bibitem{Thiru}
{\sc Vasantam, T., Mukhopadhyay, A. and Mazumdar, R.~R.}
\newblock Mean field analysis of loss models with mixed-erlang distributions
  under power-of-d routing.
\newblock Accepted for ITC~29, Genoa, Italy, Sept. 2017 2017.

\bibitem{Vvedenskaya_inftran_1996}
{\sc Vvedenskaya, N.~D., Dobrushin, R.~L. and Karpelevich, F.~I.} (1996).
\newblock Queueing system with selection of the shortest of two queues: an
  asymptotic approach.
\newblock {\em Problems of Information Transmission\/} {\bf 32,} 20--34.

\bibitem{xie}
{\sc Xie, Q., Dong, X., Lu, Y. and Srikant, R.} (2015).
\newblock Power of d choices for large-scale bin packing: A loss model.
\newblock In {\em Proceedings of the 2015 ACM SIGMETRICS}.
\newblock pp.~321--334.

\bibitem{zhang}
{\sc Zhang, J.} (2013).
\newblock Fluid models of many-server queues with abandonment.
\newblock {\em Queueing Systems\/} {\bf 73,} 147--193.

\end{thebibliography}

\end{document}